\documentclass[manuscript]{emulateapj}
\usepackage{graphicx}
\usepackage{xspace}
\usepackage{rotating}
 \usepackage{amsmath, amsthm, amssymb}

\renewcommand{\thefootnote}{\fnsymbol{footnote}}

\slugcomment{To appear in The Astrophysical Journal}

\shorttitle{Kinematics of Ultra-compact Dwarfs in M87}
\shortauthors{Zhang, Peng, C\^ot\'e et al.}

\begin{document}

\title{The Next Generation Virgo Cluster Survey. VI. The Kinematics of Ultra-compact Dwarfs and Globular Clusters in M87}

\author{
Hong-Xin Zhang$^{1,2,3,4}$, 
Eric, W. Peng$^{1,2}$,
Patrick C\^ot\'e$^{5}$, 
Chengze Liu$^{6,7}$, 
Laura Ferrarese$^{5}$,
Jean-Charles Cuillandre$^{8}$,
Nelson Caldwell$^{9}$,
Stephen D. J. Gwyn$^{5}$,
Andr\'es Jord\'an$^{10}$,
Ariane Lan\c{c}on$^{11}$,
Biao Li$^{1,2}$,
Roberto P. Mu\~noz$^{10,11}$,
Thomas H. Puzia$^{10}$,
Kenji Bekki$^{12}$,
John Blakeslee$^{5}$, 
Alessandro Boselli$^{13}$,
Michael J. Drinkwater$^{14}$,
Pierre-Alain Duc$^{15}$,
Patrick Durrell$^{16}$,
Eric Emsellem$^{17,18}$,
Peter Firth$^{14}$,
Ruben S\'anchez-Janssen$^{5}$
}
\affil{
$^{1}$Department of Astronomy, Peking University, Beijing 100871, China; hongxin@pku.edu.cn; peng@pku.edu.cn\\
$^{2}$Kavli Institute for Astronomy and Astrophysics, Peking University, Beijing 100871, China\\
$^{3}$CAS-CONICYT Fellow\\
$^{4}$Chinese Academy of Sciences South America Center for Astronomy, Camino EI Observatorio \#1515, Las Condes, Santiago, Chile\\
$^{5}$National Research Council of Canada, Herzberg Astronomy and Astrophysics Program, 5071 West Saanich Road, Victoria, BC V9E 2E7, Canada\\
$^{6}$Center for Astronomy and Astrophysics, Department of Physics and Astronomy, Shanghai Jiao Tong University, Shanghai 200240, China\\
$^{7}$Shanghai Key Lab for Particle Physics and Cosmology, Shanghai Jiao Tong University, Shanghai 200240, China\\
$^{8}$Canada--France--Hawaii Telescope Corporation, Kamuela, HI 96743, USA\\
$^{9}$Harvard-Smithsonian Center for Astrophysics, Cambridge, MA, 02138\\
$^{10}$Instituto de Astrof\'isica, Facultad de F\'isica, Pontificia Universidad Cat\'olica de Chile, Av. Vicu\~na Mackenna 4860, 7820436 Macul, Santiago, Chile\\
$^{11}$Observatoire astronomique de Strasbourg, Universit\'e de Strasbourg, CNRS, UMR 7550, 11 rue de l$'$Universite, F-67000 Strasbourg, France\\
$^{12}$School of Physics, University of New South Wales, Sydney 2052, NSW, Australia\\
$^{13}$Aix Marseille Universit\'e, CNRS, LAM (Laboratoire d$'$Astrophysique de Marseille) UMR 7326, F-13388 Marseille, France\\
$^{14}$School of Mathematics and Physics, University of Queensland, Brisbane, QLD 4072, Australia\\
$^{15}$Laboratoire AIM Paris-Saclay, CNRS/INSU, Universit\'e Paris Diderot, CEA/IRFU/SAp, F-91191 Gif-sur-Yvette Cedex, France\\
$^{16}$Department of Physics \& Astronomy, Youngstown State University, Youngstown, OH 44555\\
$^{17}$Universit\'e de Lyon 1, CRAL, Observatoire de Lyon, 9 av. Charles Andr\'e, F-69230 Saint-Genis Laval; CNRS, UMR 5574; ENS de Lyon, France\\
$^{18}$European Southern Observatory, Karl-Schwarzchild-Str. 2, D-85748 Garching, Germany
}

\begin{abstract}
We present the dynamical properties of 97 spectroscopically 
confirmed ultra-compact dwarfs (UCDs; $r_{\rm h}$ $\gtrsim$ 10 pc) and 911 globular clusters 
(GCs) associated with central cD galaxy of the Virgo cluster, M87.\ Our UCDs, of which 89\% 
have $M_{\star}$ $\gtrsim$ 2$\times10^{6}$ $M_{\sun}$ and 92\% are as blue as the classic blue GCs, 
nearly triple previously confirmed sample of Virgo UCDs, providing by far the best opportunity 
for studying global dynamics of UCDs.\ We found that
(1) UCDs have a surface number density profile that is shallower than that of blue GCs in the inner 
$\sim$ 70 kpc and as steep as that of red GCs at larger radii; 
(2) UCDs exhibit a significantly stronger rotation than GCs, and blue GCs seem to have a velocity 
field that is more consistent with that of the surrounding dwarf ellipticals than with that of UCDs; 
(3) UCDs have an orbital anisotropy profile that is tangentially-biased at radii $\lesssim$ 
40 kpc and radially-biased further out, whereas blue GCs become more tangentially-biased at 
larger radii beyond $\sim$ 40 kpc;
(4) GCs with $M_{\star}$ $\gtrsim$ 2$\times$10$^{6}$ $M_{\sun}$ have rotational properties 
indistinguishable from the less massive ones, suggesting that it is the size, instead of mass, that 
differentiates UCDs from GCs as kinematically distinct populations.\
We conclude that most UCDs in M87 are not consistent with being merely the most luminous and 
extended examples of otherwise normal GCs.\ The radially-biased orbital structure of UCDs at large 
radii is in general agreement with the ``tidally threshed dwarf galaxy'' scenario.

\end{abstract}


\section{Introduction}
Ultra-compact dwarfs (UCDs; Phillipps et al.\ 2001) were originally discovered 
(Hilker et al.\ 1999b; Drinkwater et al.\ 2000a) as compact stellar systems which 
are more than 1 mag brighter than the known brightest globular clusters 
($M_{V} \sim$ $-$11 mag; Harris 1991) but at least 2 mag fainter than the prototypical 
compact elliptical M32 ($M_{V}$ = $-$16.4 mag).\ The first five UCDs found 
in the core of the Fornax Cluster were unresolved or marginally resolved on ground-based 
arcsec-resolution images, implying effective radii of $r_{h}$ $\lesssim$ 100 pc.\ 
Subsequent {\it Hubble Space Telescope} (HST) imaging of those Fornax UCDs 
(Drinkwater et al.\ 2004) gave $r_{h}$ $\gtrsim$ 10 pc, which differs significantly from  
$r_{h}$ of $\sim$ 3 pc for conventional GCs (e.g.\ van den Bergh et al.\ 1991; 
Jord\'an et al.\ 2005).\ Since their discovery in the Fornax Cluster, similarly bright UCDs 
have been found in other clusters (Virgo: Hasegan et al.\ 2005; Jones et al.\ 2006; Abell S0740: 
Blakeslee \& Barber DeGraaff 2008; Coma: Madrid et al.\ 2010; Chiboucas et al.\ 2011; 
Centaurus: Mieske et al.\ 2009; Hydra: Misgeld et al.\ 2011; Antlia: Caso et al.\ 2013), 
groups (HCG22 and HCG90: Da Rocha et al.\ 2011; NGC1132: Madrid \& Donzelli 2013), 
and even relatively isolated galaxies (Sombrero: Hau et a.\ 2009; NGC 4546: Norris \& Kannappan 2011).\ 

Given the intermediate nature of UCDs, since their discovery, there has been ongoing debate 
about their origin.\ The few proposed formation mechanisms in the literature are: (1) they 
are merely luminous, genuine GCs (Murray 2009), or mergers of young massive 
star clusters formed in starburst regions, such as those formed during collisions between 
gas-rich galaxies (Fellhauer \& Kroupa 2002; Bruns et al.\ 2011; Renaud, Bournaud \& Duc 2014); 
(2) they are the remains of tidally-stripped nucleated galaxies (e.g.\ Bekki et al.\ 2003; Goerdt et al.\ 2008; 
Pfeffer \& Baumgardt 2013); (3) they are the remnants of primordial compact galaxies (Drinkwater et al.\ 2004).

A consensus about the primary origin of UCDs has yet to be reached.\ In fact, even the name given to this category of object, ultra-compact dwarf, has been debated. Since the UCD designation implies a galactic origin, they have also been referred to as ``dwarf-globular transition objects'' (DGTOs, Ha\c{s}egan et al.\ 2005). In this paper, we will refer to these objects as UCDs since that is the most common usage in the literature, but our usage is not meant to pre-suppose their origin.

The three properties of UCDs that make them distinct from GCs include their larger sizes 
(e.g.\ Kissler-Patig, Jord\'an \& Bastian 2006), a possible size-luminosity relation (e.g.\ C\^ot\'e et al.\ 2006; 
Dabringhausen et al.\ 2008) and slightly elevated dynamical mass-to-light ratios (e.g.\ Ha\c{s}egan et al.\ 2005; 
Mieske et al.\ 2008) above a dynamical mass of $\sim$ 2$\times$10$^{6}$ M$_{\sun}$.\
Mieske, Hilker \& Misgeld (2012) found that the number counts of UCDs, which they defined as stellar systems 
with $M_{V}$ $<$ $-$10.25, in several different environments (the Fornax cluster, Hydra cluster, Centaurus cluster 
and the Local Group) are fully consistent with them being the bright tail of the normal GC population.\ 
On the other hand, C\^ot\'e et al.\ (2006) and Brodie et al.\ (2011) found that UCDs follow dE nuclei, instead of 
GCs, on the color-magnitude diagram, suggesting that most UCDs may be a distinct population that is 
more likely to be related to tidally stripped galaxy nuclei, rather than to GCs.\ 

Recently, Seth et al.\ (2014) found strong evidence for the existence of a supermassive black hole 
(2.1$\times$10$^{7}$) in the brightest known UCD -- M60-UCD1 ($M_{V} = -14.2$ mag, Strader et al.\ 2013), 
indicating that this UCD is most probably a tidally stripped nucleus of a low-mass elliptical galaxy.\ 
Nevertheless, a spatially resolved analysis of the kinematics of the most luminous UCD in the 
Fornax cluster (UCD3, $M_{V} = -13.6$ mag; Hilker et al.\ 1999) by Frank et al.\ (2011) found that 
its internal kinematics are fully consistent with it being merely a massive star cluster, without strong 
evidence for the presence of either an extended dark matter halo or a central black hole.\ Moreover, 
there exists direct evidence that UCD-like objects can form as supermassive star clusters, such as W3 
in the merger remnant NGC 7252 ($M_{\star} \sim 7\times10^{7} M_{\sun}$, age $\sim$ a few 100 Myr: 
Maraston et al.\ 2004) and the recently discovered young ``UCDs'' (Penny et al.\ 2014) associated with 
star-forming regions in NGC 1275 (a member of the Perseus cluster).

All previous investigations of UCDs were based on either incomplete or inhomogeneous small 
samples, which hinders us from understanding the global properties of UCDs in any one galaxy or 
environment.\ Over the past 5 years, we have been collecting low-resolution ($R$ $\sim$ 1300) 
spectroscopic data for UCDs and luminous GCs toward the central regions of the Virgo cluster, 
using two multi-fiber spectrographs: the 2dF/AAOmega (Sharp et al.\ 2006) on the 3.9-m 
Anglo-Australian Telescope (AAT) and Hectospec (Fabricant et al.\ 2005) on the 6.5-m MMT.\
Our spectroscopic surveys of the Virgo UCDs and GCs have been highly efficient (in terms of contamination level 
of non-Virgo targets), thanks to an unprecedentedly clean sample of Virgo UCD and GC candidates  
selected based on the recently completed Next Generation Virgo Survey (NGVS, Ferrarese et al.\ 2012), which offers 
deep ($g_{\rm limit}$ = 25.9 mag at 10$\sigma$ for point sources) and high resolution (FWHM $\sim$ 0.6\arcsec~in 
$i$ band) $u^*giz$ (and $r$ in the cluster core) imaging data of the Virgo cluster from its core to the virial radius 
($\sim$ 104 deg$^{2}$) with the MegaCam instrument on the Canada-France-Hawaii Telescope.

Details of the spectroscopic surveys will be presented in future papers in a series.\ 
In this paper, we present a dynamical analysis of the UCDs associated with the central cD galaxy 
M87 ($D$ = 16.5 Mpc; Mei et al.\ 2007; Blakeslee et al.\ 2009), 
which hosts the majority of confirmed UCDs from our spectroscopic surveys, and thus provides the best opportunity 
for studying the dynamics and photometric properties of UCDs as a population.\ For comparison purposes, we also 
collected radial velocities of 911 GCs associated with M87, and did the dynamical analysis in parallel with the UCDs.\
Other papers in the NGVS series relevant to the topics covered here include a systematic study of photometrically-selected 
UCDs in the three Virgo giant ellipticals M87, M49 and M60 (Liu et al.\ 2015), studies of the distributions of cluster-wide 
GC populations in the Virgo cluster (Durrell et al.\ 2014),  a detailed study of the spatial, luminosity and color distributions of GCs 
selected based on various NGVS bands in the central $2^\circ\times2^\circ$ around M87 (Lancon et al.\ in preparation), 
dynamical modeling of M87 GCs (Zhu et al.\ 2014), and the physical classification of stellar and galactic sources based 
on the optical and deep $K_{s}$ imaging (Mu\~noz et al.\ 2014).\ Liu et al.\ (2015) is especially complementary to this work, 
in that it presents a thorough description of the photometry and size measurements of the UCD samples, and a detailed 
analysis of the color-magnitude relation, color distribution, specific frequencies, and spatial distribution of the UCDs.\

This paper is structured as follows.\ 
In Section~\ref{data}, we introduce the data and samples used in this work.\
A brief description of the methodology used to select a highly clean sample of Virgo UCDs and GCs 
from the spectroscopic catalogs is given in Section~\ref{sample}.\ The definition of our working subsamples is 
described in Section~\ref{definition}.\ Section~\ref{overview} provides an overview of the UCD sample, including 
the spatial distribution, completeness and surface number density profiles.\ Section~\ref{vrmsrad} presents the 
phase-space distribution and velocity dispersion profiles of the UCDs and GCs.\ Section~\ref{velhis} presents the 
velocity distributions of our samples.\ A kinematic modeling of the rotational properties of UCDs and GCs is given 
in Section~\ref{kinem}, while Section \ref{jeans} is devoted to a Jeans analysis for determining the radial 
anisotropies of UCDs and GCs.\ A brief discussion and summary of this paper follow in Section~\ref{summary}.\

\section{Data}\label{data}
\begin{figure}
\centering
\includegraphics[height=0.35\textheight]{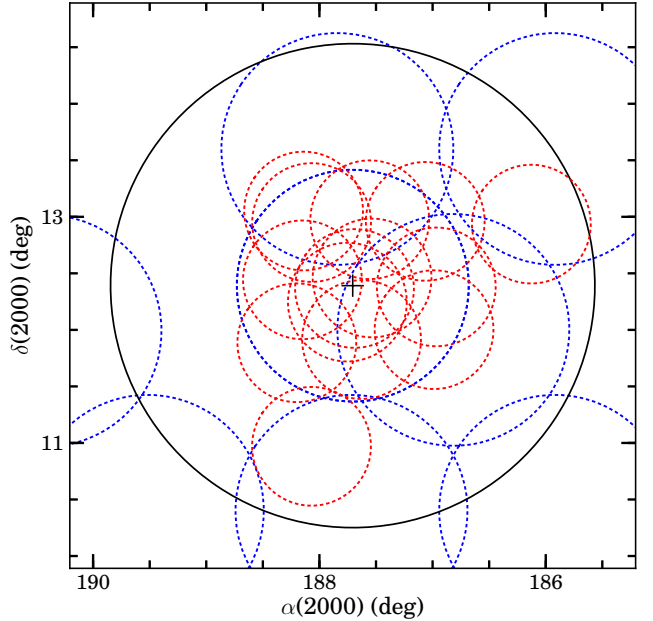}
\caption{
Location of the fields covered by our AAT ({\it blue dotted circles}, 1$^{\circ}$~in radius each) and MMT 
({\it small red dotted circles}, 0\fdg5 in radius each) surveys.\ The {\it big black solid circle} marks the scale 
radius (2\fdg143 = 0.617 Mpc) of the NFW model fitted to the surrounding dark matter halo of the Virgo 
A subcluster (McLaughlin 1999).\ The $black~plus$ marks the photometric center of M87. 
\label{radc_survey}}
\end{figure}

This paper is devoted to a detailed dynamical analysis of confirmed UCDs ($r_{\rm h}$ $\gtrsim$ 10 pc) associated 
with the cD galaxy M87.\ To this end, we compiled a sample of spectroscopically confirmed UCDs, together with GCs 
which will be used for comparison purpose, from three different sources, i.e.\ our recently finished 2dF/AAOmega AAT 
and Hectospec/MMT surveys, and the radial velocity catalogs of Virgo GCs and UCDs compiled by 
Strader et al.\ (2011, hereafter S11).\ For duplicate observations among the three sources, a weighted average of the 
individual radial velocities will be used in this work.\ The location of the pointings around M87 covered by our 
AAT ({\it blue dotted circles}) and MMT ({\it small red dotted circles}) surveys is shown in Figure \ref{radc_survey}.\ 
It can be seen that our surveys covered most of the area encompassed by one scale radius ({\it big solid circle}) 
of the NFW dark matter halo toward the Virgo A subcluster (McLaughlin 1999).\

The photometric data for the spectroscopic samples are from the NGVS.\ The reader is referred to Ferrarese et al.\ (2012) 
for a detailed description of the NGVS.\ Here we only mention that the average seeing of the NGVS $g$- and $i$-band 
imaging data in the central 4 square degrees around M87 is $\sim$ 0.7\arcsec and 0.6\arcsec, which makes it possible to 
measure (through profile modeling) the size of Virgo objects down to $r_{\rm h}$ of $\gtrsim$ 5 pc.\ 
The reader is referred to Liu et al.\ (2015, in preparation) for details about size measurements on NGVS images.\
In addition, we have obtained deep $K_{s}$-band imaging data for this region (NGVS-IR; 
Mu\~noz et al.\ 2014).\ Combining the $K_{s}$ band with the NGVS optical band photometry allows us to identify most of the foreground stars which would otherwise contaminate our spectroscopic catalogs of Virgo UCDs and GCs.\ 

All the magnitudes (in the MagaCam $u^*griz$ filters) that appear in this paper are on the AB system.\ 
In addition, a subscript of 0 denotes the magnitude in question has been corrected for the Galactic extinction.\
In the remainder of this section, we give a brief introduction to the three individual radial velocity catalogs of Virgo 
UCDs and GCs.\ In addition, we will also introduce the radial velocity catalog of early-type dwarf galaxies surrounding 
M87.\ In this work, the velocity field of surrounding early-type dwarf galaxies will be compared to that of the UCDs and GCs.\

\subsection{The AAT sample}
We have carried out (Mar 28 to Apr 1 in 2012) a systematic spectroscopic survey of compact stellar systems (GCs and 
UCDs) toward the central regions of the Virgo cluster (Virgo A subcluster), using the 2dF/AAOmega multi-fiber spectrograph 
on the 3.9-m Anglo-Australian Telescope (AAT).\ The survey consists of nine 2dF pointings, covering a total sky area 
of $\sim$ 30 deg$^{2}$.\ The 8 pointings around M87 are shown as {\it blue dotted circles} in Figure \ref{radc_survey}.\
The observations cover a wavelength range from $\sim$ 3700 ${\rm \AA}$ to 8800 ${\rm \AA}$, with a 
resolution of $R$ = 1300.\ 

The candidates for our survey were selected to fall in the region occupied by spectroscopically confirmed 
Virgo GCs and UCDs on the MegaCam $u^*-g$ vs. $g-i$ diagram\footnote{By the time we prepared input catalogs 
for our spectroscopic surveys (including the MMT survey described below), the NGVS-IR $K_{s}$ band data, 
which is very efficient in separating out the foreground stars, was not available.}, 
and have 18.5 $\leq$ $g$ $\leq$ 20.5 magnitude ($-$12.6 $\leq$ $M_{g}$ $\leq$ $-$10.6 at a distance of 
16.5 Mpc for the Virgo cluster).\ The Virgo UCD candidates were mainly selected to have 
10 $\lesssim~r_{\rm h, NGVS} \lesssim$ 30 pc, and the GC candidates have 1 $<$ $r_{\rm h, NGVS} <$ 10 pc, 
where $r_{\rm h, NGVS}$ is the half-light radius (assuming a distance of the Virgo cluster) measured based on the 
NGVS $g$ and $i$ images.\ 

The AAT survey obtained radial velocities of 55 Virgo UCDs and 52 GCs, of which 22 UCDs and 
20 GCs have no published velocities before.\ At the limiting magnitude of $g$ $\sim$ 20.5, we obtained radial 
velocities with errors of $\sim30$~km s$^{-1}$ in a typical exposure time of 1.5 hour.\ 

\subsection{The MMT sample}
In 2009 and 2010, we used the Hectospec multifiber spectrograph (Fabricant et al.\ 2005) on the 
6.5-meter MMT telescope to carry out an extensive spectroscopic survey (3650--9200${\rm \AA}$, $R$ = 1000) 
of the central $2^\circ\times2^\circ$ ($576\times576$~kpc) around M87 in three observing runs.\ The 
MMT pointings around M87 are shown as {\it small red dotted circles} in Figure \ref{radc_survey}.\

Similar to the AAT survey, the GC and UCD candidates for this survey were selected using MegaCam $u^*giz$ 
photometry from NGVS imaging.\ At the limiting magnitude ($g<22.5$) of this survey, we obtained 
radial velocities with errors of $\sim30$~km s$^{-1}$ in two hour exposures.\ This survey produced radial velocities 
for 324 GCs and 51 UCDs, of which 207 GCs and 18 UCDs (excluding the ones discovered by our AAT survey) have 
no published velocities before.\ 

\subsection{The S11 sample}
By combining radial velocities from the literature (Huchra \& Brodie 1987; Mould et al.\ 1990; Cohen 2000; 
Hanes et al.\ 2001; Jones et al.\ 2006; Hasegan et al.\ 2007; Evstigneeva et al.\ 2007; Firth et al.\ 2009; Paudel et al.\ 2010) 
and from their new observations, S11 compiled a sample of 927 radial velocities toward the central $\sim$40\arcmin~of M87.\  
Of these, 737 were classified as Virgo GCs and UCDs.\ Given the high-quality multi-band imaging data from the NGVS,
we re-classified (Section \ref{sample}) the original 927 objects from S11, and found another 5 low-velocity objects 
belonging to the Virgo cluster.\

\subsection{Earty-type Dwarf Galaxies Surrounding M87}
Our surveys extend to the Virgo intra-cluster region.\ We will compare the kinematics of UCDs and GCs in the 
outermost part of M87 to the surrounding early-type dwarf galaxies, in order to explore any possible connection.\ 
Within a geometric radius of 2$^{\circ}$ from M87, there are 326 galaxies (either with radial velocity unavailable or 
$<$ 3500 km s$^{-1}$ ) classified as either dE or dS0 galaxies in the Virgo Cluster Catalog 
(VCC, Binggeli \& Cameron 1991).\ Among the 326 galaxies, 67 were further classified 
as nucleated dE galaxies (dE, N)\footnote{One should keep in mind that the real fraction of dE, N galaxies is 
most probably much higher than 21\% (67/326), as demonstrated by C\^ot\'e et al.\ (2006) based on high-quality 
HST imaging data of 100 early-type galaxies in the Virgo cluster.}.\ 59 of the 67 dE,N galaxies and 67 of the 
non-nucleated dE/dS0 galaxies have radial velocities available in the literature, as compiled by the GOLDMine 
project (Gavazzi et al.\ 2003).\ In addition, our AAT and MMT surveys obtained the first radial velocity 
measurements for another 2 dE galaxies, i.e.\ VCC1317 ($V$ = 327$\pm$39 km s$^{-1}$) and VCC1244 
($V$ = 824$\pm$33 km s$^{-1}$).\ In this work, we will be comparing the velocity field (number density profiles) 
of the 128 (326) galaxies with that of the M87 UCDs and GCs.\ 

\section{Identification of Virgo Objects}\label{sample}
\subsection{Culling Out the Virgo Objects}
There exists contamination from both background galaxies and foreground stars in our spectroscopic catalogs.\ 
There is a well-defined gap between the Virgo galaxies and background galaxies at radial velocities $\sim$ 
3000 km s$^{-1}$, so a simple cut in radial velocity at 3000 km s$^{-1}$ can remove all background galaxies.\ 
At the low-velocity end ($V_{\rm los}$ $<$ 400 km s$^{-1}$), to cull foreground stars from the spectroscopic 
catalogs, we made use of the $u^*-i$ vs. $i-K_{s}$ color-color diagram, which has been shown to 
clearly separate nearly all foreground stars (Mu\~noz et al.\ 2014) from Virgo stellar systems (with the exception 
of some metal-poor G-type stars).\ For sources that fall inside the overlap area of the bona-fide Virgo members 
and foreground stars on the $u^*-i$ vs. $i-K_{s}$ diagram (see Mu\~noz et al.\ 2014), we further require that 
the sources should have half-light radii $r_{\rm h}$ measured (on the NGVS $g$- and $i$-band images) to 
be $>$ 0.06$''$, corresponding to a linear scale of $\sim$ 5 pc at the distance of Virgo cluster.\ The high-quality 
NGVS $g$ and (especially) $i$ imaging data can resolve Virgo sources down to $r_{\rm h}$ $\sim$ 5 -- 10 pc.\ 

The half-light radius $r_{\rm h, NGVS}$ of each source was derived as a weighted 
average of two independent measurements in the $g$ and $i$ bands by fitting PSF-convolved King (1966) 
models to NGVS images with the KINGPHOT software package (Jord\'an et al.\ 2005).\ Briefly, when using 
KINGPHOT, we adopted a fixed concentration parameter $c$ of 1.5 and a fixed fitting radius $r_{\rm fit}$ of 
1.3$\arcsec$ ($\simeq$ 105 pc at the Virgo distance).\ According to Jorda\'n et al.\ (2005), the KINGPHOT size 
measurement suffers from large biases when $r_{\rm h} \gtrsim$ $r_{\rm fit}$/2, which is however not expected 
to be a problem for our analysis because all but one (VUCD7) of previously confirmed Virgo UCDs have 
$r_{\rm h} <$ 50 pc.\

\subsection{Separating the Virgo UCDs from GCs}\label{sepucd}
A UCD is defined to have 10 $\lesssim$ $r_{\rm h}$ $\lesssim$ 100 pc in this work (see Section \ref{definition}).\
At 10 $\lesssim$ $r_{\rm h}$ $\lesssim$ 20 pc, the Virgo UCDs are only marginally resolved in the NGVS images.\ 
Assuming a Gaussian-shaped PSF, the typical FWHM of the NGVS $i$-band seeing disc (0.6\arcsec) is equivalent 
to an $r_{\rm h}$ of 0.172$\arcsec$, which corresponds to $\sim$ 14 pc at the distance of the Virgo cluster.\ 
Therefore, size measurements based on the NGVS images are especially sensitive to the S/N and possible 
inaccuracy of the PSF, and are unavoidably subject to relatively large uncertainties compared to the measurements 
based on HST images.\ Our test with the NGVS images suggests that, sources with $g$ $\gtrsim$ 21.5 -- 22 
mag are subject to relatively large bias and uncertainties ($>$ 20\%) in their size measurement, and 
thus are not suitable for our analysis.\ To pick out a clean sample of UCDs based on NGVS images, 
we require UCDs to have $r_{\rm h, NGVS}$ $\geq$ 11 pc, $\Delta r_{\rm h, NGVS}$/$r_{\rm h, NGVS}$ 
$<$ 0.1, and $g$ $\leq$ 21.5 ($M_{g} \leq -9.6$).\ In addition, sources with $r_{\rm h, HST}$ $>$ 9.5 pc 
based on measurements with existing HST images (e.g.\ S11; Jord\'an et al.\ 2005) are also included 
as UCDs, regardless of their brightness.\ All the other confirmed Virgo compact clusters are regarded to be GCs.\

By comparing our size measurements with that determined with existing HST imaging data 
(see Table \ref{ucd_table}), our size criteria of UCD selection based on the NGVS measurements 
result in zero contamination from Virgo objects with $r_{\rm h, HST}$ $<$ 10 pc.\ Among the old 
sample of 34 Virgo UCDs with $r_{\rm h, HST} >$ 9.5 pc, 3 did not have NGVS size measurements 
due to their proximity (within 10\arcsec) to saturated foreground stars, 28 have $r_{\rm h, NGVS}$ 
$\geq$ 11 pc, 1 (T15886: $g$ = 22.97 mag) has 10 $\leq$$r_{\rm h, NGVS}$$<$ 11 pc, and the 
other 2 (S6004: $g$ = 21.32; S8006: $g$ = 20.53) have $r_{\rm h}$ measured to be less than 
10 pc either in NGVS $g$ or $i$ band.\ Therefore, by selecting UCD-sized objects ($r_{\rm h}$ $\gtrsim$ 10 pc) 
based on the NGVS images, we may miss $\sim$ 6\% of genuine UCDs at $g$ $<$ 21.5 mag and $\lesssim$ 3\% 
at $g$ $<$ 20.5 mag.

\subsection{The Samples of UCDs and GCs}

Given the above selection procedure, we end up with a total number of 97 UCDs and 911 GCs, which 
fall within 1.5$^{\circ}$~of M87 and are not associated with any galaxies other than   
M87 based on spatial location and radial velocities.\ Some of these UCDs and GCs, especially those at  
the outermost radii, probably belong to the intra-cluster population.\ The full sample of UCDs is listed in 
Table \ref{ucd_table}.\ The sample of GCs has been recently used by Zhu et al.\ (2014) to 
determine the dynamical mass profile of M87.\ The full catalog of GCs will be presented elsewhere 
(Peng et al.\ 2015, in preparation).\ The 34 UCDs confirmed previously in the literature (old sample, 
Brodie et al.\ 2011) and the 63 newly confirmed UCDs (new sample) are listed separately in 
Table \ref{ucd_table}.\ For Virgo objects that were already spectroscopically confirmed (as 
compiled by S11), we follow the old naming; for the newly confirmed Virgo members, we 
adopted a naming scheme which starts with ``M87UCD-''.\ Note that, the column 
$r_{\rm h, NGVS}$ gives the weighted average half-light radius measurements in the NGVS $g$ and 
$i$ bands.\ The $r_{\rm h, HST}$ measurement (Jord\'an et al.\ 2005; Jord\'an et al.\ 2009; S11), if available, 
is also listed.\ We point out that the uncertainties of $r_{\rm h, NGVS}$ reported in Table \ref{ucd_table} 
only include the formal errors returned from the KINGPHOT fitting, and do not take into account any 
potential systemic uncertainty, such as the degree of accuracy of the PSF and suitability of King models 
for representing the UCD light profiles.

Our sample of UCDs is nearly 3 times larger than previously known (34) Virgo UCDs, and our GC sample 
is $\sim$ 20\% larger than that of S11.\ Radial velocities for 39 of the 97 UCDs were obtained from our AAT 
and MMT surveys for the first time.\ In terms of spatial coverage, one of the most important improvement of 
our GC sample is at the projected galactocentric distances larger than 30$'$ from M87.\ Specifically, 
our sample includes 63 GCs in the projected radius range from 30$'$ to 60$'$, and this is 7 times larger than that 
of S11.\ While the full catalog of M87 GCs will be presented elsewhere, we emphasize that the main results 
related to GCs in this paper would not change qualitatively if only the S11 sample of GCs was used in our 
analysis because of the already large spectroscopic sample of M87 GCs in the literature.\
Since the surface number density of UCDs is relatively low, our kinematic analysis will be 
carried out in a coarser spatial (or radial) resolution than previous studies (C\^ot\'e et al.\ 2001; S11).\ 
Therefore, whenever relevant, we refer the readers to C\^ot\'e et al.\ (2001) and 
S11 for a more detailed kinematical analysis of M87 GCs within the central 30$'$.\

\section{Definition of Working Samples}\label{definition}
The primary goal of this work is to explore the differences or similarities between UCDs and GCs.\ 
To this end, we define the following subsamples in this paper.\
\begin{itemize} 
\item {\bf UCDs}.
UCDs are distinguished from GCs as having half-light radius $r_{\rm h}$ $\gtrsim$ 10 pc (although 
in practice we require $r_{\rm h, NGVS}\geq11$ pc, as described above) in this work.\ 
The two most commonly adopted definitions of UCDs are mass (2$\times$10$^{6}$ $\lesssim$ M$_{\rm dyn}$ 
$\lesssim$ 10$^{8}$ $M_{\sun}$; e.g.\ Hasegan et al.\ 2005; Mieske et al.\ 2008) and/or $r_{\rm h}$ 
(10$\lesssim$$r_{\rm h}$$\lesssim$100 pc; e.g.\ Norris et al.\ 2011; Brodie et al.\ 2011).\ The mass definition 
is justified by the findings that  1) compact stellar systems with M $\gtrsim$ 2$\times$10$^{6}$ tend to 
have M/L significantly higher than the lower mass systems; 2) there seems to be a size-luminosity relation setting in at 
M $\gtrsim$ 2$\times$10$^{6}$ (e.g.\ Rejkuba et al.\ 2007; Evstigneeva et al.\ 2008; Dabringhausen et al.\ 2008; 
Norris \& Kannappan 2011), in contrast to the more or less constant $r_{\rm h}$ ($\sim$ 3 pc, e.g.\ van den Bergh et al.\ 1991; 
Jordan et al.\ 2005) of ``normal'' GCs.\ The size definition of UCDs differentiates them from normal GCs as dynamically 
un-relaxed stellar systems (e.g.\ Mieske et al.\ 2008).\ The two definitions may converge at the highest mass 
end.\ While we adopted the size definition in this work, we will try to explore the significance of mass in 
differentiating UCDs as stellar systems distinct from normal GCs.\ 

\item
{\bf Blue GCs and Red GCs}.
A double-Gaussian fitting to the NGVS $(g-i)_{0}$ bimodal color distribution of photometrically-selected GCs in 
M87 suggests that the blue and red components cross at $(g-i)_{0}$ = 0.89 mag.\ Therefore, we classified 
GCs as blue ($N$=683) and red ($N$=228) at a dividing $(g-i)_{0}$= 0.89 mag.\ 

\item
{\bf Bright GCs and faint GCs}.
Bright GCs and faint GCs are separated at NGVS $i_{0}$ = 20.5 magnitude, which corresponds to a stellar mass 
of $\sim$ 1.6 -- 2$\times$10$^{6}$$M_{\sun}$ at [Fe/H] ranging from $-$1.3 (the typical value for blue GCs in M87; 
Peng et al.\ 2006) to $-$0.3 (the typical value for red GCs in M87) for a 10 Gyr old stellar population with a Chabrier 
or Kroupa stellar initial mass function (IMF).\ The dividing magnitude (or mass) was chosen to roughly correspond to 
the proposed mass boundary between UCDs and GCs for the mass definition of UCDs.\ By separating the bright GCs 
from the faint GCs, we will explore the importance of mass or luminosity in differentiating UCDs from GCs.\ 

\end{itemize}

The median $(g-i)_{0}$ of our samples of UCDs and blue GCs are 0.75 and 0.74 respectively, 
and about 92\% of our UCDs fall into the color range of the blue GCs.\ So we will 
place additional emphasis on a comparison between dynamical properties of UCDs and blue 
GCs throughout this paper.\

\section{Overview of the UCD and GC Samples}\label{overview}
\subsection{2D Spatial Distribution}

\begin{figure*}
\centering
\includegraphics[height=0.38\textheight]{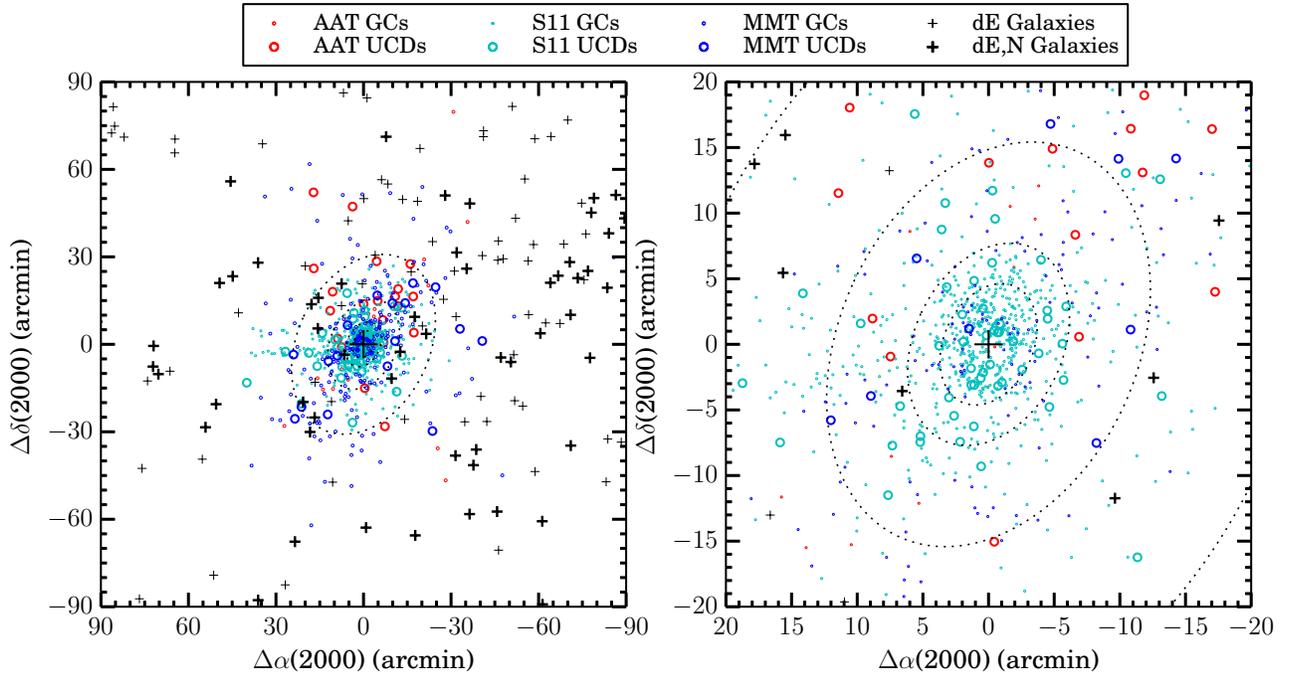}
\caption{
Spatial distribution of spectroscopically confirmed GCs, UCDs, and dE galaxies within the 
central 1.5$^{\circ}$ (left panel) and the central 20\arcmin~(right panel) of M87.\ Dotted ellipses 
represent the stellar isophotes of M87 at 3$R_{\rm e}$, 5$R_{\rm e}$, 
10$R_{\rm e}$ and 20$R_{\rm e}$, where $R_{\rm e}$ is the effective radius measured based on the 
NGVS $g$-band image.\ Different colors indicate data sets from different sources.\ 
The large black plus marks the photometric center of M87.
\label{radc_all}}
\end{figure*}

\begin{figure*}
\centering
\includegraphics[height=0.5\textheight]{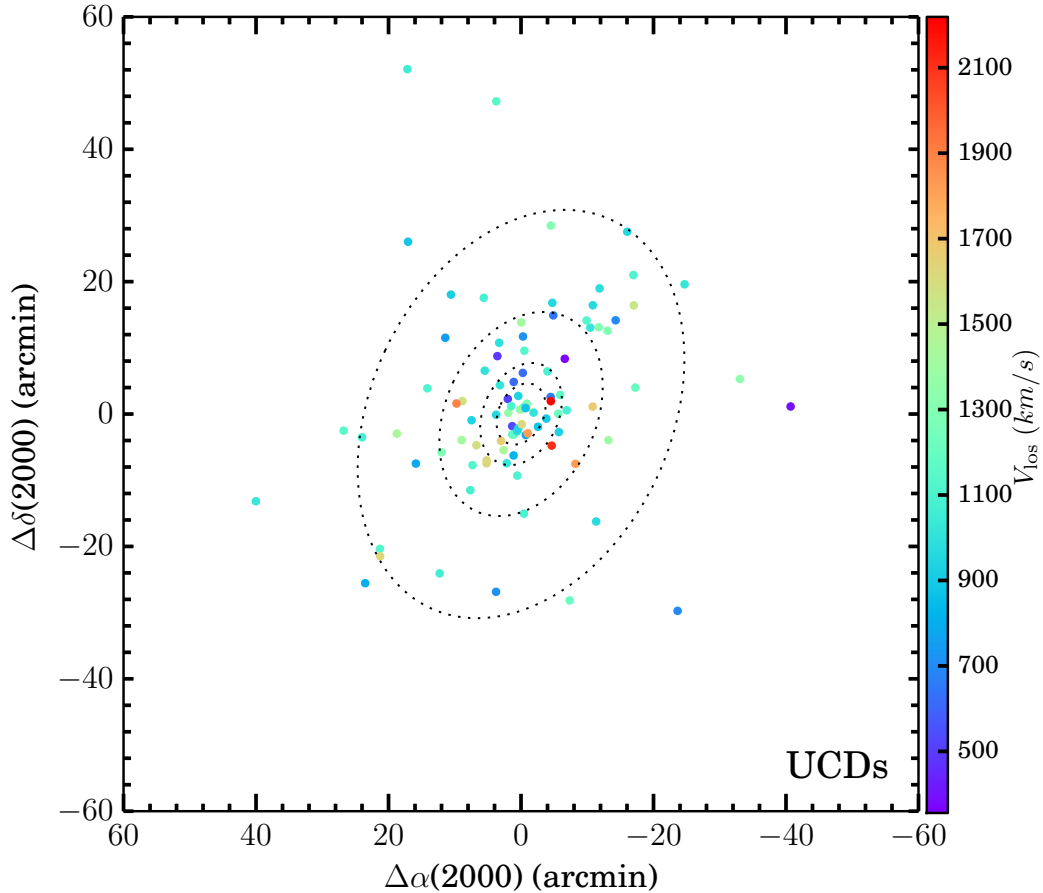}
\caption{
Spatial distribution of spectroscopically confirmed UCDs around M87.\ As in Figure \ref{radc_all}, dotted 
ellipses represent the stellar isophotes of M87 at 3$R_{\rm e}$, 5$R_{\rm e}$, 10$R_{\rm e}$ and 
20$R_{\rm e}$.\ The plotted objects are  color-coded based on their line-of-sight velocities.
\label{radc_velall}}
\end{figure*}

Figure \ref{radc_all} presents the spatial distributions of all spectroscopically confirmed UCDs, GCs and dE galaxies 
around M87.\ Data from our AAT survey, MMT survey, and S11 are represented by different colors.\ Different types 
of objects are plotted as different symbols.\ When plotting Figure \ref{radc_all}, for duplicated observations among 
the three data sources, we group them into, in order of priority, the S11 catalog, the AAT survey catalog, and the MMT 
survey catalog.\ Figure \ref{radc_velall} shows the spatial distribution of UCDs, color-coded according to their line-of-sight 
velocities.\

Like the GC system (e.g.\ McLaughlin et al.\ 1994; Forte et al.\ 2012; Durrell et al.\ 2014), 
the spatial distribution of the UCDs broadly follows the stellar diffuse light.\ Within 2$^{\circ}$ of M87, the 
outermost confirmed UCD has a projected galactocentric distance $R_{\rm p}$ of $55\arcmin$ from the 
center of M87, the outermost red and blue GCs have $R_{\rm p} = 49\arcmin$ and 85\arcmin~respectively, 
with only one GC lying within 65$\arcmin$ $<$ $R_{\rm p}$ $<$ 85\arcmin.\

An interesting ``overdensity'' of $\sim$ 11 UCDs can be seen toward the northwest of M87 between $\sim15\arcmin$ 
and $30\arcmin$.\ After checking the radial velocity distribution of UCDs belonging to this spatial ``overdensity'', 
we found that the ``members''  of this overdensity have radial velocities ranging from $\sim$ 900 to 1750 km s$^{-1}$, 
suggesting this ``overdensity'' is due to chance alignment, rather than a physical substructure.\

In what follows, we will be mostly working with the geometric average radius 
$R_{\rm av}$ when exploring various radial trends.\ $R_{\rm av}$ is defined to be 
equal to $a~\sqrt[]{1-\epsilon}$, where $a$ is the length along the semi-major axis 
(PA $\simeq$ 155$^{\circ}$) and $\epsilon$ is the ellipticity.\ Given that the spatial distribution 
of UCDs and GCs, in terms of flattening and orientation, roughly matches the stellar diffuse 
light of M87 (Durrell et al.\ 2014), we adopted the radial profiles of $\epsilon$ and PA of the 
stellar isophotes of M87 determined with the high-quality NGVS $g$-band imaging data.\ 
Our measurements of $\epsilon$, ranging from $\sim$ 0 in the central 0.5\arcmin~to 0.33 
around 10\arcmin--15\arcmin~along the semi-major axis, are in good agreement with previous 
studies (e.g.\ Ferrarese et al.\ 2006; Kormendy et al.\ 2009).\ Measurements of $\epsilon$ beyond 
$\sim$ 15\arcmin~are subject to relatively large uncertainties, so we fixed $\epsilon$ as 0.33 at $a$ $>$ 15\arcmin.\

\begin{figure}[tph]
\centering
\includegraphics[height=0.3\textheight]{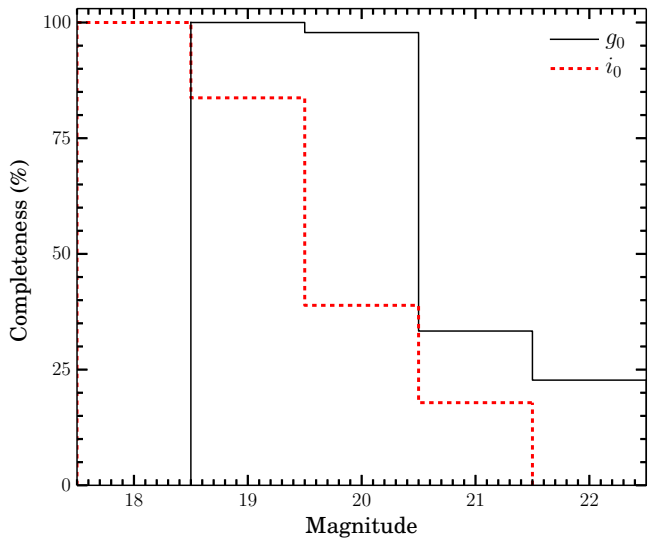}
\caption{
Completeness levels of our sample of UCDs as a function of magnitude in $g$ (black solid lines) and 
$i$ (red dotted lines) bands.
\label{ucd_compl}}
\end{figure}

\subsection{Completeness of the UCD Sample}
To quantify the completeness of our spectroscopic sample of UCDs, we selected all of the UCD 
candidates within the central $60\arcmin$ (in geometric average radius) of M87 from the NGVS photometric 
catalog (Mu\~noz et al.\ 2014), based on the $u^*-i$ vs. $i-K_{s}$ diagram (Section \ref{sample}), size 
measurement (11$\leq$$r_{\rm h, NGVS}\leq$50 pc), and isophotal shapes ({\textsc{SExtractor}} ellipticities $<$ 0.25).\
In addition, any obviously extended galaxies were further removed from our final photometric sample.\

The completeness in a given magnitude range was determined as the $\frac{N_{\rm spec}}{N_{\rm cand}}$$\times$100, 
where $N_{\rm cand}$ is the number of photometric candidates and $N_{\rm spec}$ is the number of candidates 
with radial velocity measurements.\ The completeness in different $g_{0}$ and $i_{0}$ magnitude bins is 
shown in Figure \ref{ucd_compl}.\ Overall, our sample is expected to be $\sim$ 60\% complete at $g_{0}$ $<$ 21.5 
mag, and $\sim$ 55\% complete at $i_{0}$ $<$ 20.5 mag.\ In particular, the sample is $\sim$ 98\% complete at 
$g_{0}$ $<$ 20.5, which corresponds to $M_g\leq -10.6$.\

\subsection{Magnitude vs.\ Galactocentric Distance}

\begin{figure*}
\centering
\includegraphics[height=0.35\textheight]{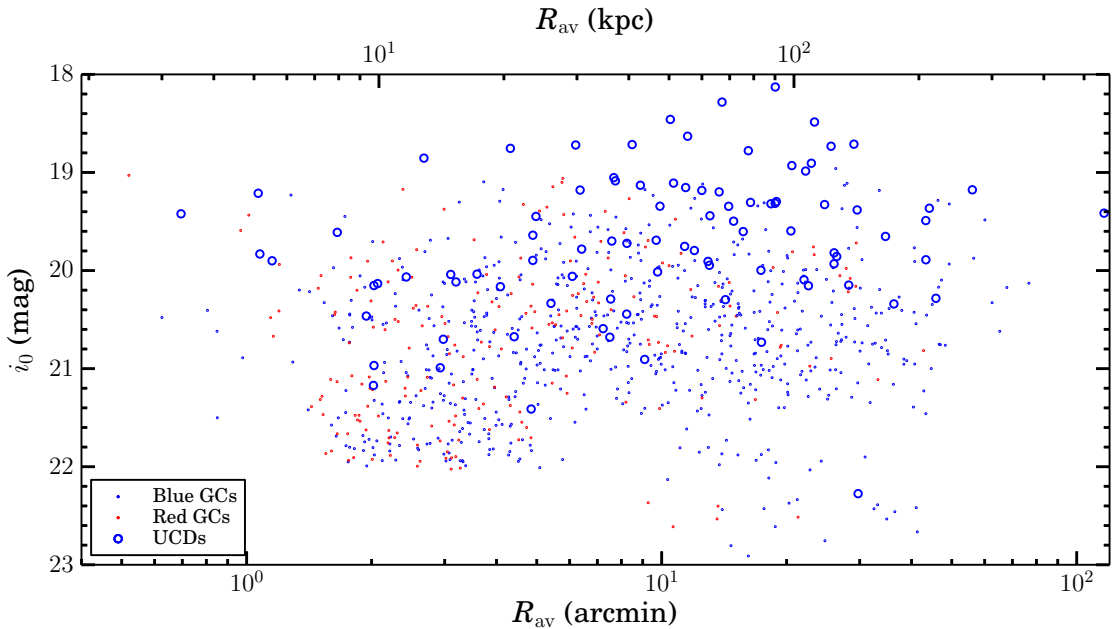}
\caption{
Radial distribution of $i_{0}$ magnitude for UCDs and GCs.\ 
\label{irad_all}}
\end{figure*}

Figure \ref{irad_all} presents the $i_{0}$ magnitude distribution as a function of $R_{\rm av}$ from M87.\ 
The UCDs, blue GCs and red GCs are plotted as different symbols in Figure \ref{irad_all}.\ 
We can see that the available observations of GCs within the central $\sim$ 5$'$ reach down to 22 mag, 
which is about 1 mag fainter than that at larger radii.\ We point out that $\sim$ 89\% of our 
UCDs have $i_{0}$ $<$ 20.5 mag.\

\begin{figure}[tb]
\centering
\includegraphics[height=0.33\textheight]{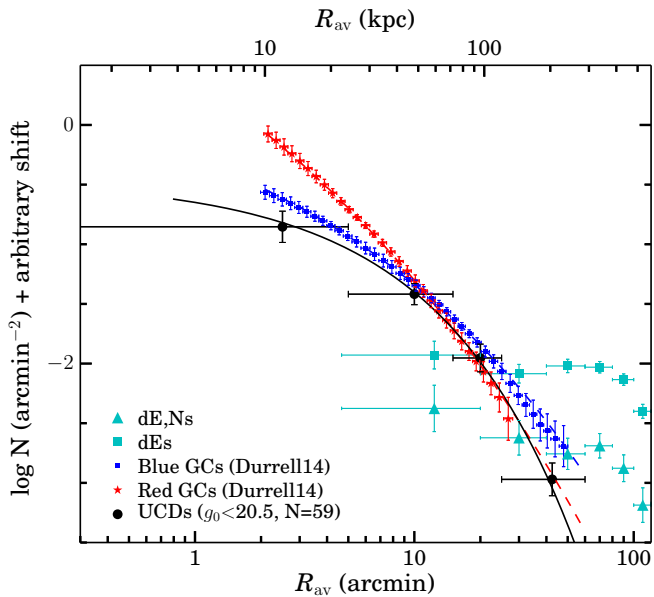}
\caption{
Radial surface number density profiles M87 UCDs (black) with $g_{0}$ $<$ 20.5 mag.\
The radial profiles of blue GCs (blue) and red GCs (red) determined by Durrell et al.\ (2014) were also plotted 
for comparison.\ Also over plotted are the number density profiles of surrounding dE galaxies.\ 
The profiles of blue GCs and red GCs have been vertically shifted arbitrarily for comparison purpose.\ 
Overplotted on the data are the best-fit S\'ersic profiles for UCDs, blue GCs and red GCs.\ 
Note that our sample of UCDs with $g_{0}$ $<$ 20.5 mag is expected to be nearly 100\% complete.
\label{ucdgc_rp}}
\end{figure}

\subsection{Surface Number Density Profiles}\label{radprof}
As is shown in Figure \ref{ucd_compl}, our sample of UCDs is expected to be nearly 100\% complete down to 
$g_{0}$ $<$ 20.5 mag.\ In Figure \ref{ucdgc_rp}, we show the radial number density profile  
of the 59 UCDs with $g_{0}$ $<$ 20.5 mag, together with profiles of the blue GCs and red GCs 
(18.5 $<g_{0}<$ 24.0 mag) determined by Durrell et al.\ (2014).\ Number density profiles for the surrounding 
dE galaxies were also shown for comparison.\ Our specific choice of radial binning for 
constructing the profile of UCDs ensures that at least 10 data points fall into each radial bin.\ 
The vertical error bars of the radial profiles represent the poisson noise.\ 
Note that radial profiles of the GCs have been vertically shifted down arbitrarily 
(2.1 for the blue GCs and 1.7 for the red GCs) for comparison purposes.\ We point out that the GC surface density 
profiles derived by Durrell et al.\ (2014) were based on an adaptive-smoothed GC density maps, with the smoothing 
kernel FWHM $\gtrsim$ 3$'$ -- 5$'$.\ Therefore, the intrinsic surface profiles of GCs in the inner radii may  
be a little steeper than those shown in Figure \ref{ucdgc_rp}.\ 

                                                                                  
%
\begin{deluxetable}{lcccc} 
\tabletypesize{\small}                                                
\tablecolumns{5}                                                           
\tablewidth{0pt}                                                           
\tablecaption{S\'ersic Profile Fitting}
\tablehead{                                                                                                                
\colhead{ID}
& \colhead{$g_{0}$}
& \colhead{$N_{\rm e}$} 
& \colhead{$R_{\rm e}$}  
& \colhead{$n$}\\
\colhead{}
& \colhead{(mag)}           
& \colhead{(arcmin$^{-2}$)}                                                                             
& \colhead{(arcmin)}
& \colhead{} \\                                              
\colhead{(1)}                                                                             
& \colhead{(2)}                                                                             
& \colhead{(3)}
& \colhead{(4)}
& \colhead{(5)}
}                                                                          
\startdata        

UCDs & 18.5 -- 20.5 & 0.03 & 12.57 & 1.43\\
Blue GCs & 18.5 -- 24.0 & 0.87 & 20.10 & 3.03\\
Red GCs & 18.5 -- 24.0 & 3.61 & 5.89 & 4.26\\

\enddata

\tablecomments{
(1) Population name.\
(2) $g$-band magnitude range of the sample.\
(3) Surface number density at the effective radius $R_{\rm e}$.\
(4) Effective radius.\
(5) S\'ersic index.
}
\label{surfprof}
\end{deluxetable}                                                          

We adopt the S\'ersic function (S\'ersic 1968; Ciotti 1991; Caon et al.\ 1993; 
Graham \& Driver 2005) to quantify the radial profiles.\ Due to the small sample size of sparsely-distributed 
UCDs, instead of simply fitting the binned profiles, we used the maximum likelihood method 
(e.g.\ Kleyna et al.\ 1998; Westfall et al.\ 2006; Martin et al.\ 2008) to estimate the UCD number density profile.\ 
Specifically, the likelihood function to be maximized is defined as 

\begin{equation}
  {\cal L}(N_{\rm e}, R_{\rm e}, n) \propto \prod_i \ell_i(R_{i}|N_{\rm e}, R_{\rm e}, n)
\label{eqspf1}
\end{equation}
where $\ell_i(r_{i}|N_{\rm e}, R_{\rm e}, n)$ is the probability of finding the datum $i$ at radius $R_{i}$ given 
the three S\'ersic parameters, i.e.\ the effective radius $R_{\rm e}$, the number density $N_{\rm e}$ at $R_{\rm e}$, 
and the S\'ersic index $n$.\ In particular, 

\begin{equation}
\ell_i(R_{i}|N_{\rm e}, R_{\rm e}, n) \propto N_{\rm e} \exp\left\{ -b_n\left[\left( \frac{R_i}{R_{\rm e}}\right) ^{1/n} -1\right]\right\}
\label{eqspf2}
\end{equation}
where $b_{n}$ is a constant that is defined as a function of $n$ such that $R_{\rm e}$ is the effective radius, and 
we adopted the formula determined by Ciotti \& Bertin (1999) to relate $b_{n}$ to $n$.\
Furthermore, $N_{\rm e}$ can be expressed as a function of $R_{\rm e}$, $n$ and the total number of UCDs 
$N_{\rm tot}$ (= 59) by integrating the S\'ersic profile over the projected area $\pi R^{2}$ to the limit radius 
$R_{\rm lim}$ = 60$\arcmin$~(e.g.\ Graham \& Driver 2005).\ Specifically, 
\begin{equation}
N_{\rm e} = \frac{N_{\rm tot}}{2\pi R_{\rm e}^2 n \frac{\exp({b_n})}{b_n^{2n}} \gamma(2n, b_n(R_{\rm lim}/R_e)^{1/n})}
\label{eqspf3}
\end{equation}
where $\gamma(2n, b_n(R_{\rm lim}/R_e)^{1/n})$ is the incomplete gamma function.\
By substituting $N_{\rm e}$ from Equation \ref{eqspf3} in Equation \ref{eqspf1}, $\cal L$ was maximized to 
find the most likely parameters $R_{\rm e}$ and $n$ (and thus $N_{\rm e}$).\
For the blue and red GCs, we directly fit the S\'ersic function to the binned radial profiles shown in 
Figure \ref{ucdgc_rp}, which is adequate given the large sample size of the photometric samples of GCs.\

The best-fit S\'ersic profiles of the three populations are overplotted in Figure \ref{ucdgc_rp}, and the 
most likely estimation of the S\'ersic parameters is listed in Table \ref{surfprof}.\
The difference between radial profiles of UCDs and GCs is significant.\ 
The UCDs have the shallowest radial profiles in the inner $\sim$ 15$'$ among the three populations, 
and in the outer radii the profile of UCDs is as steep as that of the red GCs.\ Previous studies have shown that 
the surface number density profile of the red GCs closely follow that of the diffuse stellar light 
(e.g.\ Geisler et al.\ 1996; Harris 2009; Durrell et al.\ 2014).\ In addition, the surrounding dE galaxies have 
much flatter and extended number density profiles than UCDs and GCs.\

\section{Phase-Space Distribution and $v_{\rm rms}$ Profiles}\label{vrmsrad}
\begin{figure*}
\centering
\includegraphics[height=0.4\textheight]{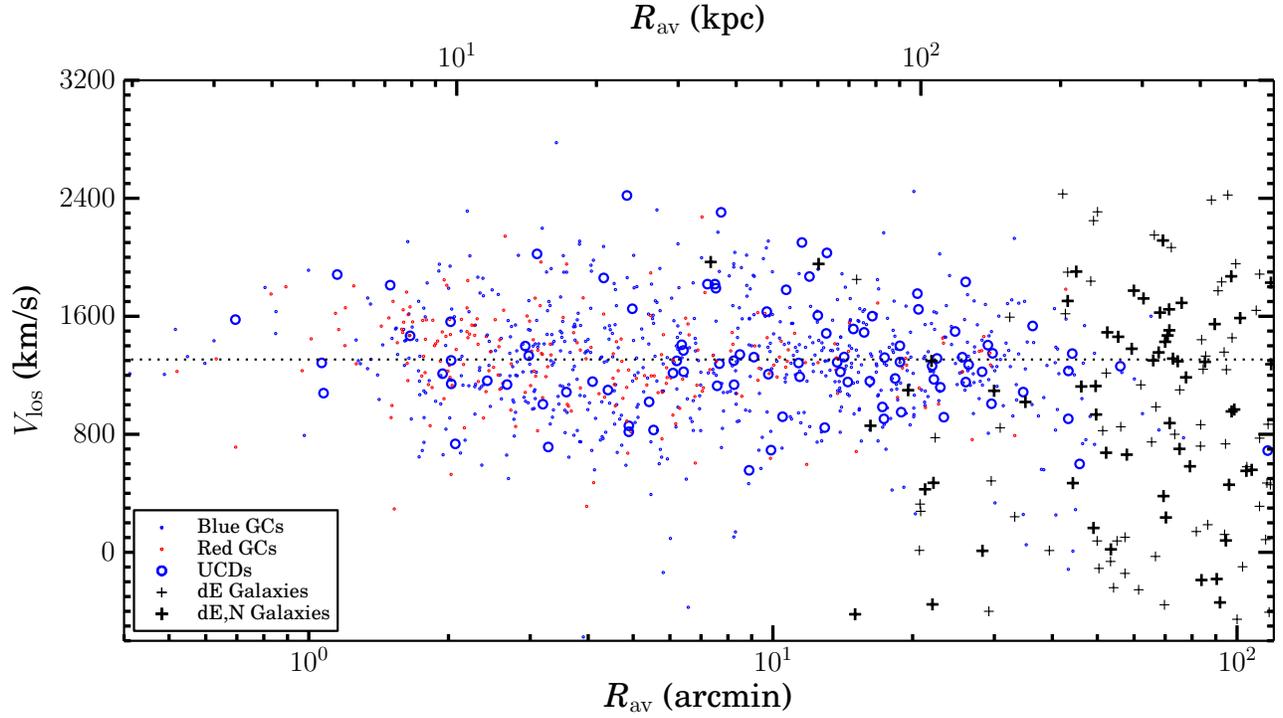}
\caption{
The line-of-sight velocities vs.\ the geometric average radius from M87.\ 
The UCDs, blue GCs, red GCs, and dE galaxies (non-nucleated and nucleated) are plotted separately as 
different symbols.\ The horizontal dotted line marks the systemic radial velocity of 1307 km s$^{-1}$ for M87 
(Binggeli et al.\ 1993).
\label{velrad_all}}
\end{figure*}

\begin{figure*}
\centering
\includegraphics[height=0.4\textheight]{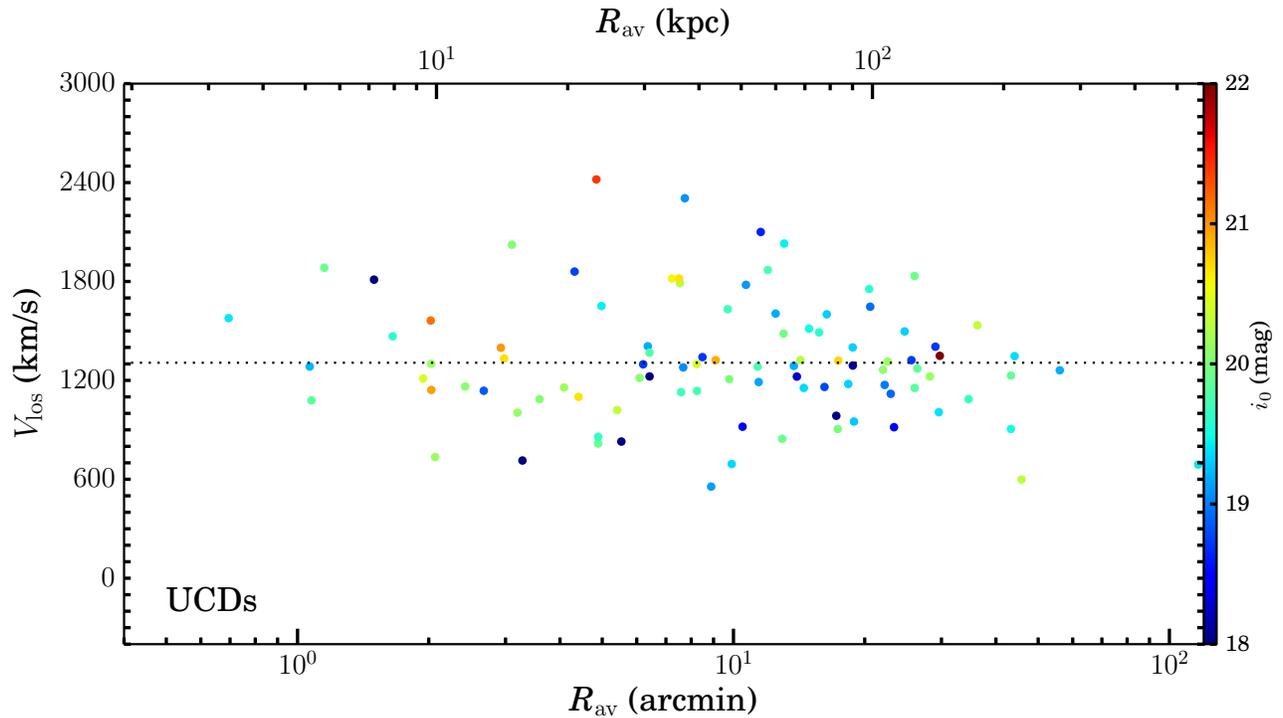}
\caption{
The UCD line-of-sight velocities vs.\ the geometric average radius from M87.\ 
Here different UCDs are color-coded with their $i_{0}$ magnitude.\
The horizontal dotted line marks the systemic radial velocity of 1307 km s$^{-1}$ for M87.\
\label{velrad_i}}
\end{figure*}

Figure \ref {velrad_all} gives the radial variation of the line-of-sight velocities of UCDs, GCs 
and dE galaxies.\ The blue and red GCs are plotted separately.\ A $i$-magnitude color-coded 
plot of $V_{\rm los}$ vs.\ $R_{\rm av}$ for UCDs is shown in Figure \ref{velrad_i}.\ 

\begin{figure*}[h]
\centering
\includegraphics[height=0.33\textheight]{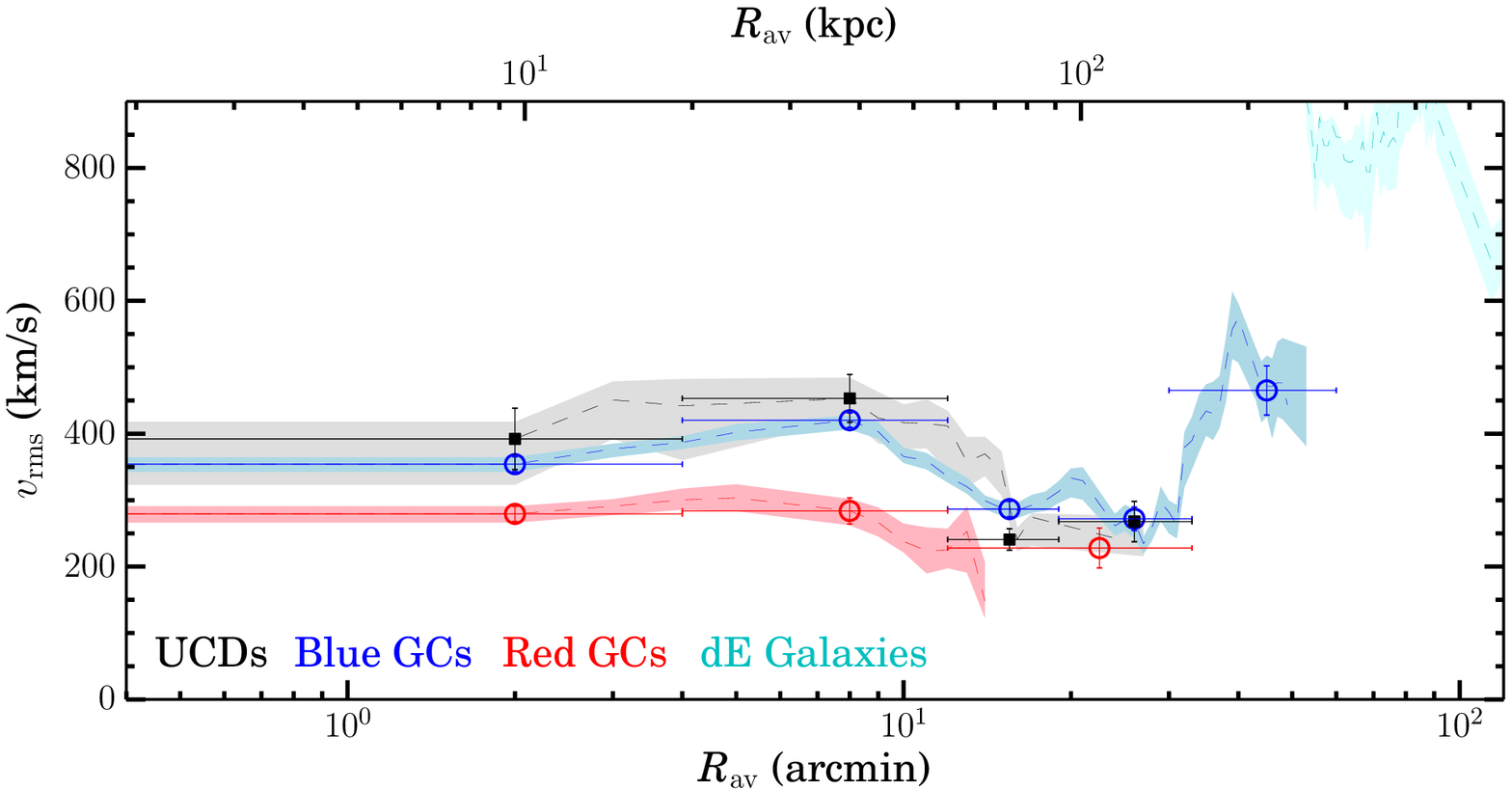}
\caption{
Radial variations of the rms velocities of UCDs, blue GCs, red GCs and dE galaxies.\ 
The profiles were constructed by sliding bins of fixed radial widths 
($\Delta R_{\rm av}$ = 4$'$ for $R_{\rm av}$ $<$ 8$'$; $\Delta R_{\rm av}$ = 8$'$ for 8$'$$<$ $R_{\rm av}$ $<$ 20$'$; 
$\Delta R_{\rm av}$ = 12$'$ for 20$'$$<$ $R_{\rm av}$ $<$ 40$'$; $\Delta R_{\rm av}$ = 16$'$ for $R_{\rm av}$ $>$ 40$'$), 
with a step of 1$'$ between adjacent sliding windows.\ The solid curves correspond to the calculated $v_{\rm rms}$, 
and the shaded regions mark the 68\% confidence limits.\ To be used in constructing the profiles, at least 19 data 
points should be available for a given radius bin.
\label{vrms_all}}
\end{figure*}

While we will present a detailed kinematic modeling of the rotation and intrinsic velocity 
dispersion for our samples in Section \ref{kinem}, it is helpful to first explore variations in the root-mean-square 
line-of-sight velocity $v_{\rm rms}$ as a function of the galactocentric distance from M87 (Figure \ref{vrms_all}).\ 
To construct the $v_{\rm rms}$ profiles, we adopted the ``sliding bin'' method.\ Specifically, bins of fixed 
radial width were slid from the center outward, with an offset of $1\arcmin$ between adjacent bins.\ 
$v_{\rm rms}$ was calculated for data points falling into each individual sliding bin.\ 
Considering the lower number density of UCDs and GCs in the larger radii we used different bin widths in 
different ranges of radii.\ The bin widths are: $\Delta R_{\rm av}$ = 4$'$ for $R_{\rm av}$ $<$ 8$'$, 
$\Delta R_{\rm av}$ = 8$'$ for 8$'$$<$ $R_{\rm av}$ $<$ 20$'$, $\Delta R_{\rm av}$ = 12$'$ for 
20$'$$<$ $R_{\rm av}$ $<$ 40$'$, and $\Delta R_{\rm av}$ = 16$'$ for $R_{\rm av}$ $>$ 40$'$.\ 

We require at least 19 data points to be available in each bin 
when constructing the profiles shown in Figure \ref{vrms_all}.\ A biweight $v_{\rm rms}$ was calculated 
for each sliding bin following the methodology of Beers, Flynn, \& Gebhardt (1990; cf.\ Equation 9).\
The biweight $v_{\rm rms}$ estimator is relatively outlier-insensitive, and has proven to be superior 
to the classical formula when dealing small samples.\ The 68\% confidence interval for $v_{\rm rms}$ 
was estimated by randomly resampling the real data sets without replacement.\ $v_{\rm rms}$ in 
some fixed radial bins is also overplotted in Figure \ref{vrms_all}.\ The dispersion profile for the surrounding 
dE galaxies is also shown for comparison.\

As shown in Figure \ref{vrms_all}, the UCDs follow a velocity dispersion profile more 
similar to that of the blue GCs than the red GCs.\ The $v_{\rm rms}$ of blue GCs 
beyond 30\arcmin~increases steeply to reach $\sim$ 500 km s$^{-1}$.\ The red GCs 
have an overall lower velocity dispersion than the UCDs and blue GCs.\ The blue GCs 
and UCDs show a slight increase in velocity dispersion between $\sim$ 4\arcmin~and 
12\arcmin, whereas the red GCs do not clearly show such a ``hot'' feature.\ This may 
suggest the existence of ``hot'' substructures which have not yet reached an dynamical 
equilibrium state, as suspected by Zhu et al.\ (2014).\

We note that the rising $v_{\rm rms}$ of the blue GCs beyond $\sim$ 30\arcmin~should not be 
regarded as signifying a transition to the general cluster potential, because the rising dispersion 
is mostly driven by a larger scatter of $V_{\rm los}$ toward the low-velocity side of the 
systemic velocity of M87 (see Figures \ref{velrad_all} and \ref{velpa_hr}), and the majority of the 
blue GCs out to $R \sim 40\arcmin$ ($\sim$ 190 kpc) are actually tightly clustered around the 
systemic velocity of M87.\ A low-velocity excess (toward the North West of M87; Figure \ref{kin_des}) 
is also present in the velocity distribution of dwarf galaxies locally surrounding M87 
(Binggeli, Popescu \& Tammann 1993).\ So an excess of low-velocity GCs beyond $\sim$ 
30\arcmin~indicates a significant contamination from the intra-cluster population of GCs projected 
along the line of sight.

Moreover, beyond the central $\sim$ 30\arcmin, the GCs that are clustered around the systemic velocity 
of M87 tend to have redder colors than those with lower velocities.\ If we only consider relatively 
``red'' blue GCs, say those with $(g-i)_{0} > 0.75$, the resultant $v_{\rm rms}$ of the 
22 GCs between 30\arcmin~and 60\arcmin~is 252$_{-59}^{+38}$ km s$^{-1}$, whereas for the 
other 32 blue GCs with $(g-i)_{0} < 0.75$ the corresponding velocity dispersion is 524$_{-34}^{+42}$ 
km s$^{-1}$.\ Since lower luminosity galaxies have on average bluer colors than 
higher luminosity galaxies for both their blue and red GC systems (e.g.\ Peng et al.\ 2006), 
it is quite plausible that the high-dispersion ``bluer'' GCs are overwhelmingly contaminated 
in projection by the intra-cluster populations that have been tidally stripped from dwarf galaxies, 
whereas the relatively ``redder'' GCs trace the underlying potential of M87 more faithfully.\ 
This suggests that the stellar halo of M87 extends beyond, instead of being truncated at 
(Dopherty et al.\ 2009), $R_{\rm av}$ $\sim$ 150 kpc.\

\section{Velocity Distribution}\label{velhis}
\begin{figure*}
\centering
\includegraphics[height=0.35\textheight]{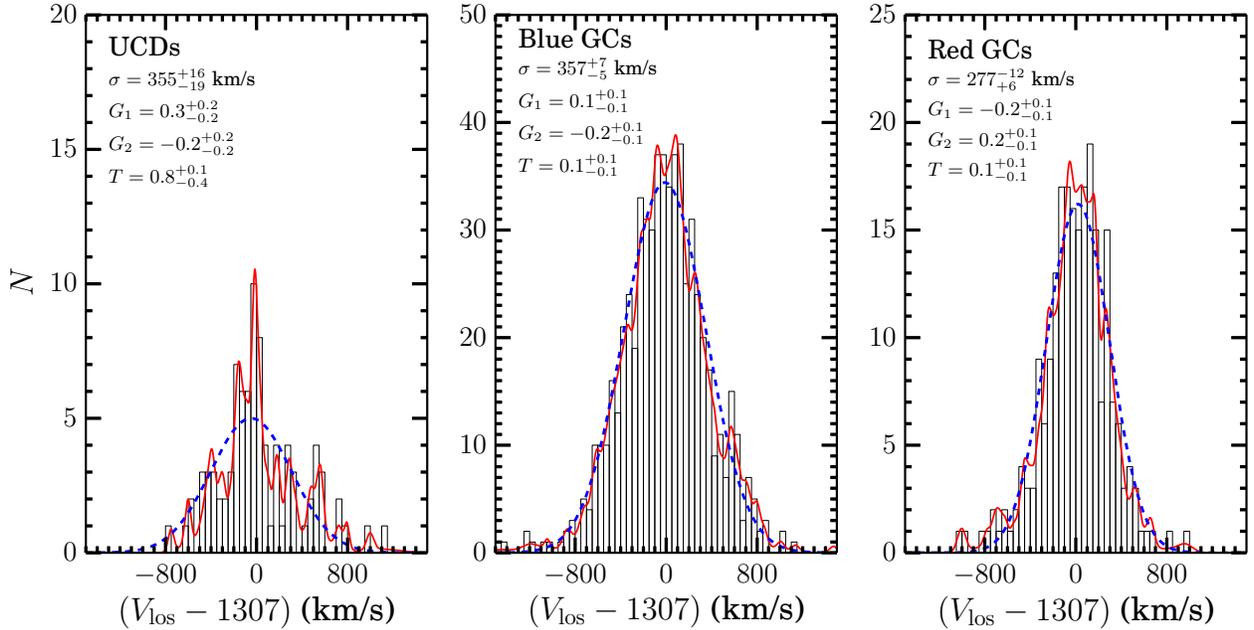}
\caption{
The line-of-sight velocity distribution of the UCDs (left), blue GCs (middle) and red GCs (right) in the inner 
30\arcmin~of M87.\ The histograms have a bin width of 50 km s$^{-1}$.\ The red solid curve overplotted on 
each histogram is the kernel density estimation of the observed distribution, and the blue dashed curve 
represents a Gaussian distribution, with the Gaussian $\sigma$ being equal to the standard deviation of the 
velocity distribution.\ Note that the Gaussian distributions have been scaled by the area under the histograms.\ 
The standard deviation $\sigma$, skewness $G_{\rm 1}$, kurtosis $G_{\rm 2}$ and $T$ parameter (see the 
Appendix for the definition) were also listed on the top of each panel.\ Note that, the UCDs have a sharper 
velocity distribution than the Gaussian, whereas the GCs (especially the blue GCs) have overall 
close-to-Gaussian distributions.
\label{velhis_all}}
\end{figure*}

\begin{figure*}
\centering
\includegraphics[height=0.35\textheight]{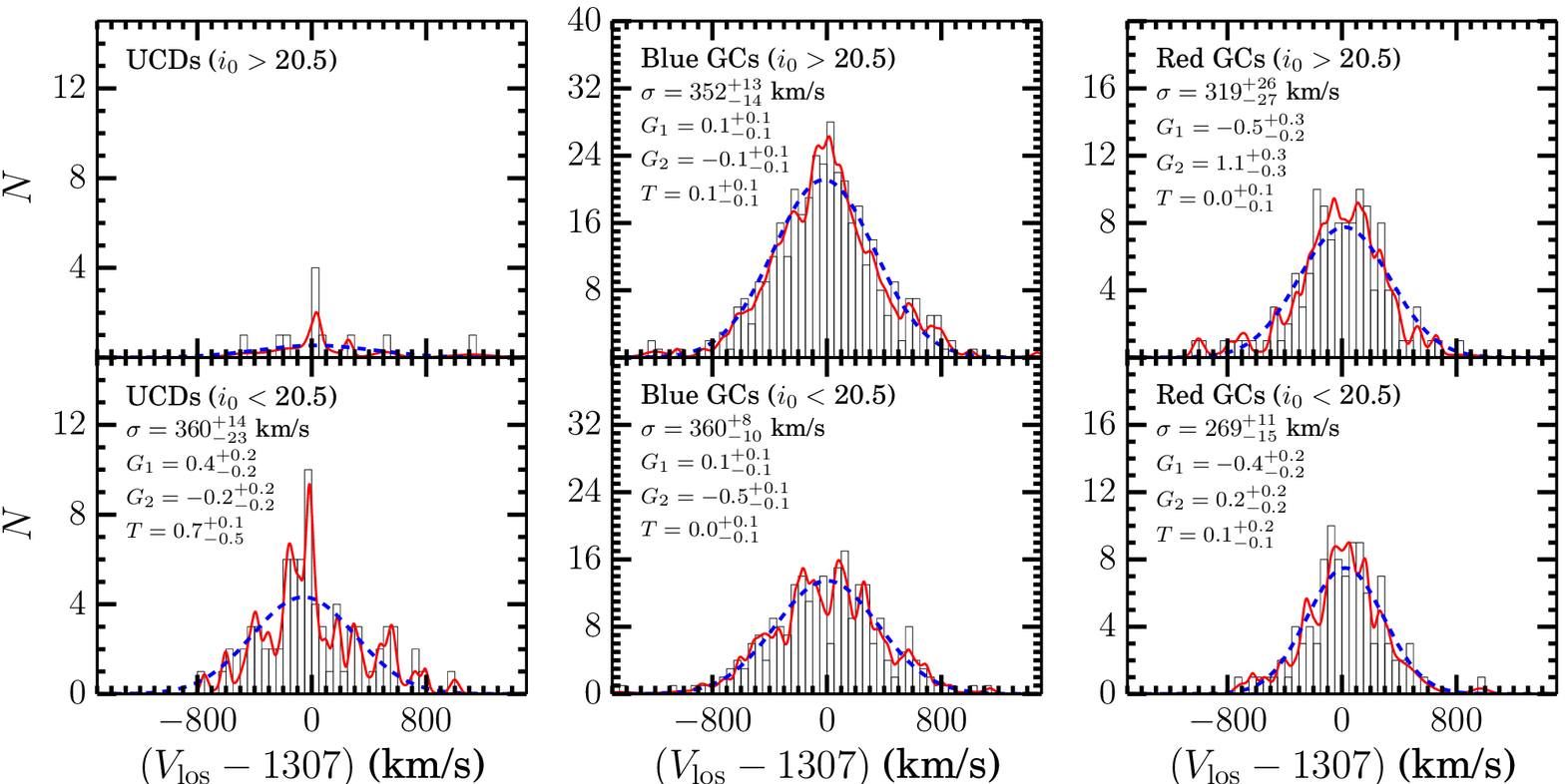}
\caption{
The line-of-sight velocity distribution of the subsamples of bright (lower panels) and faint (upper panels) UCDs (left), 
blue GCs (middle) and red GCs (right).\ The histograms have a bin width of 50 km s$^{-1}$.\ The red solid curve 
overplotted on each histogram is the kernel density estimation of the observed distribution, and the blue dashed 
curve represents a Gaussian distribution, with the Gaussian $\sigma$ being equal to the standard deviation of the 
velocity distribution.\ Note that the Gaussian distributions have been scaled by the area under the histograms.\ The 
standard deviation $\sigma$, skewness $G_{\rm 1}$, kurtosis $G_{\rm 2}$ and $T$ parameter were also listed on the top of each 
panel.
\label{velhis_sub}}
\end{figure*}

The line-of-sight velocity distribution is shaped by the global kinematics of a system.\
Figure \ref{velhis_all} presents the line-of-sight velocity histograms (with the systemic velocity 
of M87 being subtracted) of UCDs (left), blue GCs (middle) and red GCs (right).\ We show 
the bright and faint subsamples in Figures~\ref{velhis_sub}.\ 
The bin size of the histograms is 50 km s$^{-1}$, which was chosen to be larger than the typical measurement 
uncertainties.\ The red curve overplotted on each histogram is the adaptive kernel density estimation (KDE) of the 
velocity distribution.\ The KDE was constructed through a Gaussian-shaped kernel, with the Gaussian $\sigma$ 
being equal to the measurement uncertainty for each data point, and we further smoothed the resultant KDE with a 
Gaussian of $\sigma$ = 20 km s$^{-1}$ ($\sim$ 1/2.35$\times$ the bin size of the histograms).\ Also overplotted on 
each histogram as dashed blue curve is a scaled Gaussian distribution, with the Gaussian $\sigma$ being equal to 
the standard deviation of the observed distribution.\

To quantify the overall shape of the velocity distribution, we calculated the standard deviation $\sigma$, 
skewness $G_{\rm 1}$, and the kurtosis $G_{\rm 2}$, and they are listed on the top of each panel in Figures \ref{velhis_all}, 
and \ref{velhis_sub}.\ The skewness was calculated as the ratio of the third cumulant and the 1.5th power of the 
second cumulant, and the kurtosis was calculated as the ratio of the fourth cumulant and the square of the second 
cumulant.\ For small samples, skewness and kurtosis defined by cumulants are relatively unbiased compared to 
the traditional definitions with moments.\ A Normal distribution has both the skewness and kurtosis equal to zero, 
and a distribution with sharper peak and (especially) heavier tails has more positive kurtosis.\

The estimation of kurtosis is sensitive to extreme outliers.\ 
A meaningful estimation of kurtosis should reflect the overall shape of a distribution, instead 
of being driven by few extreme outliers.\ To obtain a robust estimation of kurtosis, we adopted 
a ``$\sigma$-clipping'' method in the kurtosis space.\ Specifically, for a data set of $N$ points, 
we resampled the original data set $N$ times (without replacement), with one different data point 
being taken out each time, similar to the technique of jackknife resampling.\ If the resultant kurtosis 
after taking out a given data point is more than 5$\sigma$ away from the mean, that data point is 
regarded as an outlier.\ This ``$\sigma$-clipping'' process was iterated until no further outlier was found.\
Among the full samples within $R$ $<$ 30$'$ (Section \ref{fullvel}), 2 UCD (2\%), 13 blue GCs (2\%) 
and 5 red GC (2\%) were found to be outliers for kurtosis estimation.\ 
The clipped outliers are mostly the extreme velocities in our samples, 
and it is quite possible that most of these outliers belong to the intra-cluster population of the Virgo cluster.\

As a fourth-moment measurement, it is not surprising that the standard kurtosis largely reflects 
the tail behavior.\ A complete description of the shape properties of a distribution should involve 
both the tailedness and peakedness.\ An outlier-insensitive, quantile-based alternative for the standard 
kurtosis, i.e.\ the $T$ parameter, was introduced by Moors (1998), and this alternative definition is 
expected to be more sensitive to the peakedness than $G_{\rm 2}$.\ A detailed introduction about 
$T$ is given in the Appendix.\ The $T$ parameter is defined such that a Normal distribution has a 
$T$ equal to 0, and a positive $T$ indicates heavier tails and (especially) a sharper peak than a 
Normal distribution.\ 

The $\sigma$, $G_{\rm 1}$ and $G_{\rm 2}$ reported below were calculated based on outlier-rejected 
samples.\ The 68\% confidence intervals for all the above mentioned shape parameters, including the $T$ 
parameter, were determined with by randomly resampling the real data sets.\ 
The estimated $G_{\rm 2}$ and $T$ for some specified radial bins are listed in Tables \ref{kinrad}, 
\ref{kinrad_b}, and \ref{kinrad_f}.

\subsection{The Full Samples}\label{fullvel}
The UCDs and blue GCs have similarly higher dispersion in the velocity distribution than the red GCs.\
The velocity distribution of UCDs is skewed toward the higher velocity tail, as quantified 
by a positive skewness, whereas the distribution of red GCs is skewed toward the lower velocity tail, as 
quantified by a negative skewness.\ The skewness difference between the three populations is partly 
reflected in their different systemic velocities $V_{\rm sys}$ (Table \ref{kinrad}).\
The velocity distribution of UCDs is noticeably sharper than a Gaussian (blue dashed curves in 
Figure \ref{velhis_all}), whereas the velocity distributions of GCs are only marginally sharper than a 
Gaussian.\ The different sharpness of the three distributions is well reflected in their different $T$ 
parameters.\ In addition, the UCDs and blue GCs have a similarly negative $G_{\rm 2}$, suggesting  
slightly lighter tails than a Gaussian.\ As we will show later (Section \ref{jeans}), the peaky velocity 
distribution of UCDs is consistent with a radially-biased velocity dispersion tensor at large galactocentric 
distances, whereas the lighter tails are in line with a tangentially-biased velocity dispersion tensor 
at small distances.\

\subsection{Bright and Faint Subsamples}
Since there are only 11 UCDs at $i_{0}$ $>$ 20.5 mag, the calculation of skewness and kurtosis for these 
faint UCDs is subject to large uncertainties and bias, and will not be discussed further.\ For the blue GCs, the 
bright and faint subsamples have similar velocity dispersion.\ Nevertheless, the bright blue GCs have 
much more negative $G_{\rm 2}$ and marginally lower $T$ than the faint ones.\ For the red GCs, the 
bright subsample has a significantly smaller (by $\sim$ 70 km s$^{-1}$) velocity dispersion and larger 
$G_{\rm 2}$ than the faint subsample.\ 

Most of the confirmed GCs with $i_{0} \gtrsim$ 21.5 mag were observed by S11 with the Low Resolution 
Spectrometer (LRIS) on Keck.\ The Keck/LRIS survey of S11 only covered the central $\sim$ 1.5 -- 5.5\arcmin~of 
M87, and as a result the confirmed faint GCs are primarily located in the central region (Figure \ref{irad_all}).\ 
To check if the velocity distribution differences between the bright and faint GCs are driven by the 
observational bias in spatial coverage for the sample of faint GCs, we derived the shape parameters 
of the velocity distribution for the bright and faint GCs within the central 5\arcmin.\ It turns out similar 
differences between the bright and faint subsamples still exist for both the blue and red GCs.\ 

Furthermore, previous studies (S11; Agnello et al.\ 2014) found that the M87 GCs of different colors may 
exhibit different kinematical properties.\ So we also checked for any possible color bias for the bright 
and faint subsamples.\ The median $(g-i)_{0}$ colors of the bright and faint blue GCs 
are 0.76 and 0.74 respectively, suggesting that there is no significant color bias for the bright and faint subsamples.\ 
For the red GCs, the median $(g-i)_{0}$ colors of the bright and faint subsamples are 0.95 and 1.00 respectively.\ 
When we divide the red GCs into two $(g-i)_{0}$ color groups with a division color of 0.97, the above 
mentioned difference still exist for subsamples in each color group, although there is a systematic difference 
between the ``bluer'' and ``redder'' groups, in the sense that the corresponding subsamples in the 
``bluer'' group have about $\sim$ 70 km s$^{-1}$ lower velocity dispersion than those in the the ``redder'' group.\

\subsection{Radial Trend of Kurtosis of the blue GCs}
\begin{figure}
\centering
\includegraphics[height=0.2\textheight]{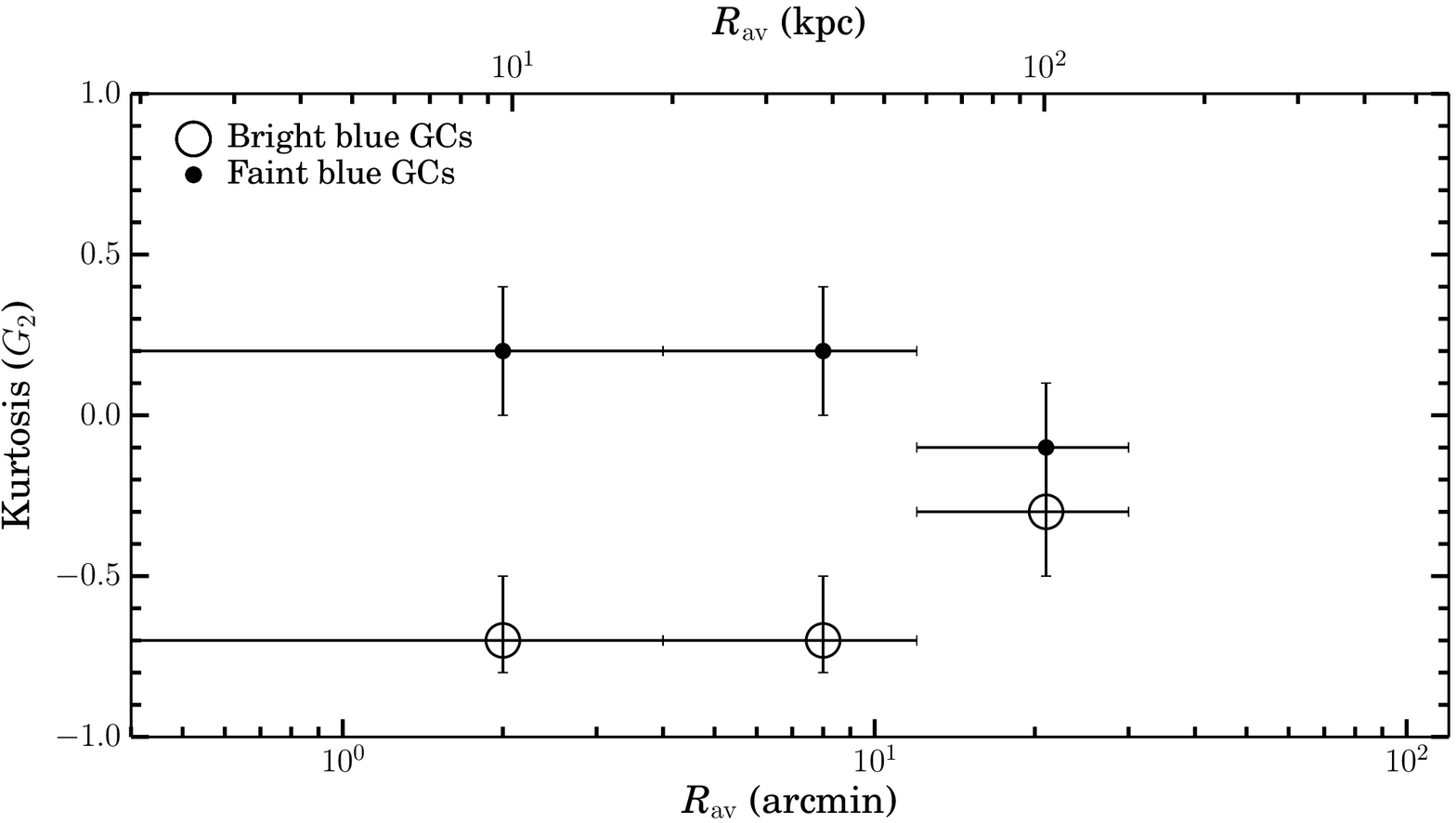}
\caption{
Radial variation of velocity Kurtosis of the bright ({\it large open circles}) and faint ({\it small filled circles}) blue GCs.
\label{kurt_i_bgc}}
\end{figure}

The large sample size of the blue GCs allows us to explore the radial trend of the 
shape parameters for the velocity distribution.\ Figure \ref{kurt_i_bgc} shows the 
kurtosis for the bright and faint blue GCs in three fixed radial bins, namely 0\arcmin--4\arcmin, 
4\arcmin--12\arcmin, and 12\arcmin--30\arcmin.\ It is significant that the bright subsamples 
have systematically lower kurtosis than the faint ones.\ 

\subsection{Interpretation: from kurtosis to orbital anisotropies}\label{orbitstr}

For a given number density profile, there is a one-to-one relationship between the kurtosis 
and the orbital anisotropy parameter $\beta$ ($\equiv$ 1$-$$\sigma_{t}^{2}$/2$\sigma_{r}^{2}$, 
where $\sigma_{t}$ and $\sigma_{r}$ are the tangential and radial components of the velocity 
dispersion ellipsoid; positive (negative) $\beta$ corresponds to radially (tangentially) anisotropic 
orbital structures), under the assumption that the velocity dispersion and $\beta$ are constant with 
galactocentric radius (Napolitano et al.\ 2009).\ Napolitano et al.\ (2009) found that $\beta$ can be directly 
estimated through deprojection of observables involving the kurtosis of $V_{\rm los}$ and volume number 
density profiles of the tracer population (see the formulae B10, B11 and B12 in Napolitano et al.\ 2009).\ 

With the kurtosis and surface density profiles 
\footnote{To deproject the observed surface number density profiles 
$N_{R}$ to volume number density profiles $n_{r}$, we assumed a spherically symmetric geometry, which gives 
\begin{equation}
n_{r} \propto \int_{r} ^{\infty}\frac{d N_{R}}{d R}\frac{d R}{\sqrt{R^{2}-r^{2}}}
\label{eq1}
\end{equation}
}
(parameterized as S\'ersic functions, Section \ref{radprof}) in hand, the average $<\beta>$ for 
UCDs, blue GCs and red GCs were estimated (through numerical integration from one effective 
radius to infinity for each of the three populations, also listed in Tables \ref{kinrad}, \ref{kinrad_b}, 
and \ref{kinrad_f}) to be $-$0.3$_{-0.4}^{+0.3}$, $-$0.3$_{-0.2}^{+0.2}$, and 0.2$_{-0.1}^{+0.1}$ 
respectively for the full samples.\ For the bright subsamples, the corresponding $<\beta>$ are 
$-$0.3$_{-0.4}^{+0.3}$, $-$1.0$_{-0.4}^{+0.3}$, and 0.2$_{-0.2}^{+0.2}$ for the UCDs, blue GCs and 
red GCs respectively; For the faint subsamples, the $<\beta>$ are estimated to be 
$-$0.1$_{-0.2}^{+0.1}$ and 0.7$_{-0.1}^{+0.1}$ for the blue and red GCs respectively.\
We emphasize that the adopted approach to calculate $\beta$ is only strictly 
applicable to a constant velocity dispersion profile which is not true for either of our three populations, so our 
estimation of $\beta$ here should be regarded at most as a zeroth-order approximation of the average 
$<\beta>$.\ Later in this paper (Section \ref{jeans}), we will solve the Jeans equations for the radial 
profiles of $\beta$.\

Taken at face value, among the three populations, the red GCs have on average the highest 
radial anisotropy, whereas the blue GCs have the lowest radial anisotropy.\ 
In addition, the faint subsamples tend to have higher radial anisotropies than the 
bright subsamples.\ Given the good correspondence between kurtosis and $<\beta>$, the 
radial kurtosis profiles shown in Figure \ref{kurt_i_bgc} indicate that the faint blue GCs 
are more radially-biased than the bright ones.\ 

\section{Kinematic Modeling}\label{kinem}

\subsection{Methodology}\label{method}
Under the assumption that the intrinsic angular velocity of a system, of either GCs or UCDs, is only 
a function of the galactocentric distance $r$, the (projected) average line-of-sight velocities vary sinusoidally
with the projected azimuth $\theta$ (e.g.\ C\^ot\'e et al.\ 2001).\ As a consequence, it is customary to fit the line-of-sight 
velocities of GC and UCD systems with sine or cosine curves as a function of projected azimuth $\theta$, 
in order to determine the rotation amplitude and rotation axis (e.g.\ Cohen \& Ryzhov 1997; Kissler-Patig \& 
Gebhardt 1998; C\^ot\'e et al.\ 2001; S11; Pota et al.\ 2013).\ Assuming Gaussian distributions 
for both the measurement uncertainties and the intrinsic velocity dispersions, the likelihood of a model fit 
to given observations (e.g.\ $v_{i}$$\pm$$\Delta v_{i}$) is:
\begin{equation}
  {\cal L} \propto \prod_i {1 \over \sqrt{\sigma_{\rm p}^2 + \Delta v_{i}^2}}
   \exp{\left[{-{1\over 2} {(v_{i} - v_{\rm mod})^2 \over \sigma_{\rm p}^2 + \Delta v_{i}^2}}\right]}.
\label{eq2}
\end{equation}
where $\sigma_{\rm p}$ is the intrinsic (projected) velocity dispersion, and 
\begin{equation}
v_{\rm mod} = v_{\rm sys} + v_{\rm rot} {\rm sin}(\theta - \theta_{0})
\label{eq3}
\end{equation}
with $\theta_{0}$ being the position angle (PA, measured east of 
north) of the rotation axis.\ Maximization of Equation \ref{eq1} is equivalent to minimizing the 
$\chi^2$ statistic (e.g.\ Bergond et al.\ 2006):
\begin{equation}
\chi^2 = \sum_i \left\{ \frac{(v_i-v_{\rm mod})^2}{\sigma_{\rm p}^2+(\Delta v_i)^2}
+ \ln\left[\sigma_{\rm p}^2 + (\Delta v_i)^2\right] \right\}
\label{eq4}
\end{equation}
By minimizing Equation \ref{eq4} for a given dataset, we can determine the systemic velocity $v_{\rm sys}$, 
rotation amplitude $v_{\rm rot}$, rotation axis $\theta_{0}$ and the intrinsic velocity dispersion $\sigma_{\rm p}$.\ 
Note that we have assumed the kinematic axis ratio $q$ to be 1 when writing Equation \ref{eq3} in order to 
be consistent with previous kinematical studies of M87 GCs (e.g.\ C\^ot\'e et al.\ 2001; S11).\
The results are essentially the same when fixing $q$ to the photometric axis ratio of the diffuse stellar light.\  
Considering the possible (unknown) inclination of the rotation axis with respect to the plane of sky, $v_{\rm rot}$ 
determined here should be regarded as a lower limit.\
To estimate the uncertainties of the fitted parameters, we randomly resampled the real dataset and repeated 
the kinematic fitting to the resamples, and then obtained the 68\% confidence intervals from the resultant parameter 
distribution.\

\subsection{Bias and Significance}\label{bias}
Kinematics fitting to discrete data points tends to overestimate the intrinsic rotation, and the degree of 
overestimation or bias depends on the sample size, the azimuthal distribution of data points, and the 
importance of rotation as compared to the dispersion (e.g.\ Sharples et al.\ 1998; Romanowsky et al.\ 2009; S11).\ 
Reliability of the fitted rotation amplitudes can be quantified in two mutually related ways, one is 
the most likely level of bias, and the other one is the significance 
(or confidence level, CL) of the fitted rotation.\

{\it Bias}. To check the level of bias for the best-fit rotation, we first constructed a series of  
kinematic models for each of the three populations, with the rotation amplitudes varying 
from 0 to 250 km s$^{-1}$ and the other kinematic parameters being fixed at their best-fit 
values from the real data sets.\ Then, starting from each of these kinematic models, velocities at 
each observed GC or UCD location were randomly drawn from a normal distribution, with 
the variance being equal to a quadrature combination of the velocity dispersion 
and measurement uncertainties.\ In particular, at each input rotation amplitude, 5000 Monte Carlo samples 
were generated for each population, and the standard kinematic fitting (Section \ref{method}) 
was carried out for these mock samples.\ Lastly, for each {\it input} rotation amplitude, we determined 
the median of the corresponding 5000 best-fit {\it output} rotation amplitudes, and this defines a 
one-to-one relation between the {\it input} and the most likely {\it output}, which can be used to estimate 
the most probable bias in our best-fit rotation for the real data sets. 

{\it Significance}. 
we follow the procedure first introduced by Sharples et al.\ (1998) to estimate the CL of our best-fit 
rotation amplitudes (see also Romanowsky et al.\ 2009).\ In particular, for each of the three populations, 
we randomly shuffled the position angles of the observed data points for 5000 times, and repeated the 
kinematics fitting to each realization.\ A random shuffling of the position angles can erase (if any) signatures 
of any possible rotation.\ Therefore, if the percentage of random realizations that lead to fitted rotation amplitudes 
greater than or equal to the best-fit value $v_{\rm rot}$ for the real data set is $p$, then the confidence level of 
$v_{\rm rot}$ can be estimated as $1-p$.\ In this paper, confidence levels $>$ 90\% are regarded to be 
significant.

\subsection{Global Kinematics}
\subsubsection{The Full Samples}
\begin{figure*}
\centering
\includegraphics[height=0.21\textheight]{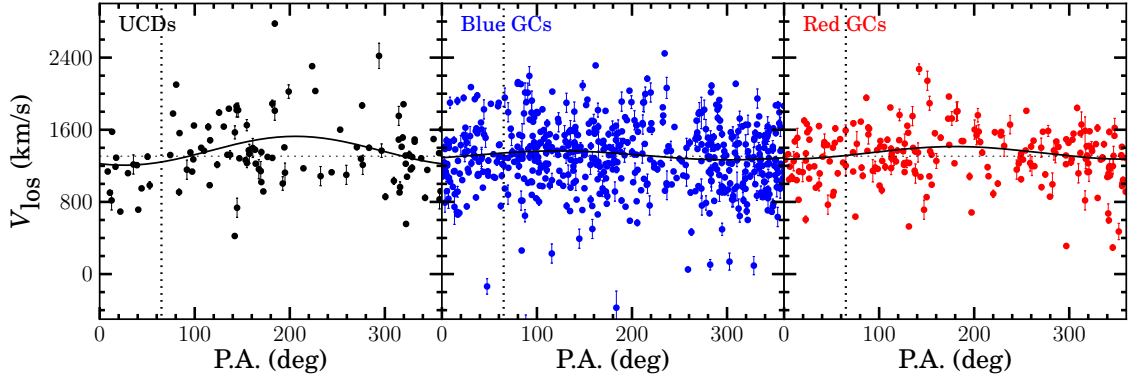}
\caption{
The line-of-sight velocity variations as a function of PA (East of North) for the full samples of UCDs (left), 
blue GCs (middle), and Red GCs (right).\ Overlaid on the data in each panel is the best-fit sine curve.\ 
The horizontal dotted lines mark the systemic radial velocity of 1307 km s$^{-1}$ of M87, and the vertical 
dotted lines mark PA of the photometric minor axis of M87.
\label{velpa_wh}}
\end{figure*}

Figure \ref{velpa_wh} shows the azimuthal variation of line-of-sight velocities for all M87 UCDs within the central $30\arcmin$, 
along with the blue GCs and red GCs for comparison.\ Overlaid on the data in each panel is the best-fit sine curves.\ 
The fitting results are summarized in Table \ref{kinrad}.\ The UCDs and blue GCs have similar intrinsic 
velocity dispersion.\ The rotation amplitude of UCDs is more than 4 (2) times stronger than that of the blue (red) GCs.\ 
Additionally, the rotation axis of UCDs is roughly orthogonal to that of blue GCs.\ It is interesting that the red GCs, which 
have a smaller velocity dispersion than the UCDs and blue GCs, have a rotation axis that is more aligned with the UCDs 
rather than the blue GCs.\ 
Our best-fit parameters for the blue and red GCs are generally consistent with those determined by 
S11 (see their Table 14) within the mutual uncertainties.\ Note that the rotation angles reported by S11  
are the direction of maximum rotation amplitude, which is 90$^{\circ}$ offset from the angular momentum vector.\ 
In addition, the kinematics parameters reported by S11 were already bias-corrected.

\begin{figure*}
\centering
\includegraphics[height=0.32\textheight]{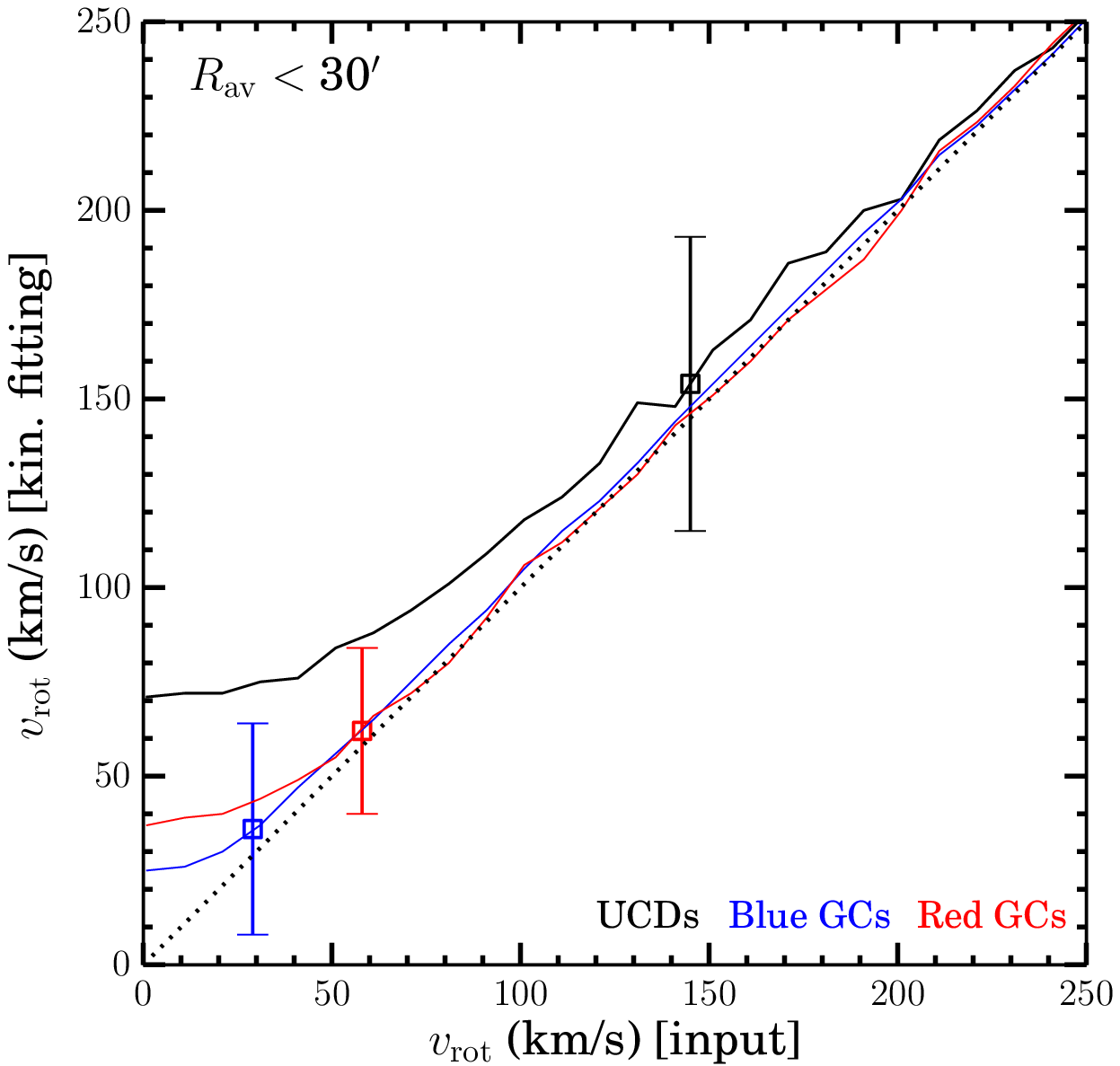}
\includegraphics[height=0.32\textheight]{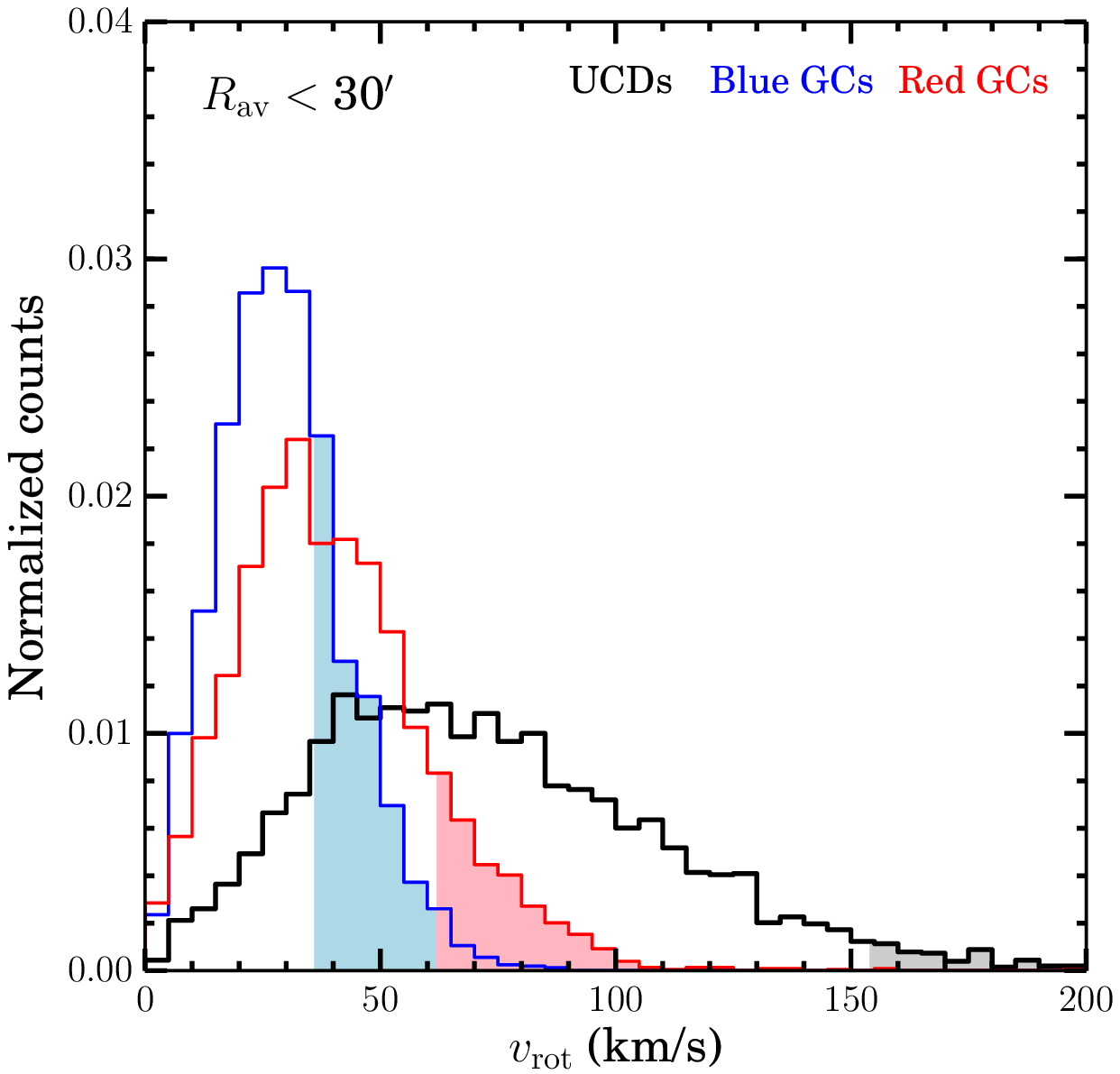}
\caption{
Estimation of the most likely bias ({\it left}) and the confidence level ({\it right}) of the best-fit rotation amplitudes 
(Figures \ref{velpa_wh} and \ref{radc_kin}) for UCDs, blue GCs, and red GCs at $R_{\rm av} < 30\arcmin$.\
In both the {\it left} and {\it right} panels, the black, blue and red colors correspond to results for the UCDs, 
blue GCs and red GCs respectively.\ In the {\it left} panel, curves of different colors correspond to the bias 
correction relation, as described in Section \ref{bias}, for the three populations, and the {\it open squares} 
mark the best-fit rotation amplitudes for the real data sets.\ In the {\it right} panel, the open histograms represent 
the normalized distribution of best-fit rotation amplitudes for randomly shuffled samples, and the {\it filled} 
region of each histogram marks the probability $p$ of obtaining a rotation amplitude greater than the 
best-fit value for the real data set just by chance, with the confidence level (CL) for the rotation being 
defined as $1-p$.
\label{kinbias}}
\end{figure*}

Following the procedure described in Section \ref{bias}, we estimated the most likely bias and the significance 
(or CL) of our best-fit rotation amplitudes, and the results are shown in Figure \ref{kinbias} and Table \ref{kinrad}.\
According to bias test, the intrinsic rotation amplitude for UCDs is most probably overestimated 
by $\sim$ 10 km s$^{-1}$, for blue GCs $\sim$ 6 km s$^{-1}$, and for red GCs $\sim$ 1--2 km s$^{-1}$.\
As to the confidence levels, the probability that we found a rotation amplitude greater than or equal to the 
best-fit value for UCDs purely by chance is $\sim$ 2\% (CL = 98\%), the probability is $\sim$ 27\% (CL = 73\%) 
for blue GCs, and $\sim$ 12\% (CL = 87\%) for red GCs.\

\begin{figure*}
\centering
\includegraphics[height=0.25\textheight]{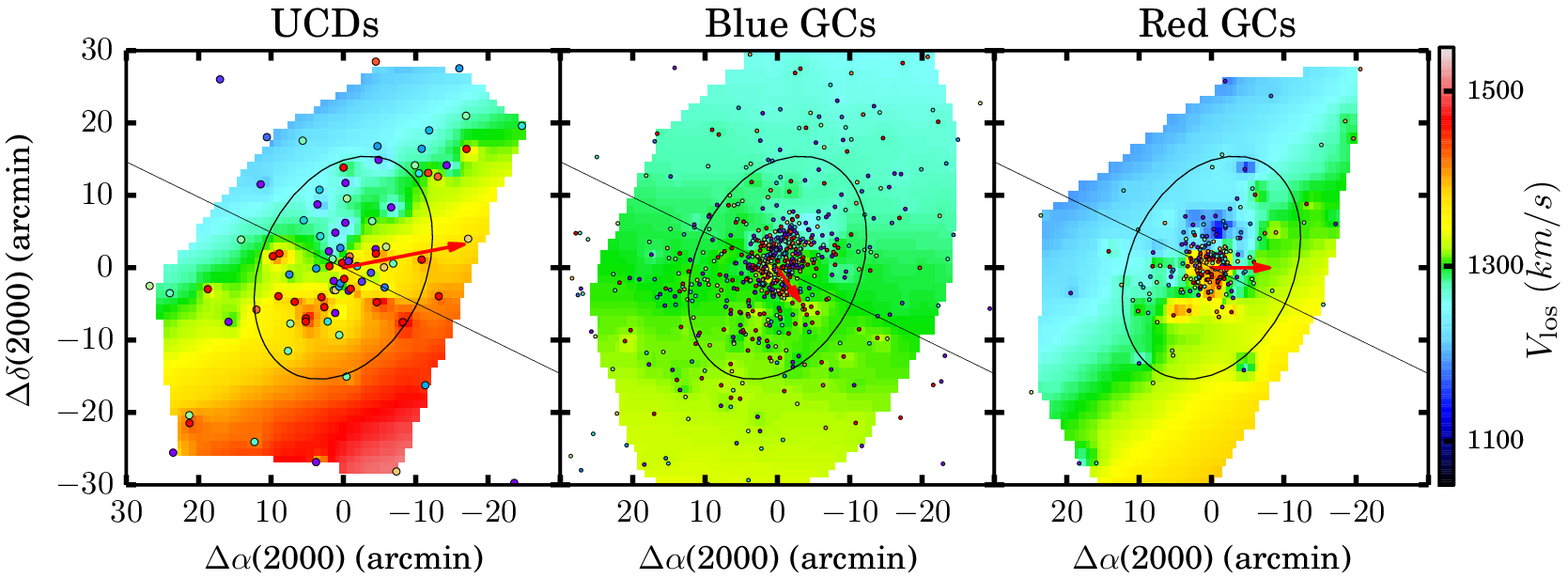}
\caption{
Spatial distribution of the UCDs ({\it left}), blue GCs ({\it middle}) and red GCs ({\it right}) are over plotted 
on their respective surface fitting ({\it the color background}) to the spatial distribution of line-of-sight 
velocities with the Kriging technique for the inner 30\arcmin~of M87.\ The {\it black ellipses} 
represent the stellar isophotes of M87 at 10$R_{\rm e}$, and the {\it black solid line} marks the 
photometric minor axis of M87 in each panel.\ The {\it red arrows}, with the length being 
proportional to the rotation amplitude, mark the direction of rotation axis from our global kinematics 
fitting to the inner 30\arcmin~(Figure \ref{velpa_wh}).\
The global kinematics fitting, which is primarily driven by the inner regions that contain most of 
the data points, matches the central velocity field from Kriging surface fitting.\
Among the three populations, the blue GCs seem to have an overall velocity field more aligned with the 
photometric major axis than the other two populations.\ See the text for details.
\label{radc_kin}}
\end{figure*}

Surface fitting to the spatial distribution (within the inner 30\arcmin) of line-of-sight velocities is 
presented in Figure \ref{radc_kin} as a color-coded background for each of the three populatons.\ 
The data points over plotted on the fitted surface are also color-coded according to their individual line-of-sight 
velocities.\ In order to bring out details of the fitted surface, individual data points with velocities $>$1550 
km s$^{-1}$ or $<$1050 km s$^{-1}$ were not distinguished in colors from those with velocities = 1550 
km s$^{-1}$ or = 1050 km s$^{-1}$ respectively.\
The surface fitting was carried out with the Kriging technique as implemented in {\sc R} 
package {\sc fields}.\ The Kriging technique has been recently used for exploring the velocity field 
of GC systems (e.g.\ Foster et al.\ 2013) and the galactic spatial distribution of metallicities 
(e.g.\ Pastorello et al.\ 2014).\ Instead of simply applying an inverse distance weighting, Kriging 
takes into account both the spatial configuration and covariances of the dataset when assigning 
weights to neighboring data points for interpolation.\ In this work, the spatial covariance is assumed 
to be an exponential function of separation distance, and the smoothing parameter $\lambda$ was 
fixed.\ The global rotation axes determined from our kinematics fitting were indicated as red arrows 
in Figure \ref{radc_kin}, and the arrow length is proportional to the best-fit rotation amplitude.\

Our global kinematics fitting is essentially driven by the clustering trend of $V_{\rm los}$ along the 
azimuthal direction, and primarily reflects the velocity field in the central region where most data 
points are located.\ Accordingly, as shown in Figure \ref{radc_kin}, there is generally a good match 
between the direction of arrows and the fitted Kriging velocity field in the central regions.\
The Kriging surface fitting is driven by the overall data configuration and spatial covariance, and 
in particular, by definition it is not influenced/biased by spatial clustering of data points.\
Of the three populations, the blue GCs have an overall velocity field that seems to be more closely 
aligned with the photometric major axis, while the UCDs have a velocity field in better agreement 
with that of the red rather than the blue GCs.\

The significantly stronger rotation of UCDs as compared to the GCs suggests that the UCDs are kinematically 
distinct from the GCs.\ The smaller velocity dispersion of red GCs is in agreement with the fact that red GCs are 
more centrally concentrated than the blue GCs and UCDs (Figure \ref{ucdgc_rp}).\ 
Similar to the blue GCs, a more or less minor axis rotation was also recently detected from integrated 
stellar-light spectra in the central one effective radius of M87 (e.g.\ Arnold et al.\ 2013).\ In addition, a twist 
of the velocity field within the central half arc minute was recently reported by Emsellem, Krajnovi\'c \& Sarzi (2014) 
based on IFU spectra of the integrated stellar light.\ The misaligned velocity field across different radii  
and among different kinematic tracers (e.g.\ UCDs vs. GCs) all suggest that the halo of M87 is most probably 
triaxial, instead of being axisymmetric (e.g.\ Schwarzschild 1979; Statler 1991; Franx, Illingworth \& de Zeeuw 1991; 
van den Bosch et al.\ 2008; Hoffman et al.\ 2009; Emsellem, Krajnovi\'c \& Sarzi 2014).\ 

\subsubsection{The Bright and Faint Subsamples}
\begin{figure*}
\centering
\includegraphics[height=0.30\textheight]{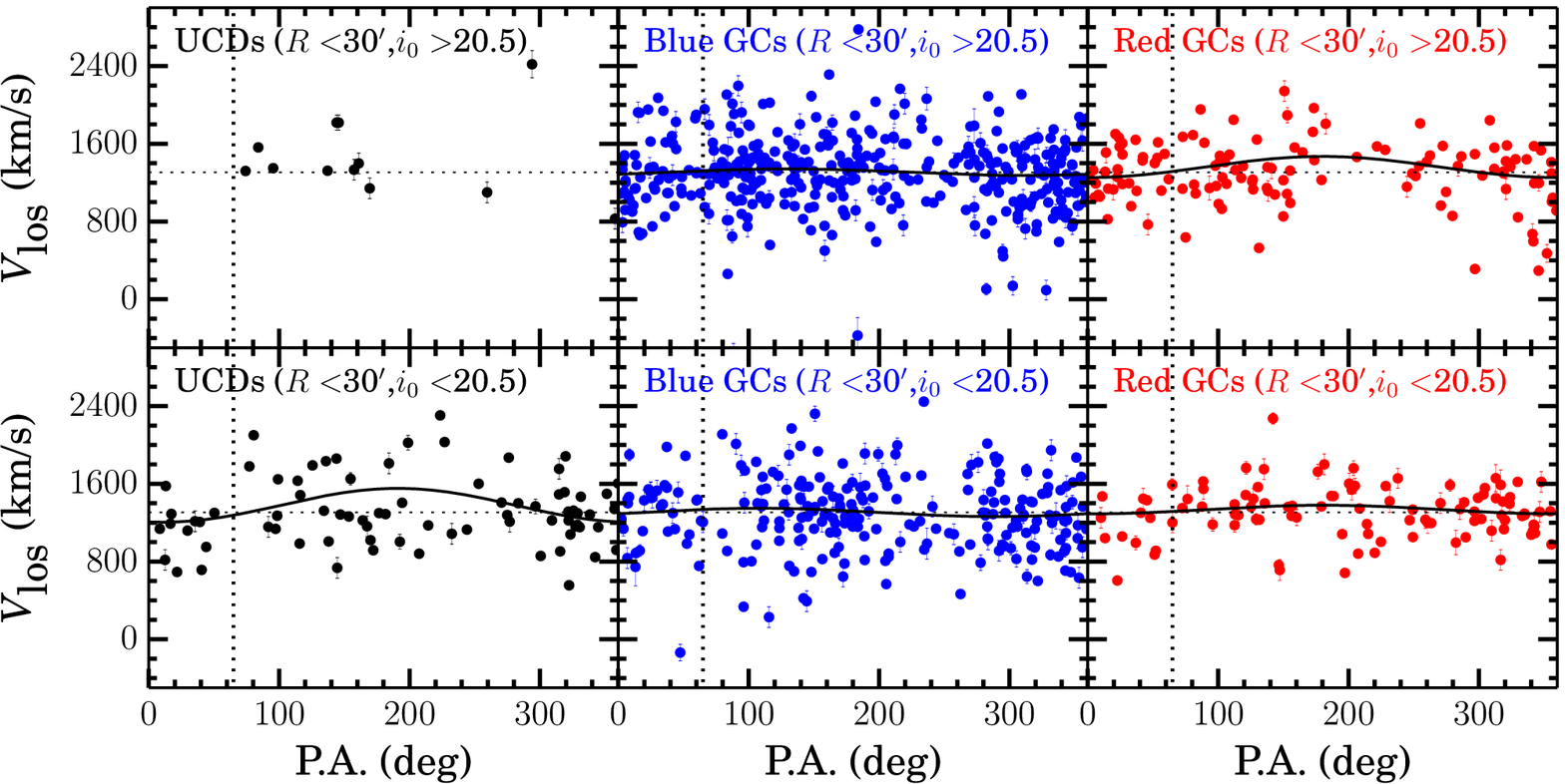}
\caption{
Radial velocity variations as a function of PA for the brighter (lower panels) 
and fainter (upper panels) subsamples of UCDs, blue GCs (middle), and Red GCs (right).\ 
Overlaid on the data in each panel is the best-fit sine curves.\ The horizontal dotted lines mark the systemic radial 
velocity of 1307 km s$^{-1}$ of M87, and the vertical dotted lines mark PA of the photometric minor axis of M87.
\label{velpa_wh_i0}}
\end{figure*}

In this subsection, we explore the possible differences in global kinematics between the bright 
and faint subsamples.\ The relevant kinematic fitting results are summarized in Tables \ref{kinrad_b} and 
\ref{kinrad_f}.\ Figure~\ref{velpa_wh_i0} shows the azimuthal variation of line-of-sight velocities for 
the bright and faint subsamples of UCDs and GCs separately.\ Again, the best-fit sine curves are 
overlaid on the data points.\ For the blue GCs, no significant rotation was found for either the bright 
or faint ones.\ The bright red GCs exhibit a significantly lower (by $\sim$ 50 km s$^{-1}$) velocity 
dispersion and much less significant rotation than the faint subsample.\ The sample size of faint 
UCDs is too small to give meaningful kinematic parameters.\ The kinematic parameters of the 
bright UCDs resemble those of the full sample.\ 

\subsection{Radial variation of Kinematics}\label{kin_rad}
\begin{figure*}[tp]
\centering
\includegraphics[height=0.55\textheight]{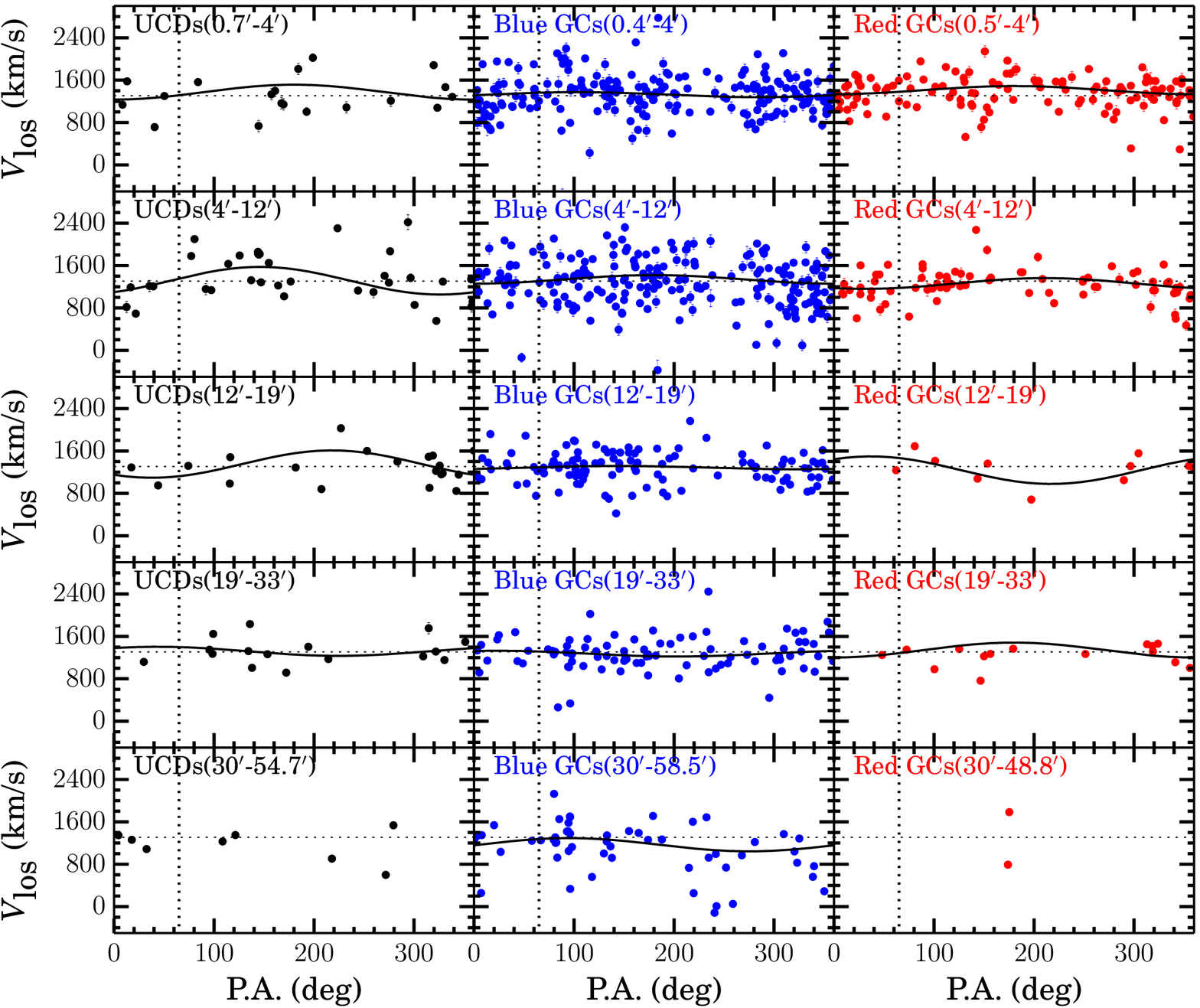}
\caption{
Radial velocity variations as a function of PA in five elliptical annuli for the full samples  
of UCDs (left), blue GCs (middle), and red GCs (right).\ The dotted lines mark the systemic radial 
velocity of 1307 km s$^{-1}$ for M87.\ The five elliptical annuli plotted here are selected to be representative 
of the primary features exhibited in the radial profiles of kinematics shown in Figure \ref{kin_rav}.
\label{velpa_hr}}
\end{figure*}

\begin{figure*}
\centering
\includegraphics[height=0.7\textheight]{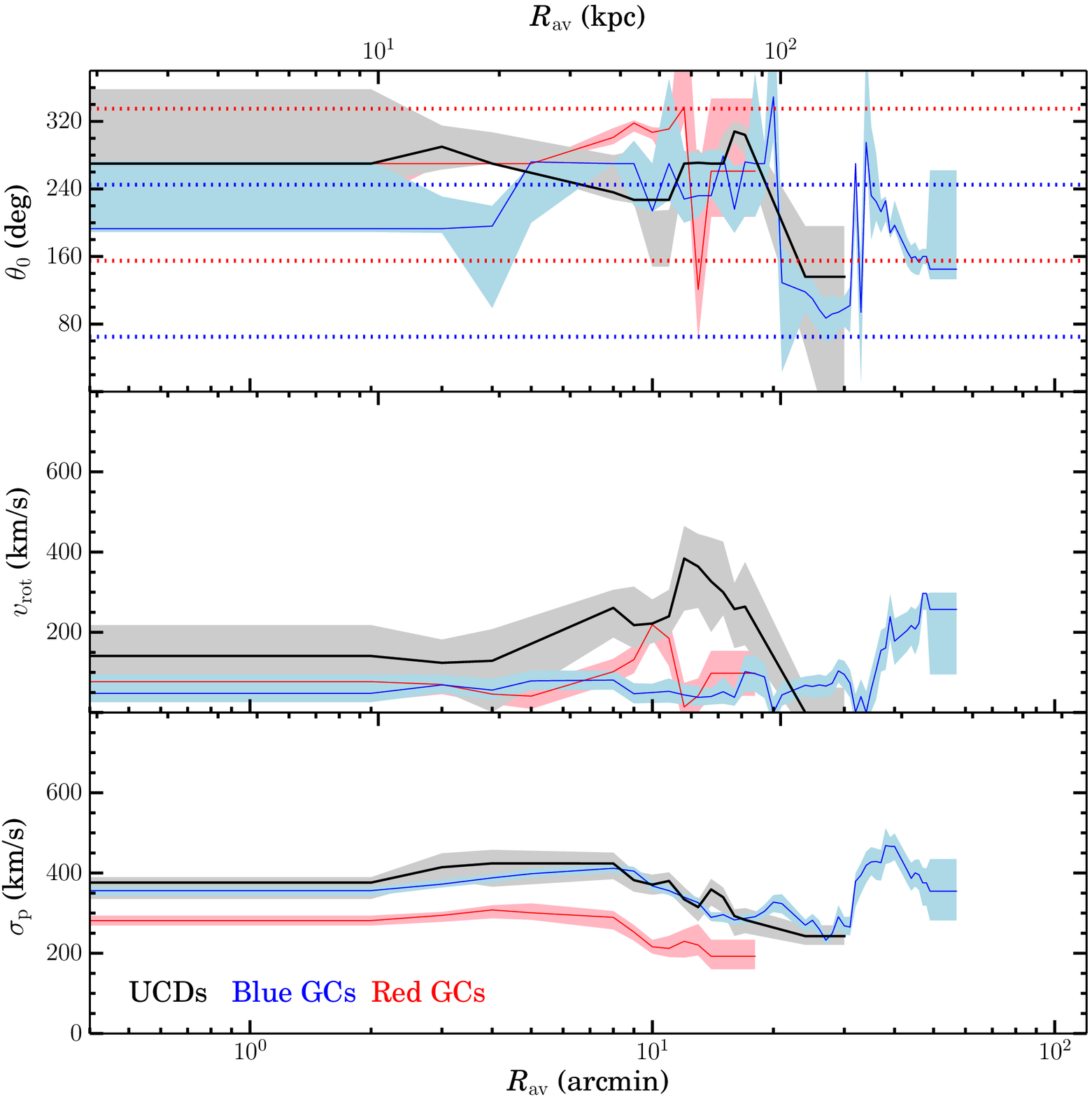}
\caption{
Radial variation of the best-fit kinematic parameters.\ 
The profiles were constructed with sliding bins of fixed radial widths, as in Figure \ref{vrms_all}.\
The red (blue) dotted horizontal lines in the top panel mark the photometric semi-major (semi-minor) axes.\ 
To be used in constructing the profiles, at least 19 data points should be available for the kinematics fitting 
in a given radial bin.\ The solid curves are the best-fit kinematics parameters, and the shaded regions 
correspond to the 68\% confidence limits.
\label{kin_rav}}
\end{figure*}

Radial variation of the kinematic parameters ($\theta_{0}$, $v_{\rm rot}$, and $\sigma_{\rm p}$) is presented in 
Figure~\ref{kin_rav}.\ Fitting results for data points falling into some fixed radial range are listed in 
Tables \ref{kinrad_b} and \ref{kinrad_f}.\ The way that we constructed the profiles in Figure~\ref{kin_rav} is 
the same as in Section~\ref{vrmsrad} which presents the $v_{\rm rms}$ profiles (Figure \ref{vrms_all}).\ 
Basically, we performed kinematic fitting to data points that fall into each individual sliding radial bin.\ 
Figure \ref{velpa_hr} shows the azimuthal variations of line-of-sight velocities of UCDs (the left column), 
blue GCs (the middle column) and red GCs (the right column) in five different elliptical annuli.\ 
The five annuli were selected to be representative of the key features in the radial profiles of the 
UCD kinematics (Figure \ref{kin_rav}).\ Note that the kinematic parameters plotted in Figure \ref{kin_rav} 
are not corrected for possible bias.\

Based on Figure~\ref{kin_rav} and Table \ref{kinrad}, the UCDs and blue GCs have similar 
$\sigma_{\rm p}$ profiles.\ $\sigma_{\rm p}$ of the red GCs is systematically lower than the other 
two populations across the full radius range.\ The strongest rotation of UCDs is found around 
$R_{\rm av}$ $\sim$ 8\arcmin -- 16\arcmin, where the rotation axis is similar to the full 
sample of UCDs.\ The blue GCs exhibit insignificant rotation across the full radius range, 
whereas the red GCs show marginally significant rotation around $\sim$ 10\arcmin.\ 
We note that, in the Kriging maps shown in Figure \ref{radc_kin}, a weak, but visible, 
gradient along the photometric major axis can be seen for blue GCs beyond the central $\sim$ 5$\arcmin$, 
suggesting that the usual one-dimensional kinematics fitting ($V_{\rm los}$ vs. PA) and the two-dimensional 
surface fitting are complementary to each other.\
The seemingly strong rotation for blue GCs beyond the central 30\arcmin~turns out to be not significant.\ 

The relevant kinematic parameters for the bright and faint subsamples of GCs in some specified 
radial bins are listed in Tables \ref{kinrad_b} and \ref{kinrad_f} separately.\ No significant 
rotation was found across the full radius range for both the bright and faint GC subsamples.\

Lastly, we checked the color-magnitude distribution of the UCDs in the radius range 
($10\arcmin$--$20\arcmin$) where the strongest rotation was found, and it turns out that these UCDs follow 
a color-magnitude relation indistinguishable from the whole sample.\ This suggests that the strong rotation 
most probably reflects the global kinematics of the UCD system.\

\subsection{Velocity Field of the Surrounding dEs}\label{kindes}
\begin{figure*}
\centering
\includegraphics[height=0.28\textheight]{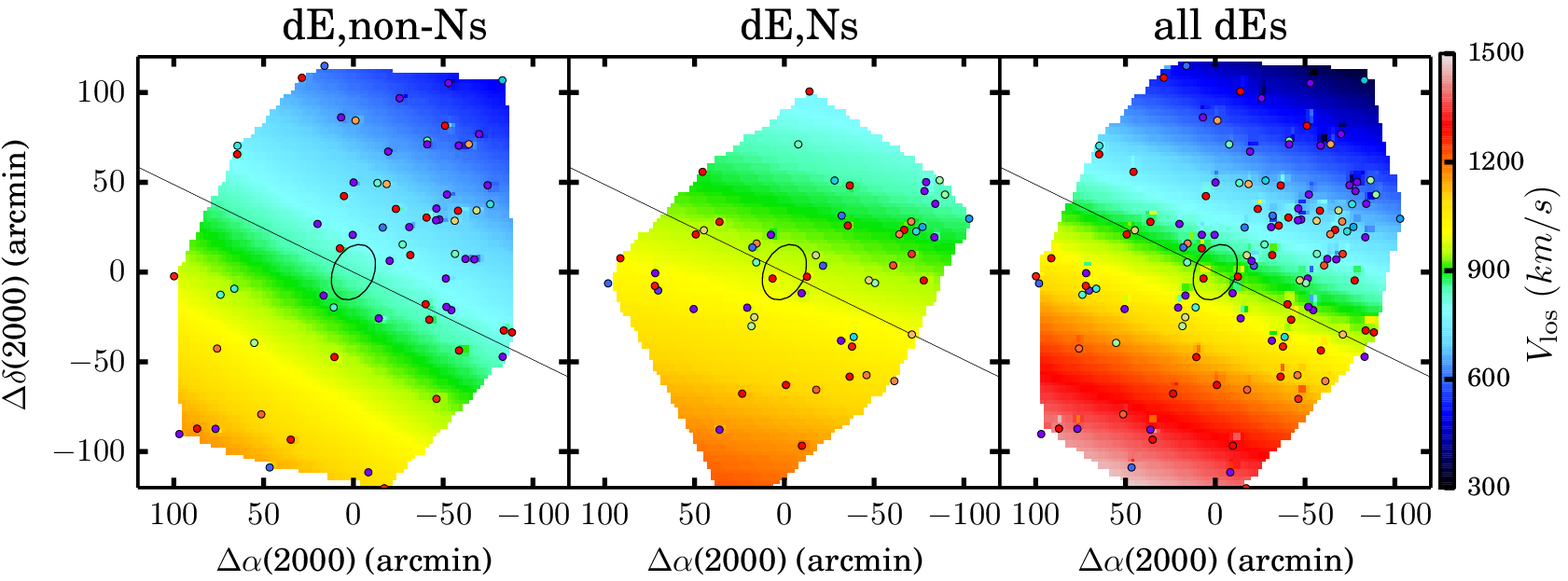}
\caption{
Spatial distribution of the non-nucleated dEs ({\it left}), nucleated dEs ({\it middle}) and 
all dEs ({\it right}) are over plotted on their respective color-coded surface fitting  
to the line-of-sight velocities with the Kriging technique for the inner 120\arcmin~of M87.\ 
The {\it black ellipses} represent the stellar isophotes of M87 at 10$R_{\rm e}$, and the {\it black 
solid line} marks the photometric minor axis of M87 in each panel.\ Note that the direction of velocity 
gradients for both the nucleated and non-nucleated dEs roughly follows the photometric major axis of 
M87.\ When compared to the velocity fields shown in Figure \ref{radc_kin}, the dEs seem to be more 
aligned with the blue GCs than the UCDs and red GCs.
\label{kin_des}}
\end{figure*}

The Virgo core region is still dynamically young (e.g.\ Binggeli et al.\ 1987).\ 
Deep wide-field optical imaging toward the core region revealed a complex network of extended 
tidal features surrounding M87 and other giant ellipticals (e.g.\ Mihos et al.\ 2005; Janowiecki et al.\ 2010), 
suggesting an ongoing hierarchical assembly of the Virgo core.\ It is thus natural to look for any connection 
between the velocity field of surrounding dwarf galaxies and stellar clusters in M87.\ Figure \ref{kin_des} 
presents the velocity field for 69 non-nucleated ({\it left panel}), 59 nucleated ({\it middle panel}) 
and all ({\it right panel}) dE galaxies within 2$^{\circ}$ of M87.\ As in Figure \ref{radc_kin}, 
Kriging maps of the mean velocity fields are color-coded in Figure \ref{kin_des}, and the 
individual data points follow the same color scheme.\ There is no significant difference between the 
direction of velocity gradients of nucleated and non-nucleated dEs.\ The direction of the velocity 
gradients of dEs more or less follows the photometric major axis of M87.\ Among the three velocity 
fields shown in Figure \ref{radc_kin}, the blue GCs seem to match the dEs best, in general agreement 
with C\^ot\'e et al.\ (2001).\ 

We note that the remarkable velocity gradient on the Kriging maps of dEs is primarily driven by an excess 
of low-velocity dEs toward the north west direction.\ These excess low-velocity dEs are most 
probably associated with a small subcluster of galaxies (e.g.\ Binggeli et al.\ 1993; 
Schindler, Binggeli, \& Bohringer 1999; Jerjen, Binggeli, \& Barazza 2004) 
centered on M86, which has a radial velocity $\simeq$ $-$244 km s$^{-1}$ 
and is about 1 Mpc more distant than M87 (Mei et al.\ 2007).\ It is no doubt that the M87 subcluster 
and M86 subcluster are moving toward each other, and an imminent merging between the two of them 
has long been speculated (e.g.\ Bohringer et al.\ 1994; Binggeli et al.\ 1993).

%
\begin{deluxetable*}{lcccccccccr} 
\tabletypesize{\footnotesize}                                                                                                             
\tablecolumns{11}                                                           
\tablewidth{0pt}                                                           
\tablecaption{Kinematics of the Full Samples}
\tablehead{
\colhead{$R_{\rm av}$}
& \colhead{$N$}
& \colhead{$v_{\rm sys}$} & \colhead{$\theta_{0}$} & \colhead{$v_{\rm rot}$} 
&  \colhead{$v_{\rm rot, bias-corr}$} & CL & \colhead{$\sigma_{\rm p}$} & \colhead{$v_{\rm rms}$} 
& \colhead{$G_{2}$} & \colhead{$T$} \\
\colhead{(arcmin)}                                                                             
& \colhead{} 
& \colhead{(km s$^{-1}$)} & \colhead{(deg)} & \colhead{(km s$^{-1}$)} 
&  \colhead{(km s$^{-1}$)} & \colhead{} & \colhead{(km s$^{-1}$)} & \colhead{(km s$^{-1}$)} 
& \colhead{} & \colhead{}\\
\colhead{(1)}                                                                             
& \colhead{(2)}                                                                             
& \colhead{(3)}
& \colhead{(4)}
& \colhead{(5)}
& \colhead{(6)}
& \colhead{(7)}
& \colhead{(8)}
& \colhead{(9)}
& \colhead{(10)}
& \colhead{(11)}
}                                                                          
\startdata                                                                                                        
 
 \multicolumn{11}{c}{UCDs}\\
 \hline
 
 0.7$'$--30$'$ & 89 & 1385$^{+31}_{-54}$ & 281$^{+13}_{-14}$ & 154$^{+39}_{-44}$ & 144  & 98\% & 344$^{+15}_{-18}$ & \
                            355$^{+16}_{-19}$ & -0.2$^{+0.2}_{-0.2}$ & 0.8$^{+0.1}_{-0.4}$ \\
 0.7$'$--4$'$ & 19  & 1374$^{+69}_{-68}$ & 270$^{+90}_{-33}$ & 141$^{+82}_{-79}$ & 0--10 & 62\% & 376$^{+13}_{-57}$ & \
                           392$^{+23}_{-61}$ & -0.1$^{+0.6}_{-0.4}$ & 0.1$^{+0.4}_{-0.2}$ \\
 4$'$--12$'$ & 34  & 1314$^{+44}_{-38}$ & 236$^{+63}_{-26}$ & 241$^{+50}_{-62}$ & 240 & 99\% & 424$^{+36}_{-53}$ & \
                          453$^{+32}_{-35}$ & -0.1$^{+0.4}_{-0.3}$ & 0.1$^{+0.3}_{-0.3}$ \\
 12$'$--30$'$ & 36  & 1352$^{+37}_{-38}$ & 304$^{+9}_{-21}$ & 150$^{+41}_{-62}$ & 130 & 87\%  & 262$^{+25}_{-24}$ & \
                         262$^{+16}_{-23}$ & -0.3$^{+0.3}_{-0.3}$ & 0.4$^{+0.2}_{-0.3}$ \\
 30$'$--60$'$ & 8 & \nodata & \nodata & \nodata & \nodata  & \nodata  & \nodata & \
                         \nodata & \nodata & \nodata \\
 
\cutinhead{Blue GCs}

 0.4$'$--30$'$ & 615  & 1306$^{+21}_{-19}$ & 214$^{+69}_{-11}$ & 36$^{+28}_{-14}$ & 30 & 73\% & 357$^{+5}_{-8}$ & \
                          357$^{+5}_{-7}$ & -0.2$^{+0.1}_{-0.1}$ & 0.1$^{+0.1}_{-0.1}$ \\
 0.4$'$--4$'$ & 207   & 1326$^{+31}_{-17}$ & 193$^{+83}_{-91}$ & 48$^{+45}_{-21}$ & 20 & 55\%  & 354$^{+9}_{-13}$ &  \
                          354$^{+11}_{-11}$ & -0.2$^{+0.1}_{-0.1}$ &  0.1$^{+0.1}_{-0.1}$ \\
 4$'$--12$'$ & 219   & 1339$^{+18}_{-26}$ & 270$^{+15}_{-35}$ & 81$^{+23}_{-24}$ &  72 & 84\%  & 412$^{+10}_{-11}$ & \
                         415$^{+7}_{-14}$ & -0.7$^{+0.1}_{-0.1}$ & 0.0$^{+0.1}_{-0.1}$ \\
 12$'$--30$'$ & 189  & 1282$^{+12}_{-13}$ & 161$^{+52}_{-58}$ & 26$^{+17}_{-27}$ & 0--20 & 22\%  & 267$^{+11}_{-9}$ & \
                         267$^{+9}_{-7}$ & -0.3$^{+0.1}_{-0.1}$ & 0.2$^{+0.1}_{-0.1}$ \\
 30$'$--60$'$ & 54   & 1168$^{+32}_{-30}$ & 185$^{+44}_{-19}$ & 125$^{+41}_{-46}$ & 91 & 80\%  & 438$^{+33}_{-27}$ & \
                         465$^{+28}_{-41}$ & \nodata & \nodata \\
 
 \cutinhead{Red GCs}
 
 0.5$'$--30$'$ &  226  & 1341$^{+15}_{-15}$ & 270$^{+21}_{-10}$ & 62$^{+22}_{-19}$ & 61  &  87\% & 268$^{+10}_{-8}$ & \
                             277$^{+6}_{-12}$ & 0.2$^{+0.1}_{-0.1}$ & 0.1$^{+0.1}_{-0.1}$ \\
 0.5$'$--4$'$ &  122   & 1404$^{+24}_{-29}$ & 270$^{+67}_{-41}$ & 77$^{+30}_{-48}$ & 71  & 80\%  & 272$^{+8}_{-14}$ &  \
                             272$^{+10}_{-11}$ & -0.2$^{+0.2}_{-0.1}$ &  -0.1$^{+0.2}_{-0.1}$ \\
 4$'$--12$'$ &  78   & 1261$^{+21}_{-19}$ & 301$^{+9}_{-9}$ & 102$^{+34}_{-28}$ & 80  & 86\%  & 284$^{+15}_{-30}$ & \
                            284$^{+17}_{-19}$ & 0.8$^{+0.4}_{-0.3}$ & 0.2$^{+0.2}_{-0.2}$ \\
 12$'$--30$'$ &  26  & 1242$^{+36}_{-26}$ & 105$^{+18}_{-36}$ & 106$^{+39}_{-50}$ & 41  & 67\%  & 216$^{+17}_{-18}$ & \
                          228$^{+20}_{-38}$ &  1.0$^{+0.7}_{-0.6}$ & 0.4$^{+0.5}_{-0.4}$ \\
 30$'$--60$'$ & 3 & \nodata & \nodata & \nodata & \nodata  & \nodata  & \nodata & \
                         \nodata & \nodata & \nodata \\
 
\enddata

\tablecomments{
(1) The range of geometric galactocentric distances from the center of M87.
(2) Number of data points used in the kinematics modeling.
(3) The best-fit systematic velocity.
(4) The best-fit azimuthal angle of the rotation axis, east of north.
(5) The best-fit rotation amplitude.
(6) The bias-corrected rotation amplitude.
(7) The confidence level of the best-fit rotation.
(8) The rotation-subtracted velocity dispersion.
(9) The biweight root-mean-square velocity.
(10) Standard kurtosis $G_{2}$ of the velocity distribution.
(11) The $T$ parameter proposed by Moors (1988).\ As a quantile-based alternative for the standard kurtosis, 
      $T$ is an increasing function of $G_{2}$.\ See the Appendix for definition of $T$.
}
\label{kinrad}
\end{deluxetable*}                                                          

                                                                                  
%
\begin{deluxetable*}{lcccccccccr} 
\tabletypesize{\footnotesize}                                                                                                             
\tablecolumns{11}                                                           
\tablewidth{0pt}                                                           
\tablecaption{Kinematics of the Subsamples with $i_{0} < 20.5$}
\tablehead{
\colhead{$R_{\rm av}$}
& \colhead{$N$}
& \colhead{$v_{\rm sys}$} & \colhead{$\theta_{0}$} & \colhead{$v_{\rm rot}$} 
&  \colhead{$v_{\rm rot, bias-corr}$} & CL & \colhead{$\sigma_{\rm p}$} & \colhead{$v_{\rm rms}$} 
& \colhead{$G_{2}$} & \colhead{$T$} \\
\colhead{(arcmin)}                                                                             
& \colhead{} 
& \colhead{(km s$^{-1}$)} & \colhead{(deg)} & \colhead{(km s$^{-1}$)} 
&  \colhead{(km s$^{-1}$)} & \colhead{} & \colhead{(km s$^{-1}$)} & \colhead{(km s$^{-1}$)} 
& \colhead{} & \colhead{}\\
\colhead{(1)}                                                                             
& \colhead{(2)}                                                                             
& \colhead{(3)}
& \colhead{(4)}
& \colhead{(5)}
& \colhead{(6)}
& \colhead{(7)}
& \colhead{(8)}
& \colhead{(9)}
& \colhead{(10)}
& \colhead{(11)}
}                                                                          
\startdata                                                                                                        
 
 \multicolumn{11}{c}{UCDs}\\
 \hline

 0.7$'$--30$'$ & 78  & 1379$^{+58}_{-57}$ & 281$^{+10}_{-24}$ & 174$^{+64}_{-70}$ & 154 & 98\% & 352$^{+11}_{-20}$ & \
                         360$^{+14}_{-23}$ & -0.2$^{+0.2}_{-0.2}$ & 0.7$^{+0.1}_{-0.5}$  \\
 0.7$'$--4$'$ & 15  & 1272$^{+127}_{-125}$ & 358$^{+54}_{-66}$ & 238$^{+135}_{-107}$ & 141 & 63\% & 371$^{+16}_{-39}$ & \
                         437$^{+12}_{-80}$ & -0.4$^{+0.7}_{-0.5}$ & 0.4$^{+0.4}_{-0.5}$ \\
 4$'$--12$'$ & 29  & 1309$^{+170}_{-86}$ & 237$^{+73}_{-69}$ & 244$^{+141}_{-102}$ & 240 & 89\% & 388$^{+32}_{-43}$ & \
                         429$^{+25}_{-39}$ &  -0.2$^{+0.4}_{-0.4}$ & 0.3$^{+0.4}_{-0.4}$ \\
 12$'$--30$'$ & 34  & 1342$^{+66}_{-53}$ & 309$^{+17}_{-36}$ & 151$^{+86}_{-104}$ & 120 & 84\%  & 282$^{+29}_{-21}$ & \
                         293$^{+19}_{-25}$ & 0.1$^{+0.5}_{-0.5}$ & 0.3$^{+0.2}_{-0.3}$ \\
 30$'$--60$'$ & 8 & \nodata & \nodata & \nodata & \nodata  & \nodata  & \nodata & \
                         \nodata & \nodata & \nodata \\
 
\cutinhead{Blue GCs}
 0.4$'$--30$'$ & 242  & 1306$^{+26}_{-23}$ & 199$^{+47}_{-103}$ & 43$^{+33}_{-34}$ & 11 & 49\%  & 360$^{+9}_{-11}$ & \
                        360$^{+8}_{-10}$ & -0.5$^{+0.1}_{-0.1}$ & 0.0$^{+0.1}_{-0.1}$  \\
 0.4$'$--4$'$ & 67   & 1336$^{+46}_{-45}$ & 188$^{+54}_{-74}$ & 38$^{+58}_{-50}$ & 0 & 17\%  & 355$^{+16}_{-15}$ &  \
                        357$^{+12}_{-17}$ & -0.7$^{+0.2}_{-0.1}$ &  0.0$^{+0.2}_{-0.2}$ \\
 4$'$--12$'$ & 95   & 1321$^{+48}_{-42}$ & 203$^{+73}_{-85}$ & 152$^{+59}_{-67}$ &  140 & 90\%  & 439$^{+16}_{-19}$ & \
                        446$^{+13}_{-19}$ & -0.7$^{+0.2}_{-0.1}$ & 0.0$^{+0.1}_{-0.2}$ \\
 12$'$--30$'$ & 80   & 1285$^{+30}_{-34}$ & 40$^{+56}_{-48}$ & 62$^{+43}_{-37}$ & 26 & 51\%  & 265$^{+12}_{-12}$ & \
                        269$^{+10}_{-14}$ & -0.3$^{+0.2}_{-0.2}$ & -0.2$^{+0.2}_{-0.1}$ \\
 30$'$--60$'$ & 22   & 1014$^{+77}_{-45}$ & 270$^{+64}_{-16}$ & 0$^{+125}_{-105}$ & 0 & 80\%  & 381$^{+41}_{-31}$ & \
                        381$^{+42}_{-33}$ & \nodata & \nodata \\
 
 \cutinhead{Red GCs}
 0.5$'$--30$'$ &  102  & 1335$^{+28}_{-29}$ & 271$^{+9}_{-37}$ & 45$^{+43}_{-38}$ & 0 & 44\%  & 264$^{+16}_{-14}$ & 
                           269$^{+11}_{-15}$ & 0.2$^{+0.2}_{-0.2}$ & 0.1$^{+0.2}_{-0.1}$  \\
 0.5$'$--4$'$ &  39   & 1402$^{+41}_{-46}$ & 270$^{+86}_{-87}$ & 92$^{+56}_{-55}$ & 50  & 66\%  & 249$^{+17}_{-19}$ &  \
                           249$^{+12}_{-18}$ & -0.9$^{+0.2}_{-0.2}$ &  -0.2$^{+0.1}_{-0.2}$ \\
 4$'$--12$'$ &  47   & 1296$^{+56}_{-47}$ & 306$^{+10}_{-52}$ & 59$^{+90}_{-64}$ & 0  & 35\%  & 261$^{+16}_{-23}$ & \
                           270$^{+12}_{-27}$ &  0.6$^{+0.4}_{-0.5}$ & 0.0$^{+0.2}_{-0.2}$ \\
 12$'$--30$'$ &  16  & 1247$^{+61}_{-65}$ & 101$^{+30}_{-70}$ & 35$^{+48}_{-34}$ & 0  & 10\%  & 173$^{+35}_{-52}$ & \
                           175$^{+26}_{-41}$ & 1.2$^{+0.6}_{-0.8}$ & 1.1$^{+0.6}_{-0.6}$ \\
 30$'$--60$'$ & 0 & \nodata & \nodata & \nodata & \nodata  & \nodata  & \nodata & \
                         \nodata & \nodata & \nodata \\
 
\enddata

\tablecomments{
(1) The range of geometric galactocentric distances from the center of M87.
(2) Number of data points used in the kinematics modeling.
(3) The best-fit systematic velocity.
(4) The best-fit azimuthal angle of the rotation axis, east of north.
(5) The best-fit rotation amplitude.
(6) The bias-corrected rotation amplitude.
(7) The confidence level of the best-fit rotation.
(8) The rotation-subtracted velocity dispersion.
(9) The biweight root-mean-square velocity.
(10) Standard kurtosis $G_{2}$ of the velocity distribution.
(11) The $T$ parameter proposed by Moors (1988).\ As a quantile-based alternative for the standard kurtosis, 
      $T$ is an increasing function of $G_{2}$.\ See the Appendix for definition of $T$.
}
\label{kinrad_b}
\end{deluxetable*}                                                          

                                                                                  
%
\begin{deluxetable*}{lcccccccccr} 
\tabletypesize{\footnotesize}                                                                                                             
\tablecolumns{11}                                                           
\tablewidth{0pt}                                                           
\tablecaption{Kinematics of the Subsamples with $i_{0} > 20.5$}
\tablehead{
\colhead{$R_{\rm av}$}
& \colhead{$N$}
& \colhead{$v_{\rm sys}$} & \colhead{$\theta_{0}$} & \colhead{$v_{\rm rot}$} 
&  \colhead{$v_{\rm rot, bias-corr}$} & CL & \colhead{$\sigma_{\rm p}$} & \colhead{$v_{\rm rms}$} 
& \colhead{$G_{2}$} & \colhead{$T$} \\
\colhead{(arcmin)}                                                                             
& \colhead{} 
& \colhead{(km s$^{-1}$)} & \colhead{(deg)} & \colhead{(km s$^{-1}$)} 
&  \colhead{(km s$^{-1}$)} & \colhead{} & \colhead{(km s$^{-1}$)} & \colhead{(km s$^{-1}$)} 
& \colhead{} & \colhead{}\\
\colhead{(1)}                                                                             
& \colhead{(2)}                                                                             
& \colhead{(3)}
& \colhead{(4)}
& \colhead{(5)}
& \colhead{(6)}
& \colhead{(7)}
& \colhead{(8)}
& \colhead{(9)}
& \colhead{(10)}
& \colhead{(11)}
}                                                                          
\startdata                                                                                                        
 
 \multicolumn{11}{c}{UCDs}\\
 \hline

 0.7$'$--30$'$ & 11 &  \nodata &  \nodata &  \nodata &  \nodata & \nodata  & \nodata & 406$^{+114}_{-93}$  & \nodata & \nodata  \\
 0.7$'$--4$'$ & 4 & \nodata & \nodata & \nodata & \nodata  & \nodata  & \nodata & \
                         \nodata & \nodata & \nodata \\
 4$'$--12$'$ & 5 & \nodata & \nodata & \nodata & \nodata  & \nodata  & \nodata & \
                         \nodata & \nodata & \nodata \\
 12$'$--30$'$ & 2 & \nodata & \nodata & \nodata & \nodata  & \nodata  & \nodata & \
                         \nodata & \nodata & \nodata \\
 30$'$--60$'$ & 0 & \nodata & \nodata & \nodata & \nodata  & \nodata  & \nodata & \
                        \nodata & \nodata & \nodata \\
 
\cutinhead{Blue GCs}
 0.4$'$--30$'$ & 373  & 1308$^{+25}_{-21}$ & 220$^{+31}_{-106}$ & 34$^{+34}_{-27}$ & 11  & 51\% & 352$^{+9}_{-8}$ & \
                        352$^{+8}_{-8}$ & -0.1$^{+0.1}_{-0.1}$ & 0.1$^{+0.1}_{-0.1}$  \\
 0.4$'$--4$'$ & 140   & 1344$^{+35}_{-33}$ & 232$^{+75}_{-85}$ & 71$^{+42}_{-37}$ & 55 & 71\%  & 357$^{+15}_{-15}$ &  \
                         361$^{+15}_{-14}$ &  0.2$^{+0.2}_{-0.2}$ &  0.2$^{+0.1}_{-0.1}$ \\
 4$'$--12$'$ & 124   & 1352$^{+38}_{-40}$ & 270$^{+60}_{-76}$ & 87$^{+57}_{-51}$ &  61 & 77\%  & 421$^{+15}_{-19}$ & \
                         421$^{+19}_{-14}$ &  0.2$^{+0.2}_{-0.2}$ & 0.1$^{+0.1}_{-0.2}$ \\
 12$'$--30$'$ & 109   & 1267$^{+30}_{-29}$ & 216$^{+71}_{-40}$ & 44$^{+37}_{-33}$ & 6 & 38\%  & 269$^{+14}_{-13}$ & \
                         269$^{+10}_{-12}$ & -0.1$^{+0.2}_{-0.2}$ & 0.4$^{+0.1}_{-0.2}$ \\
 30$'$--60$'$ & 32   & 1253$^{+53}_{-55}$ & 136$^{+50}_{-23}$ & 139$^{+84}_{-105}$ & 90 & 73\%  & 443$^{+75}_{-52}$ & \
                         487$^{+53}_{-76}$ & \nodata & \nodata \\
 
 \cutinhead{Red GCs}
 0.5$'$--30$'$ &  124  & 1361$^{+37}_{-39}$ & 270$^{+62}_{-52}$ & 109$^{+51}_{-49}$ & 108 & 94\%  & 310$^{+19}_{-13}$ & \
                       319$^{+15}_{-16}$ & 1.1$^{+0.3}_{-0.3}$ & 0.0$^{+0.1}_{-0.1}$  \\
 0.5$'$--4$'$ &  83   & 1358$^{+52}_{-45}$ & 222$^{+39}_{-100}$ & 47$^{+60}_{-51}$ & 0  & 27\%  & 320$^{+24}_{-20}$ &  \
                        329$^{+17}_{-25}$ & 1.7$^{+0.4}_{-0.5}$ &  0.2$^{+0.1}_{-0.2}$ \\
 4$'$--12$'$ &  31   & 1231$^{+45}_{-41}$ & 293$^{+25}_{-20}$ & 12$^{+84}_{-12}$ & 0  & 4\%  & 297$^{+31}_{-32}$ & \
                        299$^{+35}_{-30}$ & 0.6$^{+0.8}_{-0.6}$ & 1.2$^{+0.4}_{-0.7}$ \\
 12$'$--30$'$ &  10 &  \nodata & \nodata & \nodata & \nodata  & \nodata  & \nodata & \
                         155$^{+14}_{-22}$ & \nodata & \nodata \\
 30$'$--60$'$ & 3 & \nodata & \nodata & \nodata & \nodata  & \nodata  & \nodata & \
                         \nodata & \nodata & \nodata \\
 
\enddata

\tablecomments{
(1) The range of geometric galactocentric distances from the center of M87.
(2) Number of data points used in the kinematics modeling.
(3) The best-fit systematic velocity.
(4) The best-fit azimuthal angle of the rotation axis, east of north.
(5) The best-fit rotation amplitude.
(6) The bias-corrected rotation amplitude.
(7) The confidence level of the best-fit rotation.
(8) The rotation-subtracted velocity dispersion.
(9) The biweight root-mean-square velocity.
(10) Standard kurtosis $G_{2}$ of the velocity distribution.
(11) The $T$ parameter proposed by Moors (1988).\ As a quantile-based alternative for the standard kurtosis, 
      $T$ is an increasing function of $G_{2}$.\ See the Appendix for definition of $T$.
}
\label{kinrad_f}
\end{deluxetable*}                                                          

\subsection{Interpretation: Ongoing Accretion of Dwarf Galaxies?}
The observation that the surrounding dEs (especially the non-nucleated ones) follow a similar velocity 
field to the GCs (especially the blue ones) is consistent with the scenario that the GC systems, 
especially the metal-poor ones, may have been primarily assembled by accreting satellite dwarf galaxies 
along the photometric major axis of M87 (e.g.\ C\^ot\'e, Marzke \& West 1998).\ 
In line with this ongoing accretion or infalling picture, West \& Blakeslee (2000) found that Virgo's brightest 
ellipticals have a strong collinear arrangement in three dimensions.\ This so-called ``principal axis'', 
which appears to join a filamentary bridge of galaxies connecting the Virgo cluster to Abell 1367 and 
passes through the Virgo core, is also more or less aligned with the major axes of Virgo's ellipticals 
(including M87).\ This ``principal axis'' is thought to be the direction along which material 
flows into the cluster and forms galaxies, as seen in Cosmological $N$-body simulations 
(e.g.\ van Haarlem, Frenk \& White 1997; Hopkins, Bahcall \& Bode 2005; Faltenbacher et al.\ 2005).

\section{Orbital Anisotropies from Jeans Analysis}\label{jeans}
\subsection{Method}
In this section we will infer the orbital anisotropies of UCDs and GCs based on the spherically 
symmetric Jeans equation
\begin{equation}
-n_{r}\frac{GM(<r)}{r^{2}} = \frac{d (n_{r}\sigma_{r}^{2})}{d r} + 2\frac{\beta_{r}}{r}n_{r}\sigma_{r}^{2}
\label{eq9}
\end{equation}
where $M$($<$$r$) is the mass interior to the three-dimensional radius $r$.\
$n_{r}$ and $\sigma_{r}$ are respectively the volume number density and radial component 
of the velocity dispersion at radius $r$ for a given tracer population, and $\beta_{r}$ is the anisotropy parameter 
defined as 1$-$$\sigma_{t}^{2}$/2$\sigma_{r}^{2}$, with $\sigma_{t}$ being the tangential components 
($\sigma_{t}^{2} = \sigma_{\theta}^{2}+\sigma_{\phi}^{2}$) of the velocity dispersion (Binney \& Tremaine 2008).

In using Equation \ref{eq9}, we have assumed that the net rotation can be either ignored or 
simply folded into the velocity dispersion term.\ This approximation is reasonable given the fact 
that the overall rotation of our samples is not dynamically important ($v_{\rm rot}/\sigma_{\rm p}$
$\lesssim$ 0.4).\ In addition, as mentioned previously, the M87 system is probably triaxial in shape, 
which would caution against a spherically symmetric Jeans analysis.\ However, as (at least) a 
first order approximation, it is definitely enlightening to do a comparative study of UCDs and 
GCs under the spherically symmetric assumption.

The anisotropy parameter $\beta_{r}$ can be constrained if $M(<r)$, $n_{r}$, and $\sigma_{r}$ are 
known.\ For $M(<r)$, we adopt the most recent determination by Zhu et al.\ (2014) based on 
made-to-measure modeling (Syer \& Tremaine 1996; Long \& Mao 2010) of over 900 M87 GCs, 
which extend out to a projected radius of $\sim$ 180 kpc.\ The Zhu et al.\ (2014) mass profile is 
a combination of a stellar component and a spherical logarithmic dark matter halo model.\ 
The $n_{r}$ profile is generally related to the surface number density profile $N_{R}$ through the 
Abel integral equation.\ In particular, for deprojection of the S\'ersic surface profile that was used to 
characterize  $N_{R}$ of the UCDs and GCs in this work (Table \ref{surfprof}), we adopt the analytical 
approximation proposed by Prugniel \& Simien (1997, Eq. B6; see also Mamon \& Lokas 2005).\
In addition, treating the Jeans equation (\ref{eq9}) as a first-order linear differential equation for 
$n_{r}\sigma_{r}^{2}$, one finds (See also C\^ot\'e et al.\ 2001; Mamon \& Lokas 2005)
\begin{equation}
\sigma_{r}^{2} = -\frac{1}{n_{r}e^{[2\int \frac{\beta_{r}}{r} d r]}} 
\int_{r}^{\infty}e^{[2\int \frac{\beta_{t}}{t} d t]}n_{t} \frac{GM(<t)}{t^2} d t
\label{eq11}
\end{equation}

The line-of-sight velocity dispersion $\sigma_{{\rm los}, R}$ at projected radius $R$ is determined 
(e.g.\ Binney \& Mamon 1982) as 
\begin{equation}
\sigma_{{\rm los}, R}^{2} = \frac{2}{N_{R}}\int_{R}^{\infty}  n_{r}\sigma_{r}^{2}(1-\beta_{r}\frac{R^2}{r^2}) \frac{r}{\sqrt{r^{2}-R^{2}}} d r
\label{eq12}
\end{equation}
For the radial dependence of $\beta_{r}$, we adopt the following function form (first proposed by 
Mamon, Biviano \& Bou\'e 2013), 
\begin{equation}
\beta_{r} = \beta_{0} + (\beta_{\infty} - \beta_{0}) \frac{r}{r+r_{\beta}}
\label{eq13}
\end{equation}
where $\beta_{0}$, $\beta_{\infty}$ and $r_{\beta}$ are the three free parameters defining the radial profile.\ 
Specifically, $\beta_{0}$ and $\beta_{\infty}$ are the anisotropies at $r$ = 0 and $\infty$ respectively, 
and $r_{\beta}$ represents the scale radius of $\beta_{r}$ profile.\ This function form of $\beta_{r}$ allows 
for either a radially increasing or decreasing profile.\

The GC (the blue plus red) anisotropy profile as determined by Zhu et al.\ (2014) shows a non-monotonic 
behavior, in the sense that $\beta_{r}$ gradually increases toward the intermediate radii ($\sim$ 40 kpc) 
and then falls off in larger radii.\ Therefore, we also considered a two-part piecewise radial dependence of 
anisotropies by allowing the inner and outer radii to follow different profiles as defined by Equation \ref{eq13}, 
with the ``transition'' radius $r_{\rm tr}$ being left as a free parameter.\

To constrain the radial anisotropy profiles for each of the different samples, we first created 
a library of model $\sigma_{{\rm los}, R}$ profiles for each of them by allowing the free parameters 
that define the inner and outer anisotropy profiles to uniformly (linearly for $\beta_{0}$, $\beta_{\infty}$, 
and $r_{\rm tr}$, logarithmically for $r_{\beta}$) vary.\ Then a maximum likelihood method was 
used to fit the models to the observed line-of-sight velocities as a function of projected radii.\ 
In particular, by assuming that the observed line-of-sight velocities $v_i$ ($\pm\Delta v_i$) at a 
given projected radius $R$ follow a Gaussian distribution with 
$\sigma_i^2$ =  $(\Delta v_i)^2$ + $\sigma_{{\rm los}, R}^2$ and 
$\mu = \sum_i (\frac{1}{\sigma_i^2}v_{i})/ \sum_i (\frac{1}{\sigma_i^2})$, 
we calculated a joint probability (similar to Eq.\ \ref{eq2} in form) of each model profile for a given 
population.\ The most probable model profile is taken as the fiducial one, and the 68\% confidence 
intervals are determined by randomly resampling the real data sets, with $\sim$ 10\% of data 
points being left out for each resample.

\subsection{Results}
\begin{figure*}
\centering
\includegraphics[height=0.4\textheight]{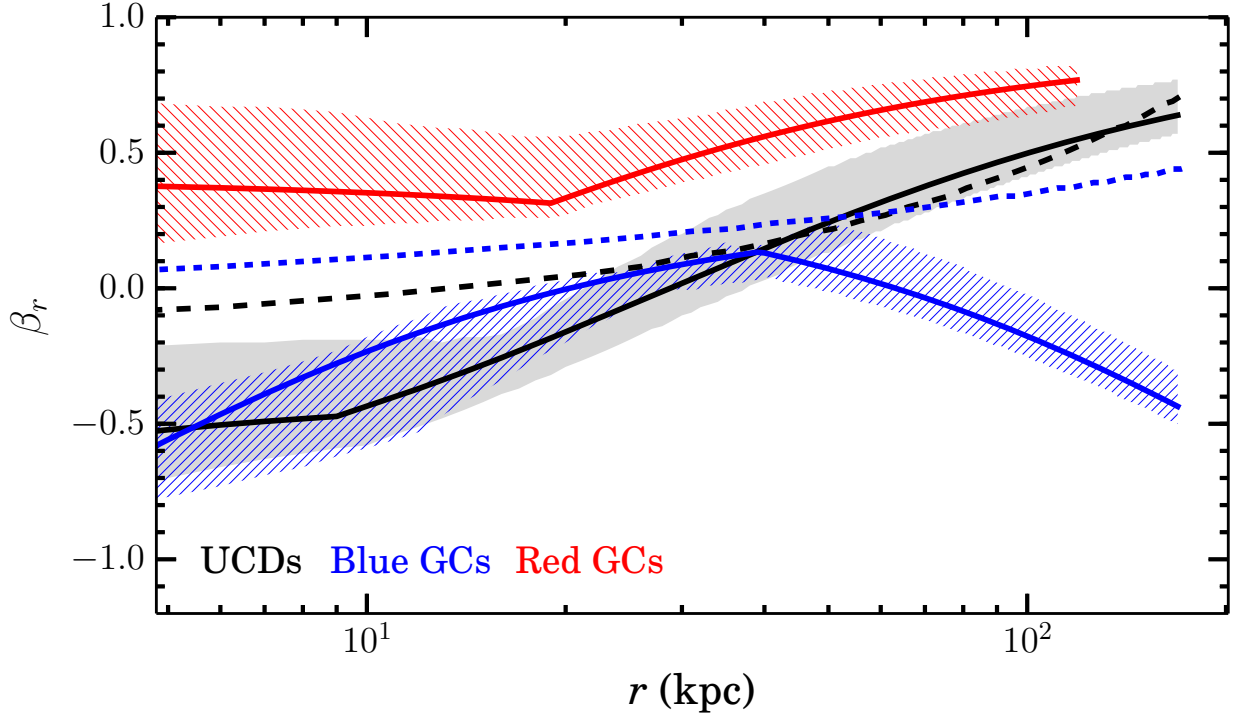}
\caption{
Variation of the anisotropy parameters as a function of the 3D radius.\
The profiles for UCDs, blue GCs, and red GCs are represented as 
black, blue, and red solid curves respectively.\ Following the same color code, 
the hatched regions of different styles mark the 68\% confidence intervals for blue GCs, 
and red GCs.\ The grey shaded region marks the 68\% confidence interval 
for the UCDs.\ The short dashed curves ($black$ for UCDs, $blue$ for blue GCs) 
represent the anisotropy profiles predicted by a universal relation between the number 
density slope and $\beta$ for relic high-$\sigma$ density peaks as found in 
cosmological simulations by Diemand et al.\ (2005).
\label{bet_rav}}
\end{figure*}

The derived anisotropy profiles for the full samples of UCDs and GCs within the central 35$\arcmin$ 
of M87 are shown in Figure \ref{bet_rav}.\ Following Zhu et al.\ (2014), we reduced the weight of the 
data points (by increasing the uncertainties) that are located in the puzzlingly ``hot'' radius range from 
$R_{\rm av}$ = 4\arcmin~to 12\arcmin~in the Jeans modeling.\ Since our Jeans analysis relies on the 
Zhu et al.\ mass profile, which was determined with made-to-measure modeling of nearly the same 
GC data set that is used in this work, we should obtain an anisotropy profile that is at least qualitatively 
consistent with Zhu et al.\ Comparing our Figure \ref{bet_rav} to the Figure 12 of Zhu et al.\ (2014), one 
can see that, although being based on different methods, our anisotropy profile for the blue GCs, which 
dominate the spectroscopic samples of M87 GCs, is in reasonably good agreement with Zhu et al.\  within 
the uncertainties.\ 

The UCD system has an anisotropy profile that becomes more radial with radius, with $\beta_{r}$ being negative 
(tangentially-biased) within the inner $\sim$ 20 -- 40 kpc and being positive (radially-biased) beyond.\
We note that a radially-biased orbital structure for UCDs at larger radii is in line with a peaky 
velocity distribution shown in Section \ref{velhis}.\ The blue GC system has a radially increasing 
$\beta_{r}$ profile in the inner $\sim$ 40 kpc but a radially decreasing profile at larger radii.\ 
Among the three samples, the red GCs exhibit the largest radially-biased velocity dispersion 
tensor across the explored radius range, which may be surprising but nevertheless in line with 
their relatively large velocity kurtosis (Figure \ref{velhis_all}; Table \ref{kinrad}).\ We note that, 
although being based on different mass models of M87, a highly radially-biased anisotropy 
($\beta$ $\sim$ 0.8 at $\sim$ 150 kpc) was also found in the outer stellar halo of M87 by 
Doherty et al.\ (2009) based on the integrated stellar absorption-line data at small radii and 
planetary nebulas (PNs; trace the stellar diffuse light, Coccato et al.\ 2009) kinematics at large 
radii ($\lesssim$ 150 kpc), and this is in agreement with our finding for red GCs.
\footnote{
Agnello et al.\ (2014) recently determined the anisotropies for 354 GCs within $\sim$ 100 kpc of M87, 
by dividing the GC system into three kinematically distinct subpopulations of different colors, 
i.e.\ blue, intermediate-color and red GCs, with intermediate GCs mostly being separated out from the classic red GCs.\
Agnello et al.\ found a mildly radially-biased anisotropy ($\sim$0.3) for their intermediate GCs and 
a slightly tangentially-biased anisotropy ($\sim$$-$0.2) for their red GCs.\ Nevertheless, a tripartition 
of M87 GCs may be still oversimplified.\ As was shown by S11 (c.f.\ their Figure 8), there is a complex 
color dependence of velocity dispersion of the classic red GCs, in the sense that the ``intermediate'' red GCs 
have a significantly higher dispersion than both the ``bluer'' and ``redder'' red GCs.\ So, there may be 
at least three kinematically distinct subpopulations for the classic red GCs alone, and the red GC 
system has not completely mixed dynamically.\ A complete understanding of the dynamics of the red GCs 
would have to wait for larger samples of radial velocities across the M87 system, in order to explore the 
full color dependence of their dynamics.\ }

\subsection{Interpretation}
The different orbital anisotropies of the different populations may be attributed to either their different origin 
or different orbital evolutionary histories.\ Our size-defined UCDs have at least an order of magnitude 
lower average density than GCs of similar luminosity, which means that UCDs are subject to a 
stronger tidal influence when approaching small galactocentric distances.\ 
In addition, at a given average orbital radius, objects with more radially-biased orbits can 
plunge deeper into the central regions of their host, and thus are prone to stronger tidal disruption.\ 
Recent simulations (Pfeffer \& Baumgardt 2013) suggest that UCD-sized clusters can be converted, 
via a continuous tidal stripping of their envelope, into GC-sized objects at small galactocentric distances.\

\begin{figure}
\centering
\includegraphics[height=0.21\textheight]{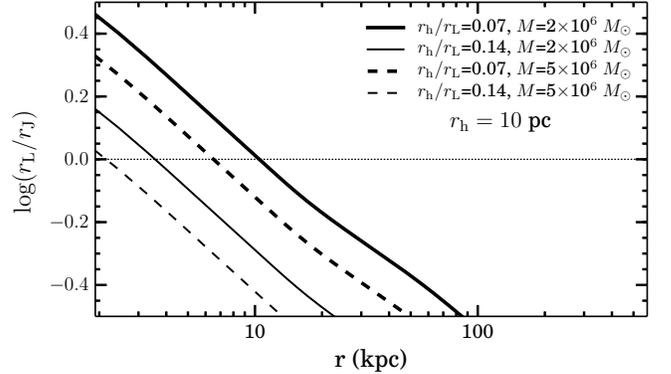}
\caption{
Radial variation of the ratio of limiting radius to Jacobi radius for M87 UCDs which have  
projected half-mass radii $r_{\rm h} = 10 $ pc and $M$ = 2$\times$10$^{6} M_{\sun}$ 
(solid curves) or 5$\times$10$^{6} M_{\sun}$ (dashed curves).\
The thick and thin curves represent results respectively for $r_{\rm h}/r_{\rm L}$ = 0.07 and 0.14, 
which straddle the range of $r_{\rm h}/r_{\rm L}$ typical of the Galactic GCs based on King (1966) 
model fitting (McLaughlin \& van der Marel 2005).\ The UCDs with log($r_{\rm L}/r_{\rm J}$) $>$ 0 
were tidally over-filling, and thus should have been tidally truncated below 10 pc in $r_{\rm h}$.
\label{rtrj_r}}
\end{figure}

\subsubsection{Depletion of Radial Orbits of UCDs at Small Galactocentric Distances}
The finding that UCDs in the central $\sim$ 40 kpc are depleted of radial orbits may be 
partly attributed to a stronger tidal influence at smaller galactocentric distances.\ 
To be quantitative, the Jacobi radius $r_{\rm J}$ of a stellar cluster moving 
in the tidal field of its host galaxy is determined by its galactocentric distance $r_{\rm gal}$, 
the galaxy mass $M_{\rm gal}$ interior to $r_{\rm gal}$ and the cluster mass $m$.\ 
Specifically, $r_{\rm J}$ = $r_{\rm gal}(m/2M_{\rm gal})^{1/3}$ (von Hoerner 1957; 
Innanen et al. 1983; Bertin \& Varri 2008; Renaud et al.\ 2011; Webb et al.\ 2013).\ 
Based on this basic tidal theory and the Zhu et al.\ mass profile of M87, we estimated 
the ratio of limiting radius $r_{\rm L}$ to $r_{\rm J}$ as a function of galactocentric distances 
for M87 UCDs ($m$ $=$ 2$\times$10$^{6}$ $M_{\sun}$ or 5$\times$10$^{6}$ $M_{\sun}$) 
with projected half-mass radius $r_{\rm h}$ = 10 pc (Figure \ref{rtrj_r}).\ $r_{\rm L}$ ideally 
represents the radius beyond which the stellar density of a cluster is zero, but in practice 
$r_{\rm L}$ is an extrapolated radius from either King or other model profile fitting (e.g.\ Harris 1996).\ 
In plotting Figure \ref{rtrj_r}, we considered two different ratios of $r_{\rm h}/r_{\rm L}$, namely, 
0.07 and 0.14, which cover the typical range for the Galactic GCs based on King (1966) model fitting 
(McLaughlin \& van der Marel 2005).\

From Figure \ref{rtrj_r}, it can be seen that the limiting radii are comparable to or greater than 
the Jacobi radii (tidally over-filling) at galactocentric distances $\lesssim$ 10 kpc for typical 
M87 UCDs if $r_{\rm h}/r_{\rm L}$ $\lesssim$ 0.07.\ 
Previous King profile fitting to some Virgo UCDs with the high-quality HST imaging 
data suggests that they have $r_{\rm h}/r_{\rm L}$ $<$ 0.1 (e.g.\ Evstigneeva et al.\ 2007).\
It is thus quite plausible that many UCDs with $r_{\rm h} >$ 10 pc at small galactocentric distances 
have been tidally truncated into GC-sized objects with $r_{\rm h}$ $<$ 10 pc, and this tidal transformation 
should be especially efficient for UCDs on plunging orbits at a given average galactocentric distance.\ 
Accordingly, UCDs that survives at smaller galactocentric distances are expected to have less 
radially-biased orbital structure, in line with our finding of an anisotropy profile that increases with 
radius for UCDs.\

\subsubsection{Different Assemby Histories of UCD and blue GC Systems}
Although the tidal transformation effect discussed above should have been playing some 
role, it is not necessarily the sole or even dominant mechanism in shaping the 
present-day anisotropy profile of our size-defined UCDs.\ It is possible that 
the progenitors of many UCDs at smaller galactocentric distances have been  
primarily accreted from less radially-biased orbits.\ 

In a $\Lambda$CDM Universe, a more radially-biased orbital structure for the outer stellar halos 
of early type galaxies seems to be a general outcome from a hierarchical structure formation (e.g.\ 
Dekel et al.\ 2005; Diemand et al.\ 2005; Hansen \& Moore 2006; Abadi, Navarro \& Steinmetz 2006).\ 
It has been conjectured that old halo populations, such as the metal-poor (blue) GCs, might form 
primarily in small dark matter halos that collapsed from high-$\sigma$ ($>$ 2.5$\sigma$; Diemand et al.\ 2005) 
peaks of the primordial density field at $z >$ 10 (e.g.\ Moore et al.\ 2006; Boley et al.\ 2009; 
Moran et al.\ 2014; see also a review by Brodie \& Strader 2006).\ One interesting finding 
from the $N$-body cosmological simulation of Diemand et al.\ (2005) is the existence of a 
universal relation between the present-day density profile slope $d\ln n/d\ln r$ and the 
anisotropy for the relic high-$\sigma$ peaks, and this relation is not sensitive to the detailed 
assembly histories.\ In particular, Diemand et al.\ (2005) found:\ $\beta$ $\simeq$ $-0.23(1.2 + d\ln n/d\ln r )$.\ 
By deprojecting the surface number density profiles presented in Section \ref{radprof} under the assumption 
of a spherically symmetric geometry, we obtained the $\beta$ profiles for M87 UCDs and blue GCs 
as expected by the above relation.\ The profiles are overplotted in Figure \ref{bet_rav} 
as dashed lines.\ 

Although being more negative in the innermost part, 
the $\beta_{r}$ profile of UCDs determined from Jeans analysis is more or less consistent with 
the Diemand et al.\ predication within 2$\sigma$ uncertainties.\ As discussed above, the 
significantly tangentially-biased orbital structure of UCDs at small radii can be partly attributed to 
a strong tidal transformation.\ The derived $\beta_{r}$ profile of the blue GCs exhibit large deviation from 
the Diemand et al.\ prediction.\ The finding that the blue GCs are tangentially-biased, rather than 
being radially-biased, at large radii may indicate\footnote{
Another possibility could be that the outer halo mass of M87 may be somehow 
underestimated, which would artificially lead to a lower $\beta$ due to the mass-anisotropy 
degeneracy.}
that the blue GC system in the outer halo of M87 has not yet established an equilibrium state 
and is still in an early and active stage of assembly, presumably through a continuous accretion 
of surrounding dwarf galaxies (e.g.\ C\^ot\'e, Marzke \& West 1998).\ Agnello et al.\ (2014) invoked 
the scenario of adiabatic contraction of the dark matter halo to explain the enhancement of 
tangentially-biased orbits of blue GCs of M87 at large radii, and this mechanism works when 
accretion of satellite galaxies happens on sufficiently slow time scales (Goodman \& Binney 1984).\

The agreement between the present-day radial anisotropy profile of UCDs and that predicted 
by cosmological simulations does not tell us the detailed accretion process of the UCD progenitors.\
Our finding that UCDs have radially-biased orbital structure at large galactocentric distances is 
in line with prediction of the ``tidal threshing'' scenario where UCDs were primarily tidally stripped 
dE nuclei (e.g.\ Bekki et al.\ 2003; Goerdt et al.\ 2008; Pfeffer \& Baumgardt 2013), 
although the primary progenitor galaxies of UCDs do not necessarily resemble the present-day 
surviving dEs.\ Previous simulations suggest that, to be tidally-threshed to a naked nucleus, 
a dE,N galaxy has to be on a highly radially-biased orbit in order to plunge deep into the central 
potential.\ If tidal threshing is indeed the dominant channel for forming UCDs, our finding 
suggests that UCDs might be preferentially but not exclusively accreted from plunging orbits.\ 
Indeed, previous simulations (e.g.\ Gill et al.\ 2004; Smith et al.\ 2013) suggest that, at small 
galactocentric distances, satellite galaxies on circular orbits may also be tidally disrupted.

\section{Summary and Discussion}\label{summary}
We have compiled a sample of 97 spectroscopically confirmed UCDs associated with 
the central cD galaxy of the Virgo cluster -- M87 (NGC~4486).\ The UCDs are defined to have 
10 $\lesssim$ $r_{\rm h}$ $\lesssim$ 100 pc.\ 89\% of our sample have $i_{0} < 20.5$ mag 
($M_{g} < -10.6$), corresponding to a stellar mass of $\sim 2\times10^{6} M_{\sun}$.\ In 
addition, 92\% of the UCDs have colors as blue as the classic blue GCs.\ Throughout 
this paper, we compared the distribution and dynamics of UCDs to that of M87 GCs.\ 
The primary results of this paper are summarized as follows.
\begin{enumerate}

\item
The M87 UCD system has a surface number density profile that is shallower than the blue GCs 
in the inner $\sim 15\arcmin$ ($\sim$ 70 kpc) and as steep as the red GCs at larger radii.\
In addition, the UCDs follow a radial velocity dispersion profile more similar to that of the 
blue rather than the red GCs.\ 
\item
Overall, the UCD system exhibits a significantly stronger rotation than the GC system.\
No significant rotation was found for either the bright ($M_{g} < -10.6$) or faint 
($M_{g} > -10.6$) blue GCs.\ Although subject to relatively large uncertainties, 
the velocity field of dE galaxies surrounding M87 seems to be more aligned with that of 
the blue GCs rather than the UCDs.\
\item
The velocity distribution of UCDs is characterized by a sharper peak and marginally 
lighter tails compared to a Gaussian distribution.\ This is in general agreement with 
results from Jeans analysis, namely, anisotropy of the velocity dispersion tensor of UCDs gradually 
increases from being tangentially-biased at inner radii to being radially-biased at larger radii.\ 
Overall, the GCs have velocity distribution similar to a Gaussian, with the blue GCs being 
slightly platykurtic and the red GCs being slightly leptokurtic.\ In addition, the bright blue GCs have 
velocity kurtosis systematically higher than the faint ones across the full range of galactocentric distances, 
indicating that the bright ones are more tangentially-biased than the faint ones.
\item
The M87 UCD system has an orbital anisotropy profile that gradually increases with galactocentric 
distances, with UCDs within $\sim$ 40 kpc being tangentially-biased while being radially-biased outward.\
In contrast to UCDs, the blue GCs beyond $\sim$ 40 kpc become gradually more tangentially-biased toward 
larger radii.\ The tangentially-biased anisotropy of UCDs in the inner region may be partly attributed to a 
continuous tidal transformation of UCD-sized objects on plunging orbits to GC-sized clusters.\

\end{enumerate}

Above all, our analysis suggests that the M87 UCDs are dynamically distinct from GCs.\
Other evidence against UCDs being the most luminous and extended tails of normal GCs 
include their different Fundamental-Plane relation (e.g.\ luminosity vs. internal velocity dispersion, 
Chilingarian et al.\ 2011).\ Our finding that UCDs have radially-biased orbital structure at large 
galactocentric distances is in general agreement with the ``tidal threshing'' scenario that UCDs 
are primarily tidally stripped dE nuclei.\ 
In an accompanying paper, Liu et al.\ (2015) find that M87 UCDs at smaller galactocentric 
distances tend to have less prominent stellar envelopes than those
lying at larger distances, providing direct evidence for tidal stripping.\
Moreover, previous studies (e.g.\ Paudel et al.\ 2010; Chilingarian et al.\ 2011; Francis et al.\ 2012) 
suggests that Virgo UCDs have metallicities that are high for their luminosity according to the 
metallicity-luminosity relation defined by early-type galaxies, which is naturally expected for 
the ``tidally threshed dwarf galaxy'' scenario.

In the context of $\Lambda$CDM hierarchical structure formation, as a more centrally 
confined population, UCDs might originate from rarer density peaks in the primordial 
density field than the more spatially distributed dwarf galaxies that have been  
presumably the main contributor of blue GCs to the outer halo of M87, and those rarer 
systems should collapse and fall into the central potential earlier.\ 
It is well known that the dE,N galaxies are strongly centrally clustered 
in galaxy clusters as compared to the non-nucleated dEs (e.g.\ van den Bergh 1986; 
Ferguson \& Sandage 1989; Lisker et al.\ 2007).\ The distinct dynamical property of the UCD 
system might owe its origin to an earlier accretion of the progenitors of UCDs, and the present-day 
surviving dE,Ns do not necessarily resemble the primary UCD progenitors.\ In fact, most UCDs 
are found to be significantly older, more metal-poor and have super-solar alpha-element 
abundances compared to the majority of present-day dE nuclei.\ Future spectroscopic 
stellar population analysis of large sample of M87 UCDs will be invaluable in further shedding 
light on the difference between the UCDs, blue GCs and dE nuclei.\

\begin{acknowledgements}
HXZ acknowledges support from China Postdoctoral Science Foundation under 
Grant No.\ 552101480582.\ HXZ also acknowledges support from CAS-CONICYT 
Postdoctoral Fellowship, administered by the Chinese Academy of Sciences South 
America Center for Astronomy (CASSACA).\ EWP acknowledges support from the 
National Natural Science Foundation of China under Grant No.\ 11173003, and from 
the Strategic Priority Research Program, "The Emergence of Cosmological Structures", 
of the Chinese Academy of Sciences, Grant No.\ XDB09000105.\ EWP thanks the staff of the 
MMT and AAT observatories for their unfailingly professional support of the spectroscopic 
observations presented in this paper.\ CL acknowledges support from the National Natural 
Science Foundation of China (Grant No.\ 11203017 and 11125313).

This work is based on observations obtained with
MegaPrime/MegaCam, a joint project of CFHT and
CEA/DAPNIA, at the Canada–France–Hawaii Tele-
scope (CFHT) which is operated by the National
Research Council (NRC) of Canada, the Institut
National des Sciences de Univers of the Centre Na-
tional de la Recherche Scientifique (CNRS) of France
and the University of Hawaii. 

This research used the facilities of the Canadian Astronomy
Data Centre operated by the National Research Council of
Canada with the support of the Canadian Space Agency.\ The authors
further acknowledge use of the NASA/IPAC Extragalactic
Database (NED), which is operated by the Jet Propulsion Laboratory,
California Institute of Technology, under contract with
the National Aeronautics and Space Administration, and the
HyperLeda database (http://leda.univ-lyon1.fr).

Observations reported here were obtained at the MMT Observatory, a joint 
facility of the University of Arizona and the Smithsonian Institution. MMT telescope 
time was granted, in part, by NOAO, through the Telescope System Instrumentation 
Program (TSIP). TSIP is funded by NSF. Data presented in this paper were obtained 
at the Anglo-Australian Telescope, which is operated by the Australian Astronomical Observatory.

\end{acknowledgements}

{\it Facilities: }\facility{CFHT}, \facility{AAT/AAOmega}, \facility{MMT/Hectospec}

\appendix
\subsection{A Quantile-based Alternative for Kurtosis}
As was detailed in the main text, velocity kurtosis is closely related to the orbital structure of a given tracer population.\
However, as a fourth moment measure, kurtosis in its standard form is more sensitive to heavy tails than to a sharper peak.\ 
A given distribution can be broadly divided into five parts, i.e.\ the peak, the two shoulders (e.g.\ $\mu$ $\pm$ $\sigma$), 
and the two tails.\ The standard kurtosis measures the peakedness and tailedness of a distribution, which is more 
or less equivalent to measuring the dispersion (i.e.\ toward the peak and tails) around the shoulders, in the sense that 
a higher kurtosis indicates a larger dispersion around the two shoulders.\ Based on this interpretation, Moors (1988) 
proposed a robust quantile alternative to the standard kurtosis $G_{2}$:
\begin{equation}
T = \frac{(E_{7}-E_{5}) + (E_{3}-E_{1})}{E_{6}-E_{2}}-1.23
\label{eq7}
\end{equation}
where $E_{i}$ is the $i$-th octile.\ A normal distribution has $T$ = 0.\
Moors (1988) showed that, although there is no simple relation 
between $T$ and $G_{2}$, $T$ is an increasing function of $G_{2}$.\  

The $T$ parameter, which is defined by quantiles, is more sensitive to peakedness than the standard kurtosis.\
The velocity distribution of our UCDs provides a good example to illustrate this point.\ 
UCDs have an obviously sharper peak than a Gaussian distribution, but this feature is only reflected 
in the high $T$ value, not in the $G_{2}$ measurement.

\LongTables
\setlength{\tabcolsep}{0.035in}
\tabletypesize{\tiny}
\begin{deluxetable*}{lllllllllllll}
\tablecolumns{13}
\tablewidth{0pt}
\tablecaption{Ultra-compact Dwarfs}
\tablehead{
\colhead{ID}
& \colhead{R.A.(J2000)}
& \colhead{Decl.(J2000)}
& \colhead{\it v$_{\rm los}$}
& \colhead{\it u$^*$}
& \colhead{\it g}
& \colhead{\it r}
& \colhead{\it i}
& \colhead{\it z}
& \colhead{\it K$_{s}$}
& \colhead{\it E(B-V)}
& \colhead{{\it r}$_{\rm h,NGVS}$}
& \colhead{{\it r}$_{\rm h,HST}$}\\
\colhead{}
& \colhead{(deg)}
& \colhead{(deg)}
& \colhead{\rm (km s$^{-1}$)}
& \colhead{(mag)}
& \colhead{(mag)}
& \colhead{(mag)}
& \colhead{(mag)}
& \colhead{(mag)}
& \colhead{(mag)}
& \colhead{(mag)}
& \colhead{(pc)}
& \colhead{(pc)}\\
\colhead{(1)}
& \colhead{(2)}
& \colhead{(3)}
& \colhead{(4)}
& \colhead{(5)}
& \colhead{(6)}
& \colhead{(7)}
& \colhead{(8)}
& \colhead{(9)}
& \colhead{(10)}
& \colhead{(11)}
& \colhead{(12)}
& \colhead{(13)}
}
\startdata

\multicolumn{13}{c}{Old Sample\tablenotemark{a}}\\
\hline

   H27916  &  187.71521  &   12.23610  &   1299$\pm$10  &  22.10$\pm$0.01  &  21.05$\pm$0.00  &  20.52$\pm$0.00  &  20.38$\pm$0.00  &  20.22$\pm$0.01  &  20.48$\pm$0.01  &    0.024  &    13.5$\pm$0.2  &            13.7\\
   H30401  &  187.82795  &   12.26247  &   1323$\pm$46  &  22.54$\pm$0.01  &  21.59$\pm$0.01  &  21.06$\pm$0.00  &  20.86$\pm$0.01  &  20.74$\pm$0.01  &  20.94$\pm$0.02  &    0.022  &    11.3$\pm$0.2  &            10.7\\
   H30772  &  187.74191  &   12.26728  &    1224$\pm$9  &         \nodata  &         20.75  &         \nodata  &         19.84  &         \nodata  &         \nodata  &    0.023  &         \nodata  &             9.7\\
   H36612  &  187.48603  &   12.32538  &    1601$\pm$3  &  21.11$\pm$0.00  &  19.99$\pm$0.00  &  19.43$\pm$0.00  &  19.16$\pm$0.00  &  19.06$\pm$0.00  &  19.16$\pm$0.00  &    0.027  &    17.5$\pm$0.3  &            10.9\\
   H44905  &  187.73785  &   12.39440  &   1563$\pm$18  &  22.99$\pm$0.02  &  21.92$\pm$0.01  &  21.37$\pm$0.01  &  21.14$\pm$0.01  &  21.05$\pm$0.01  &  21.13$\pm$0.02  &    0.023  &    22.3$\pm$1.1  &            18.5\\
   H55930  &  187.63929  &   12.49845  &    1297$\pm$4  &  20.42$\pm$0.00  &  19.24$\pm$0.00  &  18.68$\pm$0.00  &  18.45$\pm$0.00  &  18.33$\pm$0.00  &  18.41$\pm$0.00  &    0.021  &    32.9$\pm$0.5  &            35.8\\
     S417  &  187.75616  &   12.32351  &    1860$\pm$2  &  21.00$\pm$0.00  &  19.65$\pm$0.00  &  19.03$\pm$0.00  &  18.70$\pm$0.00  &  18.49$\pm$0.00  &  18.40$\pm$0.00  &    0.023  &    15.0$\pm$0.2  &            14.7\\
     S477  &  187.74961  &   12.30030  &   1651$\pm$62  &  21.01$\pm$0.00  &  20.06$\pm$0.00  &  19.56$\pm$0.00  &  19.34$\pm$0.00  &  19.22$\pm$0.00  &  19.52$\pm$0.01  &    0.023  &    23.8$\pm$0.5  &            33.5\\
     S547  &  187.73910  &   12.42903  &     714$\pm$2  &  20.46$\pm$0.00  &  18.87$\pm$0.00  &  18.16$\pm$0.00  &  17.76$\pm$0.00  &  17.48$\pm$0.00  &  17.10$\pm$0.00  &    0.022  &    20.3$\pm$2.5  &            21.6\\
     S672  &  187.72804  &   12.36065  &   735$\pm$106  &  21.85$\pm$0.01  &  20.83$\pm$0.00  &  20.30$\pm$0.00  &  20.08$\pm$0.00  &  19.96$\pm$0.00  &  20.18$\pm$0.01  &    0.024  &    19.3$\pm$3.0  &            25.9\\
     S682  &  187.72775  &   12.33962  &  1333$\pm$106  &  22.21$\pm$0.01  &  21.30$\pm$0.00  &  20.81$\pm$0.00  &  20.60$\pm$0.01  &  20.50$\pm$0.01  &  20.82$\pm$0.02  &    0.024  &    20.2$\pm$0.3  &            23.7\\
     S686  &  187.72421  &   12.47187  &   817$\pm$106  &  21.53$\pm$0.00  &  20.58$\pm$0.00  &  20.05$\pm$0.00  &  19.83$\pm$0.00  &  19.70$\pm$0.00  &  19.97$\pm$0.01  &    0.021  &    16.3$\pm$0.6  &            21.2\\
     S723  &  187.72399  &   12.33940  &  1398$\pm$106  &  22.68$\pm$0.01  &  21.74$\pm$0.00  &  21.21$\pm$0.00  &  21.01$\pm$0.01  &  20.85$\pm$0.01  &  21.12$\pm$0.02  &    0.024  &         \nodata  &            16.9\\
     S731  &  187.72452  &   12.28682  &    1020$\pm$9  &  22.24$\pm$0.01  &  21.10$\pm$0.00  &  20.56$\pm$0.00  &  20.25$\pm$0.00  &  20.13$\pm$0.01  &  20.19$\pm$0.01  &    0.024  &    20.7$\pm$0.4  &            19.0\\
     S796  &  187.71563  &   12.34815  &  1163$\pm$106  &  21.84$\pm$0.01  &  20.81$\pm$0.00  &  20.28$\pm$0.00  &  20.05$\pm$0.00  &  19.92$\pm$0.00  &  20.18$\pm$0.01  &    0.024  &    11.8$\pm$0.1  &            15.3\\
     S825  &  187.71263  &   12.35542  &  1142$\pm$106  &  22.60$\pm$0.01  &  21.63$\pm$0.00  &  21.12$\pm$0.00  &  20.91$\pm$0.01  &  20.80$\pm$0.01  &  21.17$\pm$0.02  &    0.024  &    12.8$\pm$1.0  &            13.3\\
     S887  &  187.70389  &   12.36544  &  1811$\pm$106  &         \nodata  &         21.19  &         \nodata  &         20.33  &         \nodata  &         \nodata  &    0.024  &         \nodata  &             9.8\\
     S928  &  187.69875  &   12.40845  &    1284$\pm$5  &  20.81$\pm$0.00  &  19.78$\pm$0.00  &  19.26$\pm$0.00  &  19.02$\pm$0.00  &  18.90$\pm$0.00  &  19.06$\pm$0.00  &    0.023  &    26.1$\pm$0.4  &            23.0\\
     S999  &  187.69130  &   12.41709  &    1467$\pm$5  &  21.35$\pm$0.00  &  20.30$\pm$0.00  &  19.78$\pm$0.00  &  19.52$\pm$0.00  &  19.40$\pm$0.00  &  19.58$\pm$0.00  &    0.022  &    20.6$\pm$0.6  &            21.9\\
    S1201  &  187.67423  &   12.39478  &  1211$\pm$106  &  22.16$\pm$0.01  &  21.18$\pm$0.00  &  20.66$\pm$0.00  &  20.42$\pm$0.00  &  20.30$\pm$0.01  &  20.60$\pm$0.01  &    0.023  &    14.5$\pm$0.4  &            29.9\\
    S1508  &  187.63087  &   12.42356  &  2419$\pm$140  &  22.99$\pm$0.02  &  22.01$\pm$0.01  &  21.49$\pm$0.01  &  21.31$\pm$0.01  &  21.25$\pm$0.01  &  21.53$\pm$0.03  &    0.022  &    22.2$\pm$0.4  &            42.4\\
    S1629  &  187.61066  &   12.34572  &    1129$\pm$7  &  21.44$\pm$0.00  &  20.38$\pm$0.00  &  19.81$\pm$0.00  &  19.63$\pm$0.00  &  19.51$\pm$0.00  &  19.69$\pm$0.01  &    0.023  &    18.0$\pm$0.3  &            26.4\\
    S5065  &  187.70854  &   12.40248  &    1578$\pm$3  &  21.32$\pm$0.01  &  20.21$\pm$0.00  &  19.67$\pm$0.00  &  19.40$\pm$0.00  &  19.26$\pm$0.00  &  19.33$\pm$0.00  &    0.023  &    12.4$\pm$0.3  &            13.6\\
    S6004  &  187.79259  &   12.26697  &   1818$\pm$77  &  22.43$\pm$0.01  &  21.32$\pm$0.00  &  20.87$\pm$0.00  &  20.56$\pm$0.01  &  20.43$\pm$0.01  &  20.63$\pm$0.01  &    0.022  &     0.6$\pm$8.4  &            40.3\\
    S8005  &  187.69252  &   12.40641  &    1883$\pm$5  &  21.56$\pm$0.01  &  20.51$\pm$0.00  &  19.97$\pm$0.00  &  19.74$\pm$0.00  &  19.61$\pm$0.00  &  19.71$\pm$0.01  &    0.022  &    28.3$\pm$0.8  &            25.9\\
    S8006  &  187.69436  &   12.40616  &    1079$\pm$5  &  21.61$\pm$0.01  &  20.53$\pm$0.00  &  19.99$\pm$0.00  &  19.73$\pm$0.00  &  19.62$\pm$0.00  &  19.75$\pm$0.01  &    0.023  &     1.0$\pm$7.1  &            21.2\\
   T15886  &  188.15205  &   12.34920  &   1349$\pm$13  &  23.96$\pm$0.03  &  22.97$\pm$0.01  &  22.55$\pm$0.01  &  22.24$\pm$0.02  &  22.22$\pm$0.03  &  22.35$\pm$0.05  &    0.028  &    10.1$\pm$0.3  &            11.0\\
    VUCD1  &  187.53155  &   12.60861  &    1223$\pm$2  &  20.28$\pm$0.00  &  19.05$\pm$0.00  &  18.51$\pm$0.00  &  18.21$\pm$0.00  &  18.04$\pm$0.00  &  18.05$\pm$0.00  &    0.022  &    12.3$\pm$0.1  &            12.1\\
    VUCD2  &  187.70085  &   12.58636  &     919$\pm$9  &  20.29$\pm$0.00  &  19.13$\pm$0.00  &  18.57$\pm$0.00  &  18.31$\pm$0.00  &  18.17$\pm$0.00  &  18.25$\pm$0.00  &    0.021  &    11.1$\pm$0.1  &            14.1\\
    VUCD4  &  187.76865  &   11.94347  &     916$\pm$2  &  20.30$\pm$0.00  &  19.14$\pm$0.00  &  18.64$\pm$0.00  &  18.34$\pm$0.00  &  18.20$\pm$0.00  &  18.34$\pm$0.00  &    0.028  &    17.8$\pm$1.0  &            25.1\\
    VUCD5  &  187.79950  &   12.68364  &    1290$\pm$2  &  20.44$\pm$0.00  &  19.01$\pm$0.00  &  18.38$\pm$0.00  &  18.01$\pm$0.00  &  17.82$\pm$0.00  &  17.66$\pm$0.00  &    0.025  &    19.5$\pm$0.4  &            19.2\\
    VUCD6  &  187.86816  &   12.41766  &    2100$\pm$2  &  20.47$\pm$0.00  &  19.32$\pm$0.00  &  18.76$\pm$0.00  &  18.50$\pm$0.00  &  18.35$\pm$0.00  &  18.45$\pm$0.00  &    0.023  &    13.1$\pm$0.3  &            18.8\\
    VUCD7  &  187.97040  &   12.26641  &     985$\pm$3  &  19.76$\pm$0.00  &  18.49$\pm$0.00  &  17.92$\pm$0.00  &  17.58$\pm$0.00  &  17.39$\pm$0.00  &  17.38$\pm$0.00  &    0.025  &    19.6$\pm$0.4  &           100.6\\
    VUCD9  &  188.06074  &   12.05149  &   1323$\pm$12  &  20.66$\pm$0.00  &  19.45$\pm$0.00  &  18.90$\pm$0.00  &  18.57$\pm$0.00  &  18.45$\pm$0.00  &  18.48$\pm$0.00  &    0.029  &    17.5$\pm$0.4  &            25.4\\
\cutinhead{New Sample}
       F6  &  187.69749  &   12.55047  &    1341$\pm$5  &  20.71$\pm$0.00  &  19.46$\pm$0.00  &  18.86$\pm$0.00  &  18.56$\pm$0.00  &  18.40$\pm$0.00  &  18.34$\pm$0.00  &    0.021  &    17.4$\pm$0.2  &         \nodata\\
      F12  &  187.76079  &   12.57058  &   1190$\pm$14  &  20.78$\pm$0.00  &  19.79$\pm$0.00  &  19.28$\pm$0.00  &  19.05$\pm$0.00  &  18.95$\pm$0.00  &  19.18$\pm$0.01  &    0.021  &    18.2$\pm$0.2  &         \nodata\\
      F16  &  188.37279  &   12.17151  &   1230$\pm$18  &  21.60$\pm$0.00  &  20.57$\pm$0.00  &  20.05$\pm$0.00  &  19.81$\pm$0.00  &  19.70$\pm$0.00  &  19.91$\pm$0.01  &    0.034  &    12.2$\pm$0.3  &         \nodata\\
   H18539  &  187.51687  &   12.12054  &    1172$\pm$9  &  20.84$\pm$0.00  &  19.73$\pm$0.00  &  19.20$\pm$0.00  &  18.94$\pm$0.00  &  18.85$\pm$0.00  &  19.05$\pm$0.00  &    0.028  &    11.1$\pm$0.2  &         \nodata\\
   H20718  &  187.58181  &   12.15683  &    861$\pm$9  &  22.15$\pm$0.01  &  21.08$\pm$0.00  &  20.58$\pm$0.00  &  20.38$\pm$0.00  &  20.29$\pm$0.01  &  20.58$\pm$0.01  &    0.027  &    11.0$\pm$0.2  &         \nodata\\
   H24581  &  187.83334  &   12.19947  &    1283$\pm$9  &  21.61$\pm$0.00  &  20.48$\pm$0.00  &  19.97$\pm$0.00  &  19.71$\pm$0.00  &  19.59$\pm$0.00  &  19.74$\pm$0.00  &    0.023  &    11.4$\pm$0.2  &         \nodata\\
   H51655  &  187.94149  &   12.45586  &   1320$\pm$25  &  22.30$\pm$0.01  &  21.34$\pm$0.00  &  20.86$\pm$0.00  &  20.64$\pm$0.00  &  20.52$\pm$0.01  &  20.80$\pm$0.02  &    0.025  &    12.1$\pm$0.3  &         \nodata\\
   H59533  &  187.76536  &   12.53695  &     693$\pm$8  &  21.04$\pm$0.00  &  20.01$\pm$0.00  &  19.51$\pm$0.00  &  19.27$\pm$0.00  &  19.16$\pm$0.00  &  19.38$\pm$0.01  &    0.020  &    13.0$\pm$0.2  &         \nodata\\
   H65115  &  187.48801  &   12.60081  &   1491$\pm$10  &  21.27$\pm$0.00  &  20.27$\pm$0.00  &  19.79$\pm$0.00  &  19.54$\pm$0.00  &  19.40$\pm$0.00  &  19.69$\pm$0.01  &    0.022  &    12.8$\pm$0.2  &         \nodata\\
      S41  &  187.81801  &   12.31261  &   1790$\pm$31  &  22.03$\pm$0.01  &  20.94$\pm$0.00  &  20.41$\pm$0.00  &  20.14$\pm$0.00  &  20.02$\pm$0.00  &  20.14$\pm$0.01  &    0.022  &    26.2$\pm$0.4  &         \nodata\\
     S323  &  187.76845  &   12.38923  &  1157$\pm$106  &  21.65$\pm$0.01  &  20.67$\pm$0.00  &  20.19$\pm$0.00  &  19.95$\pm$0.00  &  19.83$\pm$0.00  &  20.09$\pm$0.01  &    0.023  &    33.7$\pm$0.6  &         \nodata\\
     S376  &  187.75891  &   12.46381  &  1215$\pm$106  &  21.59$\pm$0.01  &  20.65$\pm$0.00  &  20.15$\pm$0.00  &  19.92$\pm$0.00  &  19.82$\pm$0.00  &  20.09$\pm$0.01  &    0.021  &    23.4$\pm$0.3  &         \nodata\\
     S804  &  187.71277  &   12.43663  &    1137$\pm$7  &  20.95$\pm$0.00  &  19.67$\pm$0.00  &  19.07$\pm$0.00  &  18.75$\pm$0.00  &  18.56$\pm$0.00  &  18.50$\pm$0.00  &    0.022  &    13.5$\pm$0.3  &         \nodata\\
     S991  &  187.69376  &   12.33826  &   1004$\pm$75  &  21.82$\pm$0.01  &  20.79$\pm$0.00  &  20.25$\pm$0.00  &  20.00$\pm$0.00  &  19.86$\pm$0.00  &  19.98$\pm$0.01  &    0.024  &    15.5$\pm$0.3  &         \nodata\\
    S1044  &  187.68891  &   12.34263  &   2023$\pm$75  &  21.70$\pm$0.01  &  20.68$\pm$0.00  &  20.14$\pm$0.00  &  19.91$\pm$0.00  &  19.78$\pm$0.00  &  19.99$\pm$0.01  &    0.024  &    24.1$\pm$0.4  &         \nodata\\
    S1301  &  187.66316  &   12.35901  &  1086$\pm$106  &  21.75$\pm$0.01  &  20.72$\pm$0.00  &  20.11$\pm$0.00  &  19.91$\pm$0.00  &  19.74$\pm$0.00  &  19.96$\pm$0.01  &    0.023  &    13.6$\pm$0.4  &         \nodata\\
    S1449  &  187.64218  &   12.37956  &  1100$\pm$106  &  22.37$\pm$0.01  &  21.33$\pm$0.00  &  20.80$\pm$0.00  &  20.59$\pm$0.00  &  20.46$\pm$0.01  &  20.75$\pm$0.01  &    0.022  &    17.2$\pm$0.3  &         \nodata\\
    S1504  &  187.63137  &   12.43405  &    858$\pm$33  &  21.26$\pm$0.00  &  20.30$\pm$0.00  &  19.75$\pm$0.00  &  19.57$\pm$0.00  &  19.45$\pm$0.00  &  19.75$\pm$0.01  &    0.022  &    14.7$\pm$0.3  &         \nodata\\
    S1617  &  187.61207  &   12.39196  &   1407$\pm$28  &  21.06$\pm$0.00  &  19.93$\pm$0.00  &  19.35$\pm$0.00  &  19.13$\pm$0.00  &  18.98$\pm$0.00  &  19.08$\pm$0.00  &    0.022  &    16.7$\pm$0.2  &         \nodata\\
    S1631  &  187.60730  &   12.43919  &   1368$\pm$75  &  21.52$\pm$0.00  &  20.46$\pm$0.00  &  19.93$\pm$0.00  &  19.71$\pm$0.00  &  19.59$\pm$0.00  &  19.72$\pm$0.01  &    0.022  &    13.5$\pm$0.2  &         \nodata\\
    S6003  &  187.79226  &   12.27445  &   1818$\pm$77  &  22.30$\pm$0.01  &  21.21$\pm$0.00  &  20.68$\pm$0.00  &  20.40$\pm$0.00  &  20.24$\pm$0.01  &  20.42$\pm$0.01  &    0.022  &    20.3$\pm$0.4  &         \nodata\\
    S9053  &  187.70126  &   12.49469  &   829$\pm$106  &  22.53$\pm$0.01  &  21.39$\pm$0.00  &  20.89$\pm$0.00  &  20.60$\pm$0.00  &  20.56$\pm$0.01  &  20.92$\pm$0.04  &    0.021  &    30.2$\pm$0.6  &         \nodata\\
    VUCD8  &  188.01813  &   12.34176  &    1647$\pm$3  &  20.76$\pm$0.00  &  19.64$\pm$0.00  &  19.09$\pm$0.00  &  18.82$\pm$0.00  &  18.71$\pm$0.00  &  18.82$\pm$0.00  &    0.026  &    12.4$\pm$0.3  &         \nodata\\
   VUCD10  &  187.62858  &   12.31157  &   2305$\pm$23  &  20.86$\pm$0.00  &  19.73$\pm$0.00  &  19.18$\pm$0.00  &  18.94$\pm$0.00  &  18.79$\pm$0.00  &  18.90$\pm$0.00  &    0.023  &    15.4$\pm$0.4  &         \nodata\\
 M87UCD-1  &  187.83029  &   12.37554  &   1136$\pm$21  &  21.48$\pm$0.00  &  20.34$\pm$0.00  &  19.79$\pm$0.00  &  19.53$\pm$0.00  &  19.41$\pm$0.00  &  19.51$\pm$0.00  &    0.023  &    28.4$\pm$0.3  &         \nodata\\
 M87UCD-2  &  187.69858  &   12.14034  &   1288$\pm$14  &  20.77$\pm$0.00  &  19.73$\pm$0.00  &  19.25$\pm$0.00  &  19.06$\pm$0.00  &  18.90$\pm$0.00  &  19.18$\pm$0.00  &    0.026  &    23.6$\pm$0.4  &         \nodata\\
 M87UCD-3  &  187.58354  &   11.92186  &   1404$\pm$13  &  20.44$\pm$0.00  &  19.34$\pm$0.00  &  18.77$\pm$0.00  &  18.54$\pm$0.00  &  18.37$\pm$0.00  &  18.53$\pm$0.00  &    0.030  &    22.9$\pm$0.1  &         \nodata\\
 M87UCD-4  &  187.59096  &   12.40067  &   1279$\pm$11  &  21.33$\pm$0.00  &  19.93$\pm$0.00  &  19.29$\pm$0.00  &  18.96$\pm$0.00  &  18.75$\pm$0.00  &  18.62$\pm$0.00  &    0.023  &    11.7$\pm$0.4  &         \nodata\\
 M87UCD-5  &  187.41829  &   12.45780  &   1400$\pm$27  &  21.08$\pm$0.00  &  19.96$\pm$0.00  &  19.42$\pm$0.00  &  19.18$\pm$0.00  &  19.08$\pm$0.00  &  19.30$\pm$0.00  &    0.023  &    21.0$\pm$0.2  &         \nodata\\
 M87UCD-6  &  187.51025  &   12.60938  &   1514$\pm$27  &  21.34$\pm$0.00  &  20.21$\pm$0.00  &  19.69$\pm$0.00  &  19.42$\pm$0.00  &  19.27$\pm$0.00  &  19.37$\pm$0.01  &    0.022  &    15.0$\pm$0.1  &         \nodata\\
 M87UCD-7  &  187.52525  &   12.66499  &   1160$\pm$12  &  20.65$\pm$0.00  &  19.48$\pm$0.00  &  18.95$\pm$0.00  &  18.67$\pm$0.00  &  18.51$\pm$0.00  &  18.58$\pm$0.00  &    0.021  &    17.4$\pm$0.1  &         \nodata\\
 M87UCD-8  &  187.43858  &   12.85025  &   1154$\pm$18  &  21.60$\pm$0.00  &  20.51$\pm$0.00  &  19.99$\pm$0.00  &  19.75$\pm$0.00  &  19.63$\pm$0.00  &  19.70$\pm$0.01  &    0.020  &    12.1$\pm$0.1  &         \nodata\\
 M87UCD-9  &  187.59571  &   12.53021  &    556$\pm$12  &  20.87$\pm$0.00  &  19.75$\pm$0.00  &  19.22$\pm$0.00  &  18.99$\pm$0.00  &  18.89$\pm$0.00  &  19.06$\pm$0.01  &    0.022  &    17.9$\pm$0.1  &         \nodata\\
M87UCD-10  &  187.50812  &   12.70747  &   1178$\pm$30  &  20.82$\pm$0.00  &  19.86$\pm$0.00  &  19.41$\pm$0.00  &  19.17$\pm$0.00  &  19.12$\pm$0.00  &  19.35$\pm$0.00  &    0.022  &    21.9$\pm$0.3  &         \nodata\\
M87UCD-11  &  187.62463  &   12.63949  &    845$\pm$30  &  21.42$\pm$0.00  &  20.47$\pm$0.00  &  20.00$\pm$0.00  &  19.79$\pm$0.00  &  19.72$\pm$0.00  &  20.02$\pm$0.02  &    0.021  &    19.0$\pm$0.6  &         \nodata\\
M87UCD-12  &  187.63100  &   12.86572  &   1497$\pm$10  &  21.51$\pm$0.00  &  20.13$\pm$0.00  &  19.48$\pm$0.00  &  19.17$\pm$0.00  &  19.01$\pm$0.00  &  18.85$\pm$0.00  &    0.021  &    12.4$\pm$0.5  &         \nodata\\
M87UCD-13  &  187.70533  &   12.62171  &    1605$\pm$7  &  21.07$\pm$0.00  &  19.95$\pm$0.00  &  19.38$\pm$0.00  &  19.12$\pm$0.00  &  18.98$\pm$0.00  &  19.01$\pm$0.01  &    0.021  &    12.8$\pm$0.2  &         \nodata\\
M87UCD-14  &  187.76812  &   13.17849  &    1347$\pm$7  &  21.64$\pm$0.01  &  20.26$\pm$0.00  &  19.67$\pm$0.00  &  19.31$\pm$0.00  &  19.13$\pm$0.00  &  19.03$\pm$0.01  &    0.025  &    11.2$\pm$0.3  &         \nodata\\
M87UCD-15  &  187.99162  &   13.25948  &   1261$\pm$12  &  20.71$\pm$0.00  &  19.66$\pm$0.00  &  19.15$\pm$0.00  &  18.90$\pm$0.00  &  18.83$\pm$0.00  &  19.08$\pm$0.01  &    0.027  &    40.2$\pm$0.4  &         \nodata\\
M87UCD-16  &  187.88208  &   12.69177  &   1119$\pm$12  &  20.65$\pm$0.00  &  19.49$\pm$0.00  &  18.89$\pm$0.00  &  18.62$\pm$0.00  &  18.46$\pm$0.00  &  18.62$\pm$0.00  &    0.025  &    38.1$\pm$0.2  &         \nodata\\
M87UCD-17  &  187.89638  &   12.58316  &     950$\pm$7  &  21.40$\pm$0.00  &  20.10$\pm$0.00  &  19.47$\pm$0.00  &  19.19$\pm$0.00  &  19.00$\pm$0.00  &  19.04$\pm$0.00  &    0.024  &    11.9$\pm$0.3  &         \nodata\\
M87UCD-18  &  187.85313  &   12.42384  &    1780$\pm$8  &  20.85$\pm$0.00  &  19.80$\pm$0.00  &  19.27$\pm$0.00  &  19.04$\pm$0.00  &  18.90$\pm$0.00  &  19.09$\pm$0.00  &    0.022  &    13.2$\pm$0.1  &         \nodata\\
M87UCD-19  &  187.99000  &   12.82488  &   1086$\pm$15  &  21.38$\pm$0.00  &  20.31$\pm$0.00  &  19.81$\pm$0.00  &  19.55$\pm$0.00  &  19.48$\pm$0.00  &  19.71$\pm$0.01  &    0.023  &    18.9$\pm$0.2  &         \nodata\\
M87UCD-20  &  187.42217  &   12.66457  &  1754$\pm$105  &  21.43$\pm$0.00  &  20.25$\pm$0.00  &  19.73$\pm$0.00  &  19.45$\pm$0.00  &  19.30$\pm$0.00  &  19.42$\pm$0.01  &    0.021  &    16.7$\pm$0.3  &         \nodata\\
M87UCD-21  &  187.90599  &   12.29465  &   1484$\pm$28  &  21.59$\pm$0.00  &  20.60$\pm$0.00  &  20.16$\pm$0.00  &  19.86$\pm$0.00  &  19.77$\pm$0.00  &  20.02$\pm$0.01  &    0.023  &    11.4$\pm$0.3  &         \nodata\\
M87UCD-22  &  187.46783  &   12.62716  &    905$\pm$20  &  22.09$\pm$0.01  &  20.80$\pm$0.00  &  20.21$\pm$0.00  &  19.92$\pm$0.00  &  19.72$\pm$0.00  &  19.72$\pm$0.01  &    0.021  &    11.7$\pm$0.2  &         \nodata\\
M87UCD-23  &  187.42317  &   12.74100  &   1314$\pm$23  &  22.13$\pm$0.01  &  20.95$\pm$0.00  &  20.38$\pm$0.00  &  20.10$\pm$0.00  &  19.95$\pm$0.00  &  20.04$\pm$0.01  &    0.022  &    11.1$\pm$0.2  &         \nodata\\
M87UCD-24  &  187.29442  &   12.71781  &   1224$\pm$34  &  22.01$\pm$0.01  &  20.88$\pm$0.00  &  20.32$\pm$0.00  &  20.07$\pm$0.00  &  19.97$\pm$0.01  &  20.01$\pm$0.01  &    0.022  &    12.3$\pm$0.2  &         \nodata\\
M87UCD-25  &  187.91054  &   11.98996  &   1264$\pm$26  &  21.88$\pm$0.01  &  20.78$\pm$0.00  &  20.26$\pm$0.00  &  20.01$\pm$0.00  &  19.89$\pm$0.00  &  20.13$\pm$0.01  &    0.029  &    19.4$\pm$0.4  &         \nodata\\
M87UCD-26  &  187.56892  &   12.26579  &   2030$\pm$18  &  21.05$\pm$0.00  &  20.04$\pm$0.00  &  19.55$\pm$0.00  &  19.30$\pm$0.00  &  19.17$\pm$0.00  &  19.39$\pm$0.00  &    0.023  &    17.3$\pm$0.3  &         \nodata\\
M87UCD-27  &  188.10566  &   12.33246  &   1272$\pm$23  &  21.56$\pm$0.00  &  20.52$\pm$0.00  &  20.01$\pm$0.00  &  19.76$\pm$0.00  &  19.69$\pm$0.00  &  19.89$\pm$0.01  &    0.027  &    11.7$\pm$0.3  &         \nodata\\
M87UCD-28  &  187.52553  &   12.40982  &   1870$\pm$29  &  21.78$\pm$0.00  &  20.59$\pm$0.00  &  20.03$\pm$0.00  &  19.74$\pm$0.00  &  19.62$\pm$0.00  &  19.68$\pm$0.01  &    0.026  &    12.9$\pm$0.2  &         \nodata\\
M87UCD-29  &  187.02759  &   12.41012  &    599$\pm$33  &  21.97$\pm$0.01  &  20.94$\pm$0.00  &  20.42$\pm$0.00  &  20.21$\pm$0.00  &  20.11$\pm$0.01  &  20.32$\pm$0.01  &    0.028  &    12.6$\pm$0.3  &         \nodata\\
M87UCD-30  &  187.15505  &   12.47934  &   1534$\pm$28  &  22.06$\pm$0.01  &  21.01$\pm$0.00  &  20.48$\pm$0.00  &  20.25$\pm$0.00  &  20.14$\pm$0.01  &  20.31$\pm$0.02  &    0.024  &    12.6$\pm$0.3  &         \nodata\\
M87UCD-31  &  187.73054  &   12.41109  &   1301$\pm$30  &  22.03$\pm$0.01  &  20.95$\pm$0.00  &  20.40$\pm$0.00  &  20.16$\pm$0.00  &  20.01$\pm$0.01  &  20.08$\pm$0.01  &    0.023  &     9.7$\pm$0.1  &            10.4\\
M87UCD-32  &  187.85521  &   12.32549  &   1632$\pm$34  &  21.21$\pm$0.00  &  20.22$\pm$0.00  &  19.73$\pm$0.00  &  19.48$\pm$0.00  &  19.39$\pm$0.00  &  19.65$\pm$0.00  &    0.023  &    33.2$\pm$0.5  &         \nodata\\
M87UCD-33  &  188.06020  &   12.03307  &   1833$\pm$22  &  21.81$\pm$0.01  &  20.65$\pm$0.00  &  20.09$\pm$0.00  &  19.79$\pm$0.00  &  19.66$\pm$0.00  &  19.70$\pm$0.01  &    0.029  &    16.3$\pm$0.4  &         \nodata\\
M87UCD-34  &  187.31196  &   11.89551  &    905$\pm$19  &  21.25$\pm$0.00  &  20.19$\pm$0.00  &  19.65$\pm$0.00  &  19.40$\pm$0.00  &  19.27$\pm$0.00  &  19.45$\pm$0.00  &    0.027  &    11.7$\pm$0.3  &         \nodata\\
M87UCD-35  &  188.09787  &   11.96527  &   1007$\pm$16  &  21.36$\pm$0.00  &  20.19$\pm$0.00  &  19.57$\pm$0.00  &  19.34$\pm$0.00  &  19.19$\pm$0.00  &  19.23$\pm$0.00  &    0.032  &    11.5$\pm$0.3  &         \nodata\\
M87UCD-36  &  187.79709  &   12.50030  &   1207$\pm$21  &  22.03$\pm$0.01  &  20.76$\pm$0.00  &  20.18$\pm$0.00  &  19.88$\pm$0.00  &  19.74$\pm$0.00  &  19.79$\pm$0.01  &    0.020  &    12.1$\pm$0.4  &         \nodata\\
M87UCD-37  &  187.54091  &   12.62679  &   1324$\pm$38  &  22.15$\pm$0.01  &  21.04$\pm$0.00  &  20.52$\pm$0.00  &  20.26$\pm$0.00  &  20.12$\pm$0.01  &  20.20$\pm$0.01  &    0.022  &    11.3$\pm$0.1  &         \nodata\\
M87UCD-38  &  187.62717  &   12.67106  &   1154$\pm$18  &  21.31$\pm$0.00  &  20.07$\pm$0.00  &  19.49$\pm$0.00  &  19.20$\pm$0.00  &  19.09$\pm$0.00  &  19.12$\pm$0.01  &    0.022  &    22.1$\pm$0.3  &         \nodata\\
M87UCD-39  &  188.44296  &   11.95064  &   1351$\pm$23  &  22.32$\pm$0.01  &  21.00$\pm$0.00  &  20.38$\pm$0.00  &  20.03$\pm$0.00  &  19.85$\pm$0.01  &  19.77$\pm$0.01  &    0.036  &    11.2$\pm$0.2  &         \nodata\\
\enddata
\tablecomments{
Col.(1):\ Object ID;
Col.(2):\ Right ascension in decimal degrees (J2000);
Col.(3):\ Declination in decimal degrees (J2000);
Col.(4):\ Heliocentric radial velocity;
Cols.(5--9):\ MegaCam $u^*griz$ five-band 3\arcsec-aperture (in diameter) AB magnitudes   (not corrected for Galactic extinction);
Col.(10):\ WIRCam $K_{s}$-band 3\arcsec-aperture (in diameter) AB magnitude    (not corrected for Galactic extinction);
Col.(11):\ The Galactic reddening determined by Schlegel et al.\ (1998);
Col.(12):\ Half-light radius (in units of pc) measured on NGVS images;
Col.(13):\ Half-light radius (in units of pc) measured on HST images.
}
\tablenotetext{a}{The UCDs have been compiled by Brodie et al.\ (2011).\ 
In addition, $g$ and $i$ band photometry for H30772 and S887 is from Brodie et al.\ (2011)}

\label{ucd_table}
\end{deluxetable*}

\end{document}